\providecommand{\forArxiv}[1]{}
\providecommand{\forACM}[1]{#1}
\providecommand{\forThesis}[1]{}
\providecommand{\notforThesis}[1]{#1}

\documentclass[format=acmsmall, review=false, screen=true]{acmart}

\makeatletter                   %1
\def\mdseries@tt{m}             %1
\makeatother                    %1

\pdfoutput=1

\renewcommand\footnotetextcopyrightpermission[1]{} % removes footnote with conference information in first column
\newcommand{\mypar}[1]{\shortonly{\vspace{-1ex}}\vspace{1.5ex}\noindent{\bf #1.~}}
\newenvironment{myeqn}{\shortonly{\vspace{-1ex}}\begin{equation}\shortonly{\small\setlength{\lineskip}{-1ex}}}{\end{equation}}

\newcommand{\tlaplus}{TLA\textsuperscript{+}}
\newcommand{\msg}[1]{\ifmmode\text{\textcolor{blue}{\textsf{#1}}}\else\textcolor{blue}{\textsf{#1}}\fi}
\newcommand{\msgtype}[1]{\ifmmode\text{\textcolor{blue}{\textsf{``#1''}}}\else\textcolor{blue}{\textsf{``#1''}}\fi}
\newcommand{\figref}[1]{Fig.~\ref{#1}}
\newcommand{\notes}[1]{}
\newcommand{\fullonly}[1]{#1}
\newcommand{\shortonly}[1]{}
\newif\iftmi
\tmifalse
\usepackage{amsmath}
\usepackage{tla2}
\usepackage{amssymb}
\usepackage{multirow}

\usepackage{dsfont}
\usepackage{xcolor}
\usepackage[T1]{fontenc}
\usepackage{letltxmacro}
\usepackage{scalefnt}
\usepackage{latexsym}
\usepackage{ifthen}
\usepackage{verbatim}
\usepackage{xfrac}
\usepackage{url}
\usepackage{pdfpages}
\usepackage{enumitem}
\usepackage{makecell}
\ifx\documentclass\undefined
\else
  \DeclareSymbolFont{tlaitalics}{\encodingdefault}{cmr}{m}{it}
  \let\itfam\symtlaitalics
\fi
\begin{document}
%\tracingall

\title{Formal Verification of Multi-Paxos for Distributed Consensus}%
\author{Saksham Chand}
\affiliation{%
	\department{Computer Science Department}
	\institution{Stony Brook University}
	\city{Stony Brook}
	\state{New York}
	\postcode{11794}
	\country{USA}
}
\email{schand@cs.stonybrook.edu}

\author{Yanhong A. Liu}
\affiliation{%
	\department{Computer Science Department}
	\institution{Stony Brook University}
	\city{Stony Brook}
	\state{New York}
	\postcode{11794}
	\country{USA}
}
\email{liu@cs.stonybrook.edu}

\author{Scott D. Stoller}
\affiliation{%
	\department{Computer Science Department}
	\institution{Stony Brook University}
	\city{Stony Brook}
	\state{New York}
	\postcode{11794}
	\country{USA}
}
\email{stoller@cs.stonybrook.edu}

\begin{abstract}
Paxos is an important algorithm for a set of distributed processes to agree on a single value or a sequence of values, for which it is called Basic Paxos or Multi-Paxos, respectively. 
Consensus is critical when distributed services are replicated for fault-tolerance, because non-faulty replicas must agree on the state of the system or the sequence of operations that have been performed.
Unfortunately, consensus algorithms including Multi-Paxos in particular are well-known to be difficult to understand, and their accurate specifications and correctness proofs remain challenging, despite extensive studies ever since Lamport introduced Paxos.

This article describes formal specification and verification of Lamport's
Multi-Paxos algorithm 
for distributed consensus.  The specification is written in
\tlaplus{}, Lamport's Temporal Logic of Actions.  The proof is written and
automatically checked using TLAPS, the \tlaplus{} Proof System. The proof is for the safety property of the algorithm. Building on Lamport, Merz, and Doligez's specification and proof for Basic Paxos, we aim to 
facilitate the
understanding of Multi-Paxos and its proof by minimizing the difference
from those for Basic Paxos, and to demonstrate a general way of proving
other variants of Paxos and other sophisticated distributed algorithms.
We also discuss our general strategies and results for proving complex invariants using invariance lemmas and increments, for proving properties about sets and tuples to help the proof check succeed in significantly reduced time, and for overall proof improvement leading to considerably reduced proof size. 
\end{abstract}

%\keywords{Distributed Algorithms, Formal Methods, Verification}

%\acmJournal{TOCL}
%} % NFT: Title, authors and abstract

% Cannot put CCSXML into a command for some reason MANUAL


%\notforThesis{% Introduction, Preliminaries
\maketitle
\thispagestyle{empty}
% old: We present a formal specification and machine-checked proof of safety of Lamport's Multi-Paxos for distributed consensus using TLAPS, a proof system for TLA. The proof extends earlier work for Basic Paxos by adding slots in the specification. We then add preemption as an explicit phase in the specification and machine-check the expanded specification.

\section{Introduction}
\label{sec-intro}
%\section{Introduction}
%\label{sec-intro}

Distributed consensus is a fundamental problem in distributed systems, which are increasingly important in today's interconnected world.
% systems and services.
Distributed consensus requires that a set of processes agree on
%a value or a continual sequence of values
some values proposed by some processes.
% , for example, ...
It is essential when distributed services are replicated for
fault-tolerance, because non-faulty replicas must agree.  Examples include %leader election, atomic broadcast, and 
%state machine replication in 
replicated data storage services like Google File System~\cite{ghemawat2003gfs}, Apache ZooKeeper~\cite{hunt2010zookeeper}, Amazon DynamoDB~\cite{dynamodb2019}, etc.  Unfortunately, consensus is difficult when processes or communication channels may fail.  %todo: add refs, sc: done

%From wiki:
%Examples of applications of consensus include whether to commit a transaction to a database, agreeing on the identity of a leader, state machine replication, and atomic broadcasts. The real world applications include clock synchronization, PageRank, opinion formation, smart power grids, state estimation, control of UAVs (and multiple robots/agents in general), load balancing and others.
%Some real-world applications include Supervisory Control And Data Acquisition (SCADA) systems which underlie many critical services like electricity transmission and distribution, water treatment and traffic control. Due to the presence of failures, these systems are replicated and therefore require consensus to ensure correct operation~\cite{kirsch2013survivable}. Recently, consensus-based approaches have been successfully investigated for the rendezvous problem and the formation problem in Unmanned Aerial Vehicles (UAVs)~\cite{jamshidi2011cyber}. Consensus is required for leader election, atomic broadcast, and state machine replication by all replicated data storage services including Google File System, Apache ZooKeeper, Amazon DynamoDB, etc.
%

Paxos~\cite{lamport1998part} is an important algorithm, developed by Lamport, 
for solving distributed consensus.
Basic Paxos is for agreeing on a single value, % among the processes
such as whether to commit a database transaction.
% , or who is the leader process.
Multi-Paxos is for agreeing on a continuing sequence of values, for example,
a stream of commands to execute.
Multi-Paxos has been used in many important distributed services, for example,
Google's Chubby~\cite{burrows2006chubby,chandra2007paxos} and Microsoft's
Autopilot~\cite{isard2007autopilot}.
%said used in MS Live Search in paper, %said used in Bing in van14vive 
%
There are other Paxos variants, %as extensions or optimizations,
for example, variants that reduce a message delay~\cite{lamport2006fast} or add
preemption~\cite{lamport2001paxos}, but Multi-Paxos is the most important in
making Paxos practical for distributed services that must execute a continuing sequence of operations.

%can remove this paragraph from intro if there space is tight:
%Even agreeing on a single value is challenging, 
Paxos handles processes that run concurrently without shared memory,
where processes may crash and may later recover, and
messages may be lost, delayed, reordered, or duplicated.
% by proposing, possibly repeatedly, some value to agree on, 
% and waiting, possibly indefinitely, for replies from other processes. 
In Basic Paxos, each process may repeatedly %attempt to be the leader and
%, by soliciting votes, 
propose some value, and wait for appropriate replies from appropriate
subsets of the processes while also replying appropriately to other
processes; consensus is reached if eventually enough processes and channels
are non-faulty to vote on some proposal thus agreeing on the proposed value.
In Multi-Paxos, many more different attempts, proposals, and replies may
happen in overlapping fashions to reach consensus on values in different
slots in the continuing sequence.

% Despite well-known significant effort in understanding and verifying
% Paxos, especially most recently various works given various
% specifications of various variants of Paxos in various different
% languages and systems, no

Paxos % and its correctness
has often been difficult to understand since it was first introduced~\cite{lamport1998part}.  
%created in the late 1980s~\cite{lamport2016writings}.
Lamport later wrote a simpler
description of the phases of the algorithm but only for Basic
Paxos~\cite{lamport2001paxos}.  
Lamport et al.~\cite{basicpaxos2014} wrote a formal
specification of Basic Paxos in \tlaplus{}, Lamport's Temporal Logic of Actions~\cite{lamport2002specifying},
and a proof of the safety property of the algorithm in TLAPS, the \tlaplus{} Proof System~\cite{tlaps}.
Many efforts, especially in recent years, have been spent on formal
specification and verification of Multi-Paxos, %e.g.,~\cite{}
but they use more restricted or less direct language models,
% e.g., psync (psync did multi-paxos?) or dafny
some with reformulated algorithms, and 
some mixed in large systems, % with many unrelated functionalities, 
% e.g., iron fleet
% or handle other variants of Paxos than Multi-Paxos, 
as discussed
in Section~\ref{sec-related}.
% e.g., fast paxos
What is lacking is formal specification and proof of the exact phases of
Multi-Paxos, in a most direct way in a general language like
\tlaplus{}~\cite{lamport2002specifying}, with a complete proof that is mechanically checked,
%like Lamport, Merz, and Doligez did for Basic Paxos.
and a general method for doing such specifications and proofs in a more
feasible way.

This article addresses this challenge.  We describe a formal specification of
Multi-Paxos written in \tlaplus{}, % Lamport's temporal logic for actions,
and a complete proof written and automatically checked using TLAPS.
%a proof system for TLA\textsuperscript{+}.
Building on Lamport et al.'s specification and proof for Basic
Paxos, we aim to facilitate the understanding of Multi-Paxos and its proof
by minimizing the difference from those for Basic Paxos.
The key change in the specification is to replace
% , for each of a set of processes, % acceptors
operations involving two numbers with those involving a set of 3-tuples,
for each process, %of a set of processes, 
exactly capturing the minimum conceptual
difference between Basic Paxos and Multi-Paxos.  However, the proof becomes
significantly more difficult because of the handling of sets and tuples in
place of two numbers.

This work also aims to show the minimum-change approach as a general way of
specifying and verifying other variants of Paxos, and more generally of
specifying and verifying other sophisticated algorithms by starting from the
basics.  We demonstrate this by further showing the extension of the
specification and proof of Multi-Paxos to Multi-Paxos with Preemption---letting
processes abandon proposals that are already preempted by other
proposals~\cite{lamport2001paxos,van2015paxos}.  We also extended the
specification and proof of Basic Paxos with Preemption, which is
even easier. Our specifications and proofs can be found online at
\url{github.com/DistAlgo/proofs}. %todo: fix on git. sc: done

Additionally, we discuss a general method that we followed to prove complex invariants using invariance lemmas and increments, to tackle
tedious and difficult proof obligations involving sets and tuples,
instead of just scalars in the proof of Basic Paxos, and to perform overall proof improvement leading to considerably reduced proof size.
%a well-known significant complication in general.
%
%say something general/systematic about doing induction... 
For difficult properties involving unbounded sets, we use induction and direct the prover to focus on the increments to the set. %todo: but said limited to finite for remedial proof.
%
%anything about proof by contradiction?...
For properties involving tuples, we change the ways of accessing and testing the elements to yield significantly reduced proof-checking time.
%
%We summarize the results of our specifications and proofs at the end.
Overall, we were able to keep the specification minimally changed for adding slots for Multi-Paxos and adding preemption, and keep
the proof-checking time to about 3 minutes for both specifications while the prover checks the proofs for over 750 obligations for Multi-Paxos and over 800 obligations for Multi-Paxos with Preemption.
%summarize some of the results: 
%sizes of specs and proofs for Basic Paxos
%and Multi-Paxos; number of defs, lemmas, inductions, proofs by
%contradiction for Basic and Multi; number of these items on sequences and
%tuples; number of seconds to check automatically...

This article is a corrected, improved, and extended version of~\cite{chand2016formal}. The main changes are as follows. 
\begin{enumerate}

%annie: also updated the results table and description.

\item The claim of a complete proof in~\cite{chand2016formal} was incorrect, and the problem discovered is now fixed.  The problem was due to an undocumented bug~\cite{tlapsbug} in TLAPS that we discovered after the work in~\cite{chand2016formal}, which made us realize that the proof of Multi-Paxos with Preemption was incomplete.
    % annie: just incomplete?  or something else wrong besides being incomplete?
    % sc: i wrote incorrect because (1) the proof that TLAPS generated for the theorem was incorrect, and (2) I wanted to support the bullet name.
    We added the missing proof.  This is described in the new
    Section~\ref{secMadeupProof}.
    
\item The proofs are improved throughout, shortened by 19\% for Multi-Paxos and 17\% for Multi-Paxos with Preemption, even with about 4-6\% added %annie new: lines or perhaps better percentages sc: done
proof to overcome the TLAPS bug. In fact, it was during improvement of the proof that we discovered the bug. The improvements are described in the new Section~\ref{sec-improve}. %, and the results are improved  and updated in Section~\ref{secResults}.
    % annie: what does "cleaned and stremlined" mean?  refined?  simplfied?
    % annie: where in paper body is this simplifiation described?
    % annie: details about the simplification (including changes in line numbers) should be described in paper body, and here we just say something like "simplified, shortened about 20%, as described in ...". 
    %
    % annie: Our simplification also allowed us to discover ... bug. ...
An overall proof summary is given in the new Section~\ref{secProofSummary}.

\item Sections~\ref{secAuxiliary},~\ref{secAccInv}, and~\ref{secMsgInv} are extended to define and explain all auxiliary predicates, process invariants, and message invariants, respectively, that are used in the proofs. Sections~\ref{sec-strategy} to~\ref{sec-setstuples} on proof strategies
are newly structured, and Section~\ref{sec-invinc} is extensively revised and simplified.  The new Section~\ref{sec-valid} discusses specifying and proving a validity condition.

\item Section~\ref{sec-pre} is extended with the new Sections~\ref{secTLA+} and~\ref{secTLAPS} on \tlaplus{} and TLAPS, respectively.
Section~\ref{secSpec} is extended with the new Section~\ref{sec-faultimpl} on fault model and implementation issues. % after the complete algorithm specification. 
Section~\ref{secResults} is extended with detailed results about the different versions of proofs.
    \iffalse
    %annie: move this to body
    In particular, the significance of predicate $SafeAt$ has been highlighted in Section~\ref{secAuxiliary}. The invariant about this predicate is critical in proving safety of Multi-Paxos and in fact most existing consensus algorithms. This added explanation helps understand Paxos better as it exposes the crux of the protocol.
    \fi
    Section~\ref{sec-related} is expanded with additional related works and details about other proofs.
    %the proof size and checking time for IronFleet~\cite{hawblitzel2015ironfleet} and Raft in Verdi~\cite{wilcox2015verdi}, and with new work~\cite{padon2017paxos} on specification and verification of Paxos.
    %% annie: revised Sections A, B, and C on ..., added formal presentations of the most important invariants, A, B, and C in Sec ..., especially including explanations about the significance of SafeAt in Section ...
    
    % annie: do not say "hope..." here or anywhere except perhaps for future work.
    % "we hope that A happens" -> we did ... to make A happen (or to help make A happen)"
    % sc: done. I initially wrote hope in that sentence because I can only hope that the reader actually gets helped by the description.
    
    % annie: no other changes?  what about added related work?
    % check all changes and make sure that all are covered.

\item The complete resulting %, revised, and improved 
\tlaplus{} specification and TLAPS-checked proof of Multi-Paxos with Preemption are given in the new Appendices \ref{appendix:spec} to \ref{appendix:proof}, including about 1.5 pages of specification, 1 page of invariants, and 8.5 pages of proof, including comments. 
%annie new: reduce line spacing in appendix.  would need to update page numbers.
%sc - i didn't get this. You mean remove some empty lines?

%annie new: maybe those "OBVIOUS"s do not need a new line.  saving say 20 lines.  would need to update lines numbers.
% sc - did that and also moved some other things to fit it in page numbers 19-32.
% sc - i think we can leave the line count numbers in the results unchanged because otherwise, I should also do the same change(s) in Lamport's proof for a fair comparison, right?

\end{enumerate}

The rest of the article is organized as follows. Section~\ref{sec-pre} covers preliminaries: distributed consensus, % (Section~\ref{secDistCons}),
Paxos, % (Section~\ref{secPaxosEng}), 
\tlaplus{}, % (Section~\ref{secTLA+}), 
and TLAPS. % (Section~\ref{secTLAPS}). 
Section~\ref{secSpec} presents the \tlaplus{} specification of Multi-Paxos and compares it with Lamport et al.'s specification of Basic Paxos~\cite{basicpaxos2014}. Section~\ref{secProve} presents 
%the auxiliary predicates and functions used, % (Section~\ref{secAuxiliary}), 
the invariants used and proved, % (Sections~\ref{secTypeInv} to~\ref{secMsgInv}), 
and our general proof strategies. % (Sections~\ref{sec-strategy} to~\ref{sec-setstuples}). 
Section~\ref{secPreemption} describes the changes made for specifying and verifying Multi-Paxos with Preemption and the overall proof improvements. Section~\ref{secResults} summarizes the results from our specifications and proofs. Section~\ref{sec-related} discusses related work and concludes. Appendices~\ref{appendix:spec} %, ~\ref{appendix:prop}, and~
 to \ref{appendix:proof} contain the complete resulting specification, invariants, and proof %, respectively, 
for Multi-Paxos with Preemption.

\section{Preliminaries}
\label{sec-pre}

\subsection{Distributed consensus}
\label{secDistCons}

A distributed system is a set of processes that process data locally and communicate with each other by sending and receiving messages. 
The processes may crash and may later recover, and
the messages may be lost, delayed, recordered, and duplicated.

%\mypar{Single-value consensus}
The basic consensus problem, called single-value consensus, is for a set of processes to agree on a single value.  An algorithm for single-value consensus is said to be safe if it satisfies the following conditions~\cite{lamport2001paxos}:
\begin{enumerate}
    \item[C1.] Only a value that has been proposed may be chosen,
    \item[C2.] Only a single value is chosen, and
    \item[C3.] A process never learns that a value has been chosen unless it actually has been.
\end{enumerate}
%The basic consensus problem, called single-value consensus, is to ensure that at most a single value is chosen from among the values proposed by the processes.
Conditions C1 and C3, called validity condition and learner condition, respectively,
are straightforward and easy to prove.  We do not include them in the formal proof, following Lamport et al.~\cite{basicpaxos2014}.  For completeness, we discuss C3 in Section~\ref{sec-faultimpl} and C1 in Section~\ref{sec-valid}.

The specifications and proofs presented in this article focus on C2 because it is the central and really challenging of the three conditions. 
C2 is formally defined as
\begin{myeqn}
Safe_{basic} == \A v1, v2 \in \mathcal{V} : Chosen(v1) /\ Chosen(v2) => v1 = v2
\end{myeqn}%
where $\mathcal{V}$ is the set of possible proposed values, and
$Chosen$ is a predicate that given a value $v$ evaluates to true iff $v$ was chosen by the algorithm. The specification of $Chosen$ is part of the algorithm.

%\mypar{Multi-value consensus}
The more general consensus problem, called multi-value consensus, is to agree on a sequence of values, instead of a single value. Here we have
\begin{myeqn}\label{eqnsafe}
Safe == \A v1, v2 \in \mathcal{V}, s \in \mathcal{S} : Chosen(s, v1) /\ Chosen(s, v2) => v1 = v2
\end{myeqn}%
where $\mathcal{V}$ is as above, $\mathcal{S}$ is a set of \textit{slots} used to index the sequence of chosen values, and
$Chosen(s,v)$ is true iff for slot $s$, value $v$ was chosen by the algorithm.

\subsection{Basic Paxos and Multi-Paxos}
\label{secPaxosEng}
Paxos solves the problem of consensus. Three main roles of the algorithm are performed by three kinds of processes:
\begin{itemize}

\item $\mathcal{P}$, the set of proposers that propose values that can be chosen.

\item $\mathcal{A}$, the set of acceptors that vote for proposed values.  A value is chosen when there are enough votes for it.

\item $\mathcal{L}$, the set of learners that learn chosen values. A learner learns a value when it receives enough votes for it.
\end{itemize}
These roles can be co-located, that is, a single process can take on more than one role.

A set $\mathcal{Q}$ of subsets of the acceptors, that is, $\mathcal{Q} \subseteq 2^{\mathcal{A}}$, is used as a quorum system.  It must satisfy the property that any two quorums in $\mathcal{Q}$ overlap, that is, $\forall Q1, Q2 \in \mathcal{Q} : Q1 \cap Q2 \neq \emptyset$. The most commonly used quorum system $\mathcal{Q}$ takes any majority of acceptors as an element in $\mathcal{Q}$.

Basic Paxos solves the problem of single-value consensus.  It defines predicate $Chosen$ as
\begin{myeqn}
Chosen(v) == \E Q \in \mathcal{Q} : \A a \in Q : \E b \in \mathcal{B} : sent(\textcolor{blue}{"2b"}, a, b, v)
\end{myeqn}%
where $\mathcal{B}$ is the set of proposal numbers, also called ballot numbers, which is any set that can be totally ordered. $sent(\textcolor{blue}{"2b"}, a, b, v)$ means that a message of type $\textcolor{blue}{\textsf{2b}}$ with ballot number $b$ and value $v$ was sent by acceptor $a$.  An acceptor votes (for value $v$) by sending such a message.

Multi-Paxos solves the problem of multi-value consensus.  It extends predicate $Chosen$ to decide a value for each slot $s$ in $\mathcal{S}$:
\begin{myeqn}\label{voting}
Chosen(s, v) == \E Q \in \mathcal{Q} : \A a \in Q : \E b \in \mathcal{B} : sent(\textcolor{blue}{"2b"}, a, b, s, v)
\end{myeqn}%
To satisfy the $Safe$ property, $\mathcal{S}$ can be any set.  In practice, $\mathcal{S}$ is usually the set of natural numbers.
%To make sense in practice, however, one would want $\mathcal{S}$ to be an abelian monoid to establish concept of successor and predecessor.

\newcommand\m[1]{\mbox{$#1$}} %things in math display
%
% from Paxos Made Simple
\begin{figure}[htb]
  \centering

\fbox{
\begin{tabular}{@{}p{0.96\textwidth}@{}}

\mbox{\hspace{2.5ex}} 
Putting the actions of the proposer and acceptor together, we see that
the algorithm operates in the following two phases.
\begin{description}[labelindent=1.9ex]

\item {\bf Phase 1.} (a) A proposer selects a proposal number \m{n} and sends
  a \m{prepare} request with number \m{n} to a majority of acceptors.

  \m{\hspace{-12pt}}(b) If an acceptor receives a \m{prepare} request with number \m{n}
  greater than that of any \m{prepare} request to which it has already
  responded, then it responds to the request with a promise not to accept
  any more proposals numbered less than \m{n} and with the highest-numbered
  proposal (if any) that it has accepted.

\item {\bf Phase 2.} (a) If the proposer receives a response to its
  \m{prepare} requests (numbered \m{n}) from a majority of acceptors, then
  it sends an \m{accept} request to each of those acceptors for a proposal
  numbered \m{n} with a value \m{v}, where \m{v} is the value of the
  highest-numbered proposal among the responses, or is any value if the
  responses reported no proposals.

  \m{\hspace{-12pt}}(b) If an acceptor receives an \m{accept} request for a proposal numbered
  \m{n}, it accepts the proposal unless it has already responded to a
  \m{prepare} request having a number greater than \m{n}.

\end{description}
%\\\hline
%\vspace{0ex}
A proposer can make multiple proposals, so long as it follows the algorithm
for each one.  
... It is probably a good idea to abandon a proposal if some proposer has 
begun trying to issue a high-numbered one.
Therefore, if an acceptor ignores a \m{prepare} or \m{accept} request
because it has already received a \m{prepare} request with a higher number,
then it should probably inform the proposer, who should then abandon 
its proposal.  This is a performance optimization that does not affect 
correctness.\vspace{0.8ex}\\
\hline\vspace{-1.2ex}

\mbox{\hspace{2.5ex}} 
To learn that a value has been chosen, a learner must find out that a proposal
has been accepted by a majority of acceptors. The obvious algorithm
is to have each acceptor, whenever it accepts a proposal, respond to all
learners, sending them the proposal.
\end{tabular}
}\vspace{-.5ex}

  \caption{Lamport's description of Basic Paxos in English~\cite{lamport2001paxos}.}
  \label{fig-lapaxos-paper}
\end{figure}

Multi-Paxos can be built from Basic Paxos by carefully adding slots as detailed in Section~\ref{secSpec}.
However, in order to present these modifications, we need to first describe Basic Paxos.
To this end, we present Lamport's description of Basic Paxos~\cite{lamport2001paxos} in \figref{fig-lapaxos-paper}.
It uses any majority of acceptors as a quorum. %In Phase 2a, it instructs the $accept$ request be sent to each of those acceptors that replied with the proposer's ballot number $n$, but it is sufficient for safety to send the $accept$ request to any subset of $\mathcal{A}$. For liveness, however, the set of receivers should contain at least one quorum, which is allowed to be different from the quorum that responded to $n$.
Following Lamport et al.~\cite{basicpaxos2014}, in the specifications presented in this article, the $prepare$ requests and responses have been renamed to $\textcolor{blue}{\textsf{1a}}$ and $\textcolor{blue}{\textsf{1b}}$ messages, respectively, the $accept$ requests and responses have been renamed to $\textcolor{blue}{\textsf{2a}}$ and $\textcolor{blue}{\textsf{2b}}$ messages, respectively, and the number $n$ is renamed to $b$ and $bal$.
Modifications to be made to build Multi-Paxos are as follows:
%old: Multi-Paxos can be built from Basic Paxos by carefully adding slots as detailed in Section~\ref{secSpec}. In Basic Paxos, acceptors cache the value they have accepted with the highest ballot number. In Multi-Paxos, we have a sequence of these values indexed by slot.
\begin{enumerate}

\item
Phase 1a is essentially unchanged.

\item
In Phase 1b, the acceptors now respond with a set of triples in $\mathcal{B} \times \mathcal{S} \times \mathcal{V}$ as opposed to just one ballot in $\mathcal{B}$ and one value in $\mathcal{V}$.

\item
In Phase 2a, the proposers now propose a set of pairs in $\mathcal{S} \times \mathcal{V}$ instead of just one value in $\mathcal{V}$. Similar to Basic Paxos, a proposer executes Phase 2a once it has received a set of responses for its \textcolor{blue}{\textsf{1a}} message from a quorum of acceptors and picks the value with highest ballot number. But this is now performed separately for each slot in the set of triples received in the responses.

\item
In Phase 2b, the acceptors now respond with a set of pairs in $\mathcal{S} \times \mathcal{V}$ as opposed to just one value in $\mathcal{V}$.

%sc- Removing this as we have not specified learning
%\item
%Learning, as described in the last part of Figure \ref{fig-lapaxos-paper}, is changed to consider different slots separately---a process learns that a value is chosen for a slot if a quorum of acceptors accepted it for that slot in \textcolor{blue}{\textsf{2b}} messages. %added lost?
%
%This works because there is at least one acceptor in common between any two quorums, thus for an unsafe execution, some acceptor must defect. 
%Defection is impossible in non-Byzantine environment and hence not discussed in this text. 
%Trivially for Byzantine environments if the system has knowledge of maximum number of defectors, that number can be adjusted in with the sizes of quorum sets.
\end{enumerate}

\subsection{\texorpdfstring{\tlaplus{}}{TLA+}}
\label{secTLA+}
The specifications presented in this article are written in the language \tlaplus{}~\cite{lamport2002specifying,merz2008spec,merz2003logic}, which is based on the Temporal Logic of Actions (TLA)~\cite{lamport1994temporal}, a logic for specifying concurrent and distributed systems and reasoning about their properties. In TLA, a \textit{state} is an assignment of values to the variables of the specification. An \textit{action} is a relation between a current state and a new state, specifying the effect of executing a sequence of instructions. For example, the instruction $x := x + 1$ is represented in TLA and \tlaplus{} by the action $x' = x + 1$. An action is represented by a formula over unprimed and primed variables where unprimed variables refer to the values of the variables in the current state and primed variables refer to the values of the variables in the new state.

A system is specified by its actions and initial states. Formally, a system is specified as $Spec == Init /\ [][Next]_{vars}$ where $Init$ is a predicate that holds for initial states of the system, $Next$ is a disjunction of all actions of the system, and $vars$ is the tuple of all variables. The expression $[Next]_{vars}$ is true if either $Next$ is true, implying some action is true and therefore executed, or $vars$ stutters, that is, the values of the variables are same in the current and new states. $[]$ is the temporal operator \textit{always}. Thus, $Spec$ defines a set of infinite sequences of steps where in each step either an action is true and the state changes or $vars$ stutters. Such a sequence is called a \textit{behavior}.
\def\laclkexample {}

\def\clkH {\relax\ifmmode H0\else $H0$\fi}
\def\clkh {\relax\ifmmode h\else $h$\fi}
\def\clkmdm {\relax\ifmmode meridiem\else $meridiem$\fi}

\def\laclkc {\relax\ifmmode c\else $c$\fi}

\newcommand{\mytlakw}[1]{\textcolor{purple}{\footnotesize\textsc{\scalefont{1.15}{#1}}}}
\ifdefined\clkexample
    As a simple example, consider this specification of the hour hand of a clock:

    \begin{myeqn}\label{specclk}\begin{aligned}
        \begin{tabular}{@{}l@{}l@{}l@{}}
            \multicolumn{3}{@{}l@{}}{$\CONSTANT \clkH{}$}\\
            \multicolumn{3}{@{}l@{}}{$\VARIABLE \clkh{}, \clkmdm{}$}\\
            $Init$ &$==$ &$/\ \clkh{} = \clkH{}$\\
            & &$/\ \clkmdm{} \in \{0, 1\}$\\
            $Incr$ &$==$ &$/\ \clkh{} < 12$ \\
            & &$/\ \clkh{}' = \clkh{} + 1$\\
            & &$/\ \UNCHANGED <<\clkmdm{}>>$\\
            $Wrap$ &$==$ &$/\ \clkh{} = 12$ \\
            & &$/\ \clkh{}' = 1$\\
            & &$/\ \clkmdm{}' = 1 - \clkmdm{}$\\
            $Next$ &$==$ &$Incr \/ Wrap$\\
            $Spec$ &$==$ &$Init /\ [][Next]_{<<\clkh{}, \clkmdm{}>>}$\\
        \end{tabular}
    \end{aligned}\end{myeqn}
    
    The variable \clkh{} stores the current hour value and \clkmdm{} tells whether it is am/pm. Initially, the hour hand reads \clkH{}. Actions $Incr$ and $Wrap$ specify how the values of \clkh{} and \clkmdm{} change. The first predicate in both actions is the guard.
\else
    \ifdefined\laclkexample
        As a simple example, consider the following specification of a clock based on Lamport's logical clock~\cite{lamport1978time} but on a shared memory system:
        \begin{myeqn}\label{speclaclk}\begin{aligned}
            \begin{tabular}{@{}l@{}l@{}l@{}}
                \multicolumn{3}{@{}l@{}}{$\VARIABLE \laclkc{}$}\\
                $Max(S)$ &$==$ &$\CHOOSE e \in S: \A f \in S: e \geq f$\\
                $Init$ &$==$ &$\laclkc{} = [p \in \{0, 1\} |-> 0]$\\
                $LocalEvent(p)$ &$==$ &$\laclkc{}' = [\laclkc{} \EXCEPT![p] = \laclkc{}[p] + 1]$\\
                $ReceiveEvent(p)$ &$==$ &$\laclkc{}' = [\laclkc{} \EXCEPT![p] = 
                Max(\{\laclkc{}[p], \laclkc{}[1-p]\}) + 1]$\\
                $Next$ &$==$ &$\E p \in \{0, 1\}: LocalEvent(p) \/ ReceiveEvent(p)$\\
                $Spec$ &$==$ &$Init /\ [][Next]_{<<\laclkc{}>>}$\\
            \end{tabular}
        \end{aligned}\end{myeqn}
        
        The system has two processes numbered 0 and 1. Variable \laclkc{} stores their current clock values as a function from process numbers to clock values. Both processes start with clock value 0, as specified in $Init$. $LocalEvent(p)$ specifies that process $p$ has executed some local action and therefore increments its clock value. The expression $\laclkc{}' = [\laclkc{} \EXCEPT![p] = \laclkc{}[p] + 1]$ means that function $\laclkc{}'$ is the same as function \laclkc{} except that $\laclkc{}'[p]$ is $\laclkc{}[p] + 1$. $ReceiveEvent(p)$ specifies that process $p$ updates its clock value to 1 greater than the higher of its and the other process' clock value. We define operator $Max$ to obtain the highest of a set of values. \CHOOSE 
        %denotes Hilbert's $\epsilon$ operator that 
        returns an arbitrarily chosen value satisfying the body of the \CHOOSE expression if one exists, or an arbitrary value otherwise.
    \fi
\fi

%An execution of the system is viewed as a sequence of these states. The set of all possible executions provide the semantics of the system. Reasoning about the system is therefore equivalent to reasoning about these sequence of states.

\subsection{TLAPS}
\label{secTLAPS}

\tlaplus{} Proof System (TLAPS)~\cite{chaudhuri2008tlaps,cousineau2012tlaproofs,tlaps} is a tool that mechanically checks proofs of properties of systems specified in \tlaplus{}. Proofs are written in a hierarchical style~\cite{lamport2012write}, and are transformed to individual proof obligations that are sent to backend theorem provers. An obligation is a logical formula of the form $P \protect\Rightarrow Q$. For proving an obligation, the default behaviour of TLAPS is to try three backend provers in succession: CVC3 (an SMT solver), Zenon, and Isabelle~\cite{merz2012harnessing,merz2012automatic,tlaps2019provers}. If none of them find a proof, TLAPS reports a failure on the obligation. Other SMT solvers supported by TLAPS are Z3, veriT, and Yices. Temporal formulas are proved using LS4, a PTL (Propositional Temporal Logic) prover. Users can specify which prover they want to use by using its name and can specify the timeout for each obligation separately.

\def\clkRange {\relax\ifmmode \{1, \ldots, 12\}\else $\{1, \ldots, 12\}$\fi}
\providecommand{\prfstep}[2]{\langle#1\rangle#2.\,}
\providecommand{\prfstepnum}[2]{\langle#1\rangle#2}

\ifdefined\clkexample
    As an example, we present the proof of a simple type invariant about the clock specification in~(\ref{specclk}) - Assuming $\clkH{} \in 
    \clkRange{}$, it is always the case that $\clkh{} \in \clkRange{}$:
    
    \begin{myeqn}\label{prfclk}\begin{aligned}
    &\ASSUME ParamAssumption == \clkH{} \in \clkRange{}\\
    &TypeOK == \clkh{} \in \clkRange{}\\
    &\THEOREM Inv == Spec => [](TypeOK)\\
    &\prfstep{1}{} \USE \DEF TypeOK\\
    &\prfstep{1}{1} Init => TypeOK\, \BY ParamAssumption\, \DEF Init\\
    &\prfstep{1}{2} TypeOK /\ [Next]_{<<\clkh{}, \clkmdm{}>>} => TypeOK'\, \BY \DEF Next, Incr, Wrap\\
    &\prfstep{1}{} \QED \BY \prfstepnum{1}{1}, \prfstepnum{1}{2}, \texttt{PTL}\, \DEF Spec
    \end{aligned}\end{myeqn}
    
    The proof of theorem $Inv$ is written in a hierarchical fashion. It is proved by two steps, named $\langle1\rangle1$ and $\langle1\rangle2$, and RuleINV1~\cite{lamport1994temporal}. Proof steps in TLAPS are typically written as:
    \begin{myeqn}
        \prfstep{x}{y}\, Assertion\,\, \BY e_1, \ldots, e_m\, \DEF d_1, \ldots, d_n
    \end{myeqn}
    which states that step number $\prfstepnum{x}{y}$ proves $Assertion$ by assuming $e_1, \ldots, e_m$, and expanding the definitions of $d_1, \ldots, d_n$. For example, step $\prfstepnum{1}{1}$ asserts that $Init => TypeOK$ by assuming $ParamAssumption$ and expanding the definition of operator $Init$. The \QED step for $\langle1\rangle$ requires us to invoke \texttt{PTL} - Propositional Temporal Logic prover because $Inv$ is a temporal formula.
\else
    \ifdefined\laclkexample
        As an example, we present the proof of a simple type invariant about the clock specification in~(\ref{speclaclk}) --- It is always the case that $\laclkc{} \in [\{0, 1\} -> \mathds{N}]$:
        
        \begin{myeqn}\label{prflaclk}\begin{aligned}
        &TypeOK == \laclkc{} \in [\{0, 1\} -> \mathds{N}]\\
        &\THEOREM Inv == Spec => [](TypeOK)\\
        &\prfstep{1}{} \USE \DEF TypeOK\\
        &\prfstep{1}{1} Init => TypeOK\, \BY \DEF Init\\
        &\prfstep{1}{2} TypeOK /\ [Next]_{<<\laclkc{}>>} => TypeOK'\\%\, \BY \DEF Next, LocalEvent, ReceiveEvent\\
        &\phantom{\prfstep{1}{2}}\prfstep{2}{} \ASSUME TypeOK, [Next]_{<<\laclkc{}>>}\, \PROVE TypeOK'\\
        &\phantom{\prfstep{1}{2}}\prfstep{2}{1} \CASE \E p \in \{0, 1\}: LocalEvent(p)\, \BY \prfstepnum{2}{1}\, \DEF LocalEvent\\
        &\phantom{\prfstep{1}{2}}\prfstep{2}{2} \CASE \E p \in \{0, 1\}: ReceiveEvent(p)\, \BY \prfstepnum{2}{2}\, \DEF ReceiveEvent\\
        &\phantom{\prfstep{1}{2}}\prfstep{2}{3} \CASE \UNCHANGED <<\laclkc{}>>\, \BY \prfstepnum{2}{3}\\
        &\phantom{\prfstep{1}{2}}\prfstep{2}{} \QED \BY \prfstepnum{2}{1}, \prfstepnum{2}{2}, \prfstepnum{2}{3}\, \DEF Next\\
        &\prfstep{1}{} \QED \BY \prfstepnum{1}{1}, \prfstepnum{1}{2}, \texttt{PTL}\, \DEF Spec
        \end{aligned}\end{myeqn}
        
        The proof of theorem $Inv$ is written in a step-by-step fashion. It is proved by two steps, named $\prfstepnum{1}{1}$ and $\prfstepnum{1}{2}$, and the PTL solver.
        Proof steps in TLAPS are typically written as:
        \begin{myeqn}
            \prfstep{x}{y}\, Assertion\,\, \BY e_1, \ldots, e_m\, \DEF d_1, \ldots, d_n
        \end{myeqn}
        which states that step number $\prfstepnum{x}{y}$ proves $Assertion$ by using $e_1, \ldots, e_m$, and expanding the definitions of $d_1, \ldots, d_n$. For example, step $\prfstepnum{1}{1}$ proves $Init => TypeOK$ by expanding the definition of $Init$. If TLAPS does not know if $e_i$ is true, it would try to prove $e_i$ using $e_1, \ldots, e_{i-1}$ and the current context. If TLAPS is unable to prove $e_i$, it would display both $e_i$ and $Assertion$ as failed obligations. The step ``$\prfstep{1}{}$ \USE \DEF $TypeOK$'' instructs the prover to expand the definition of $TypeOK$ in all proof steps till the \QED step for $\prfstepnum{1}{}$. The \QED step for $\langle1\rangle$ instructs TLAPS to invoke a \texttt{PTL} prover because $Inv$ is a temporal formula.
        
        To demonstrate the hierarchical proof style advocated in TLAPS, we break down the proof of step $\prfstepnum{1}{2}$. Step $\prfstepnum{2}{} \ASSUME \ldots \PROVE$ specifies the assumptions and goal to be proved in the current proof level, which is level 2. The next two steps $\prfstepnum{2}{1}$ and $\prfstepnum{2}{2}$ prove the goal for the two actions specified in $Next$. Finally, $\prfstepnum{2}{3}$ proves the goal for the case of stuttering. Together, $\prfstepnum{2}{\text{1--3}}$ cover all cases of $[Next]_{<<\laclkc{}>>}$, thus concluding the proof.
    \fi
\fi
%}

\section{Specification of Multi-Paxos}
\label{secSpec}

We develop a formal specification of Multi-Paxos by minimally extending that of Basic Paxos by Lamport et al.~\cite{basicpaxos2014}. Lamport et al.'s specification of Basic Paxos formally specifies the phases executed by proposers and acceptors described by Lamport in~\cite{lamport2001paxos} and shown in \figref{fig-lapaxos-paper}; 
%While Lamport describes a third set of processes, called learners in~\cite{lamport2001paxos}, the \tlaplus{} specification of Basic Paxos does not specify this set of processes.
it does not specify preemption, that is, abandoning proposals, and learners 
that are also in \figref{fig-lapaxos-paper}.
We discuss learners at the end of this section, %in~\ref{sec-faultimpl}, 
and add preemption in Section~\ref{secPreemption}.

\subsection{Constants and variables}

\mypar{Constants} The specification of Multi-Paxos has six global constants. It assumes that the sets of proposers, acceptors, and quorums are constant and are an input to the algorithm.
\begin{description}
\item $\mathcal{P}$: the set of proposers.
\item $\mathcal{A}$: the set of acceptors.
\item $\mathcal{Q}$: the set of quorums.
\item $\mathcal{V}$: the set of values that can be proposed.
\item $\mathcal{B}$: the set of ballots. This is defined to be the set of natural numbers.
\item $\mathcal{S}$: the set of slots. This is defined to be the set of natural numbers.
\end{description}

We do not specify how the quorum set is constructed, rather provide the property that it satisfies as an assumption:
\begin{myeqn}\label{quorumassumption}\begin{aligned}
& \ASSUME QuorumAssumption == \mathcal{Q} \subseteq \SUBSET
 \mathcal{A} /\ \A Q1 , Q2 \in \mathcal{Q} : Q1 \cap Q2
 \neq \emptyset
\end{aligned}\end{myeqn}
This allows for specifying different kinds of quorums like majority, tree-based~\cite{agrawal1992generalized}, matrix-based~\cite{maekawa1985algorithm}, etc. Upon doing so, all one has to do is to prove that the specified quorum system satisfies $QuorumAssumption$ and the rest of the proof needs no change.

\mypar{Variables}  The specification of Multi-Paxos has four global variables.
\begin{description}
\item $msgs$: the set of messages that have been sent in the system.  Processes read from or add to this set.  This is the same as in the specification of Basic Paxos except that the contents of messages are more complex.
\item $pBal$: per proposer, the current ballot number of the proposer.
%, i.e., the logical clock value of the proposer. ?
This is not in the specification of Basic Paxos; it is added to support preemption.
\item $aBal$: per acceptor, the highest ballot number seen by the acceptor. This is named $maxBal$ in the specification of Basic Paxos.
\item $aVoted$: per acceptor, a set of triples in $\mathcal{B} \times \mathcal{S} \times \mathcal{V}$ voted by the acceptor. For each slot only the triple with the highest ballot number is stored. This contrasts with two numbers per acceptor, in two variables, $maxVBal$ and $maxVal$, in the specification of Basic Paxos, which respectively store the highest ballot in which the acceptor has voted, and the value that the acceptor voted for in this highest ballot.
\end{description}

\subsection{Algorithm steps and complete specification}
\label{secSpecAlgo}
The algorithm consists of repeatedly executing two phases.

\begin{description}
\item \textbf{Phase 1a.} \figref{fig-spec-p1a} shows the specifications of Phase 1a for Basic Paxos and Multi-Paxos,
%for Basic Paxos and Multi-Paxos side by side, 
%with corresponding lines lined up horizontally.
which are in essence the same.  Parameter ballot number $b$ in Basic Paxos is replaced with proposer $p$ executing this phase in Multi-Paxos, to allow extensions such as preemption that need to know the proposer of a ballot number; %uses of $b$ are changed to  $pBal[p]$; and
$from |-> p$ is added in $Send$ and $pBal[p]$ is updated to $b$.
$Send$ is a predicate for adding its argument to $msgs$, i.e., $Send(m) \mathrel{\smash{\triangleq}} msgs' = msgs \cup \{m\}$. In this specification, $\textcolor{blue}{\textsf{1a}}$ messages do not have a receiver, making them accessible to all processes. However, this is not required. For safety, it is enough to send this message to \textit{any} subset of $\mathcal{A}$, even $\emptyset$.  For liveness, the receiving set should contain at least one quorum. Because this work aims at proving safety, we do not specify any constraints on receiving messages to keep the specifications concise and to the point.
%This detail is marginalized by the definitions of Phase 2a and Learning. %where is Learning defined
\begin{figure}
  \centering

\begin{tabular}{|l|l|}
    \hline
    Basic Paxos & Multi-Paxos \\
    \hline
    \begin{tabular}{@{}l@{}}
        $Phase1a(b \in \mathcal{B}) ==$\\
        $\quad /\ \nexists\, m \in msgs : /\ m.type = \textcolor{blue}{"1a"}$\\
        $\hspace{76.7pt} /\ m.bal = b$\\
        $\quad /\ Send([type |-> \textcolor{blue}{"1a"},$\\ $\qquad bal |-> b])$\\
        \\
        $\quad /\ \UNCHANGED <<maxVBal, maxBal, maxVal>>$\\
    \end{tabular}
 &
    \begin{tabular}{@{}l@{}}
        $Phase1a(p \in \mathcal{P}) == \E b \in \mathcal{B}:$\\
        $\quad /\ \nexists\, m \in msgs : /\ m.type = \textcolor{blue}{"1a"}$\\
        $\hspace{76.7pt} /\ m.bal = b$\\
        $\quad /\ Send([type |-> \textcolor{blue}{"1a"}, from |-> p,$ \\ $\qquad bal |-> b])$\\
        $\quad /\ pBal' = [pBal \EXCEPT![p] = b]$\\
        $\quad /\ \UNCHANGED <<aBal, aVoted>>$\\
    \end{tabular}\\
    \hline
\end{tabular}

 \caption{Specifications of Phase 1a of Basic Paxos and Multi-Paxos}
 \label{fig-spec-p1a}
\end{figure}

\item \textbf{Phase 1b.} \figref{fig-spec-p1b} shows the specifications of Phase 1b.  Parameter acceptor $a$ executes this phase.  The only key difference between the specifications is the set $aVoted[a]$ of triples in $Send$ of Multi-Paxos vs.\ the two numbers $maxVBal[a]$ and $maxVal[a]$ in Basic Paxos.
\begin{figure}
  \centering

\begin{tabular}{|l|l|}
    \hline
    Basic Paxos & Multi-Paxos \\
    \hline
    \begin{tabular}{@{}l@{}}
        $Phase1b(a \in \mathcal{A}) ==$\\
        $\E m \in msgs :$\\
        $\quad /\ m.type = \textcolor{blue}{"1a"}$\\
        $\quad /\ m.bal > maxBal[a]$\\
        $\quad /\ Send([type |-> \textcolor{blue}{"1b"},$\\
        $\qquad acc |-> a,$\\
        $\qquad bal |-> m.bal,$\\
        $\qquad maxVBal |-> maxVBal[a],$\\ 
        $\qquad maxVal |-> maxVal[a]])$\\
        $\quad /\ maxBal' = [maxBal \EXCEPT ![a] = m.bal]$\\
        $\quad /\ \UNCHANGED <<maxVBal, maxVal>>$\\
    \end{tabular}
 &
    \begin{tabular}{@{}l@{}}
        $Phase1b(a \in \mathcal{A}) ==$\\
        $\E m \in msgs :$\\
        $\quad /\ m.type = \textcolor{blue}{"1a"}$\\
        $\quad /\ m.bal > aBal[a]$\\
        $\quad /\ Send([type |-> \textcolor{blue}{"1b"},$\\
        $\qquad from |-> a,$\\        
        $\qquad bal |-> m.bal,$\\
        $\qquad voted |-> aVoted[a]])$\\
        \\
        $\quad /\ aBal' = [aBal \EXCEPT ![a] = m.bal]$\\
        $\quad /\ \UNCHANGED <<pBal, aVoted>>$\\
    \end{tabular}\\
    \hline
\end{tabular}

 \caption{Specifications of Phase 1b of Basic Paxos and Multi-Paxos}
 \label{fig-spec-p1b}
\end{figure}

\item \textbf{Phase 2a.} \figref{fig-spec-p2a} shows the specifications of Phase 2a. The key difference is, in $Send$, the bloating of a single value $v$ in $\mathcal{V}$ in Basic Paxos to a set of pairs in $\mathcal{S} \times \mathcal{V}$ given by $PropSV$ in Multi-Paxos. A proposal is a $\langle s, v \rangle$ pair.
The operation of finding the value with the highest ballot in Basic Paxos is performed for each slot by $MaxSV$ in Multi-Paxos; %The specification presents this with no ambiguity.
$MaxSV$ takes a set $T$ of triples in $\mathcal{B} \times \mathcal{S} \times \mathcal{V}$ and returns a set of pairs in $\mathcal{S} \times \mathcal{V}$.  $NewSV$ generates a set of pairs in $\mathcal{S} \times \mathcal{V}$ where values are proposed for slots not in $MaxSV$.
This is significantly more sophisticated than running Basic Paxos for each slot, because the ballots are shared and changing for all slots, and slots are paired with values dynamically where slots that failed to reach consensus values earlier are also detected and reused.
\newcommand{\tlasubsetnote}{\ensuremath{^\dagger}}
\begin{figure}[ht!]
  \centering

\begin{tabular}{|l|l|}
    \hline
    Basic Paxos & Multi-Paxos \\\hline
    \begin{tabular}{@{}l@{}}
    $Phase2a(b \in \mathcal{B}) ==$\\
    $/\ \nexists m \in msgs : /\ m.type = \textcolor{blue}{"2a"} $\\
    $\hspace{65pt} /\ m.bal = b$\\ 
    $/\ \E v \in \mathcal{V} :$\\
       $\quad /\ \E Q \in \mathcal{Q}, S \subseteq\tlasubsetnote{}\!\! \{m \in msgs \!\!:\!\! m.type = \textcolor{blue}{"1b"} /\ $\\
            $\qquad m.bal = b \} :$\\
               $\qquad /\ \A a \in Q : \E m \in S : m.acc = a$\\
               $\qquad /\ \/ \A m \in S : m.maxVBal = -1$\\
                  $\qquad \quad \/ \E c \in 0..(b-1) :$\\ 
                        $\qquad \qquad /\ \A m \in S : m.maxVBal \leq c$\\
                        $\qquad \qquad /\ \E m \in S : (m.maxVBal = c)$\\
                                        $\qquad \qquad \qquad /\ m.maxVal = v$\\
       $\quad /\ Send([type |-> \textcolor{blue}{"2a"}, bal |-> b, val |-> v])$\\
    $/\ \UNCHANGED <<maxBal, maxVBal, maxVal>>$\\
    \\
    \\
    \\
    \\
    \\
    \\
    \\
    \\
    \\
    \smallskip
    \smallskip
    \smallskip
    \end{tabular}
 &
    \begin{tabular}{@{}l@{}}
        $Phase2a(p \in \mathcal{P}) ==$\\
        $/\ \nexists m \in msgs : /\ m.type = \textcolor{blue}{"2a"} $\\
        $\hspace{65pt} /\ m.bal = pBal[p]$\\
        \\
        $/\ \E Q \in \mathcal{Q}, S \subseteq \{m \in msgs \!\!:\!\! m.type = \textcolor{blue}{"1b"} /\ $\\
        $\quad m.bal = pBal[p]\} :$\\
        $\quad /\ \A a \in Q : \E m \in S : m.from = a$\\
        $\quad /\ Send([type |-> \textcolor{blue}{"2a"},$\\
        $\qquad from |-> p,$\\
        $\qquad bal |-> pBal[p],$\\
        $\qquad propSV |-> PropSV(\UNION $\\
        $\qquad \quad \{m.voted : m \in S\})])$\\ 
        \\
        $/\ \UNCHANGED <<pBal, aBal, aVoted>>$
        \\
        where,\\
        $MaxBSV(T) \!\!==\!\! \{t \in T : \A t2 \in T :$\\
        $\quad t2.slot = t.slot => t2.bal \leq t.bal\}$\\
        $MaxSV(T) \!\!==\!\! \{[slot |-> t.slot,$\\
        $\quad val |-> t.val] \!\!:\!\! t \in MaxBSV(T)\}$
        \smallskip\\
        $UnusedS(T) \mathrel{\smash{\triangleq}} \{s \!\in\! \mathcal{S} \!\!:\! \nexists t \!\in\! T \!\!:\! t.slot = s\}$
        \smallskip\\
        $NewSV(T) \!\!==\!\! \CHOOSE D \subseteq [slot :$\\
        $\quad UnusedS(T), val \!:\! \mathcal{V}]\!\!:\!\! \A d1, d2 \!\in\! D \!\!:$\\
        $\qquad d1.slot = d2.slot => d1 = d2$
        \smallskip\\
        $PropSV(T) \!\!==\!\! MaxSV(T) \cup NewSV(T)$
    \end{tabular}\\
    \hline
\end{tabular}

 \caption[.]{Specifications of Phase 2a of Basic Paxos and Multi-Paxos.\\
 \tlasubsetnote{}\tlaplus{} forbids quantifier expressions of the form $Q\ S \subseteq T$ where $Q \in \{\A, \E, \CHOOSE, \PICK\}$, instead allowing $Q\ S \in \SUBSET T$. For convenience of presentation and spacing, we use the former in this article. % our machine-checked specifications use the latter.
 }
 \label{fig-spec-p2a}
\end{figure}

\item \textbf{Phase 2b.} \figref{fig-spec-p2b} shows the specifications of Phase 2b. %Like Phase 2a, it is easy to extend this phase to Multi-Paxos. 
In Basic Paxos, the acceptor replies with the value received in the \textcolor{blue}{\textsf{2a}} message whereas in Multi-Paxos, it replies with a set of pairs in $\mathcal{S} \times \mathcal{V}$ received in the \textcolor{blue}{\textsf{2a}} message. Also, in Basic Paxos, the acceptor updates its voted pair $maxVBal[a]$ and $maxVal[a]$ upon receipt of a $\textcolor{blue}{\textsf{2a}}$ message of the highest ballot; in Multi-Paxos, this is performed for each slot. The acceptor updates $aVoted$ to have all proposals in the received $\textcolor{blue}{\textsf{2a}}$ message and all  previous values in $aVoted$ 
%which were not heard about in this message with respect to slot number.
for slots not mentioned in that message.
\begin{figure}[ht!]
  \centering

\begin{tabular}{|l|l|}
    \hline
    Basic Paxos & Multi-Paxos \\
    \hline
    \begin{tabular}{@{}l@{}}
        $Phase2b(a \in \mathcal{A}) ==$\\
        $\E m \in msgs :$\\
        $\quad /\ m.type = \textcolor{blue}{"2a"}$\\
        $\quad /\ m.bal \geq maxBal[a]$\\
        $\quad /\ Send([type |-> \textcolor{blue}{"2b"},$\\
        $\qquad acc |-> a,$\\
        $\qquad bal |-> m.bal,$\\
        $\qquad val |-> m.val])$\\
        $\quad /\ maxBal' = [maxBal \EXCEPT ![a] = m.bal]$\\
        $\quad /\ maxVBal' =$\\ $\qquad [maxVBal \EXCEPT ![a] = m.bal]$\\
        $\quad /\ maxVal' =$\\ $\qquad [maxVal \EXCEPT ![a] = m.val]$\\
        \\
        \\
    \end{tabular}
 &
    \begin{tabular}{@{}l@{}}
        $Phase2b(a \in \mathcal{A}) ==$\\
        $\E m \in msgs :$\\
        $\quad /\ m.type = \textcolor{blue}{"2a"}$\\
        $\quad /\ m.bal \geq aBal[a]$\\
        $\quad /\ Send([type |-> \textcolor{blue}{"2b"},$\\
        $\qquad from |-> a,$\\
        $\qquad bal |-> m.bal,$\\
        $\qquad propSV |-> m.propSV])$\\
        $\quad /\ aBal' = [aBal \EXCEPT ![a] = m.bal]$\\
        $\quad /\ aVoted' = [aVoted \EXCEPT ![a] =$\\
        $\qquad \cup \{[bal |-> m.bal, slot |-> d.slot,$\\ $\qquad \quad val |-> d.val] : d \in m.propSV\}$\\
        $\qquad \cup \{e \in aVoted[a] :$\\ $\qquad \quad \nexists\, r \in m.propSV : e.slot = r.slot \}]$\\
        $\quad /\ \UNCHANGED <<pBal>>$
    \end{tabular}\\
    \hline
\end{tabular}

 \caption{Specifications of Phase 2b of Basic Paxos and Multi-Paxos}
 \label{fig-spec-p2b}
\end{figure}
\end{description}

% Last location of preemption 

\mypar{Complete algorithm specification}
To complete the algorithm specification, we specify the global constants of the system---the set of proposers, acceptors, quorums, and values, and define $\mathcal{B}$, $\mathcal{S}$, $vars$, $Init$, $Next$, and $Spec$, denoting the set of ballots, slots, variables, the initial state, possible actions leading to the next state, and the system specification, respectively:
\begin{myeqn}\label{initdefs}\begin{aligned}
%&\CONSTANTS\, {\mathcal{P}} ,\, {\mathcal{A}} ,\, {\mathcal{Q}} ,\, {\mathcal{V}}\\
%&\mathcal{B} == \mathds{N}\\
%&\mathcal{S} == \mathds{N}\\
&vars == <<msgs, pBal, aBal, aVoted>>\\
&Init ==
        msgs = \emptyset /\ pBal = [p \in \mathcal{P} |-> 0] /\ aBal = [a \in \mathcal{A} |-> -1] /\ aVoted = [a \in \mathcal{A} |-> \emptyset]  \\
&Next == (\E p \in \mathcal{P} : Phase1a(p) \/ Phase2a(p)) \/ (\E a \in \mathcal{A} : Phase1b(a) \/ Phase2b(a))\\
&Spec == Init /\ [][Next]_{vars}
\end{aligned}\end{myeqn}

We have specified ${\mathcal{P}}$, ${\mathcal{A}}$ and ${\mathcal{Q}}$ as constant. This means that the algorithm specified in this article does not include reconfiguration in which processes join and leave the system dynamically. \iffalse
It may be argued that our specifications are already specifying this by saying that a process entering the system can be viewed as a process that was always a part of the system but did not perform any action. Similarly, a process leaving the system stops performing all actions. Such behaviors are already specified by our specifications however this is not equivalent to reconfiguration. The following should be noted:
\begin{itemize}
    \item If an acceptor joins or leaves the system, an algorithm implementing reconfiguration would change $\mathcal{Q}$ as well, whereas in our specifications, $\mathcal{Q}$ is constant. This affects Phase 2a and $Chosen$, and would therefore result in different behaviors.
    \item If a proposer joins or leaves the system, $\mathcal{A}$ and $\mathcal{Q}$ remain unchanged, and would therefore not result in different behaviors.
\end{itemize}
\fi

The complete specification of Multi-Paxos with Preemption is given in Appendix~\ref{appendix:spec}. Preemption is discussed in Section~\ref{secPreemption}. We only provide specification of Multi-Paxos with Preemption because it is an extension of Multi-Paxos, and giving also specification of Multi-Paxos would be redundant.

\subsection{Fault model and implementation issues}
\label{sec-faultimpl}

\mypar{Fault model} 
We explain how the fault model described in Section~\ref{secDistCons} is naturally specified due to the definition of $Spec$, especially due to $msgs$ being a set and $\E m \in msgs$ being nondeterministic.  $Spec$ specifies the set of allowed behaviors of Multi-Paxos. Recalling the meaning of the definition of $Spec$ from Section~\ref{secTLA+}, in every allowable behavior of Multi-Paxos the next state is obtained from the current state either by a process performing one of the phases or by stuttering where $vars$ remains unchanged.
\begin{description}
    \item[\quad \quad Message loss, delay, reordering, and duplication.] Because our model exhibits infinite stuttering steps, and choosing a next message to handle is nondeterministic, the model can always stutter and avoid handling any particular message in $msgs$, allowing for arbitrary message delays and loss.  Also, the model can pick messages in any order because $msgs$ is a set and pick a message any number of times because it never removes a message from $msgs$, allowing for arbitrary message reordering and duplication.
    %Also, because $msgs$ is a set, any message can be received by a process at any time. This means that even though some message is being indefinitely delayed, during that time, another message could be received.

    %\item \textit{Message reordering and duplication}. $Send(m)$ adds message $m$ to the set of sent messages, $msgs$. Once a message is added to $msgs$, it is never removed. This allows for unbounded duplication.
    %\item \textit{Messages out of order}. Because $msgs$ is a set, any message can be received by a process at any time.
    %sc: i haven't defined "receive" really

    \item[\quad \quad Process crash and recovery.] We view a crashed process as one that cannot send or receive messages from any other process in the system. Then, similar to message delay, process crash is modeled by having stuttering steps instead of the crashed process executing some phase. 
    Note that this way of modeling process crash requires that all state ($vars$) is stored in stable storage;
    %, including the set $msgs$; %of sent messages. %Future work aims to address this issue.
    use of stable storage is standard for recovery from crashes.
\end{description}

\mypar{Specification vs.\ implementation}
Following Lamport et al.~\cite{basicpaxos2014}, our specification of Multi-Paxos abstracts from certain implementation issues. 
\begin{itemize}
    \item The first conjunct of $Phase1a$ in Lamport et al.'s and our specifications states that no \msg{1a} message has been sent with ballot $b$. Similarly for $Phase2a$ and \msg{2a} messages. To implement this in practice, we need a mechanism that lets proposers choose ballots from disjoint sets, the union of which is a total order.
    
    A common way of doing this is by defining a ballot as a pair in $\mathds{N} \times \mathcal{P}$, and having $\mathcal{P} \subseteq \mathds{N}$, as for instance in~\cite{liu2019moderately}. $(\mathds{N} \times \mathcal{P}, \preceq)$ is a totally ordered set such that $\A (n1, p1), (n2, p2) \in \mathds{N} \times \mathcal{P}: (n1, p1) \preceq (n2, p2) \iff (n1 < n2 \/ (n1 = n2 /\ p1 \leq p2))$.
    \item The purpose of our specification is to provide the core of Multi-Paxos, following Lamport et al.~\cite{basicpaxos2014} for Basic Paxos. Therefore, we omit extensions and optimizations like reconfiguration, state reduction, and failure detection, which would distract from the essence of the algorithm. 
    
    For example, in our specifications, a \msg{1b} message contains votes on all the slots in which the acceptor has ever voted, following van Renesse and Altinbuken~\cite{van2015paxos}.  In real implementations, an optimization would be used to reduce the size of \msg{1b} messages.
\end{itemize}

\mypar{Learners and learner condition}
Learners learn chosen values when they receive votes in \msg{2b} messages from a quorum of acceptors. 
Our specification of Multi-Paxos omits learners because our goal is to minimally extend~\cite{basicpaxos2014}, 
which omits learners.

Adding specification of learners  
%does not merit enough scientific labor because of its 
is straightforward because it would be the same as $Chosen$. 
As a result, 
the learner condition C3 in Section~\ref{secDistCons}, which states that only a chosen value can be learned, holds straightforwardly.
%we cannot specify property C3 presented in Section~\ref{secDistCons} which states that only a chosen value can be learnt.

\section{Verification of Multi-Paxos and general proof strategies}
\label{secProve}

We first define the auxiliary predicates and invariants used, by extending those for the proof of Basic Paxos with slots, and then describe our proof strategy, which proves $Safe$ of Multi-Paxos.

We define and prove
%use %annie new: i thought Lamport uses those, sc: yes, done
three kinds of invariants, following Lamport et al.~\cite{basicpaxos2014}: type invariants, process invariants, and message invariants. These are sufficient because a distributed algorithm handles two kinds of data: messages communicated and process local data.  Correspondingly, we obtain message invariants for the messages passed between processes and process invariants over the local data that these processes maintain. Type invariants ensure that the specification always uses data with correct types.

\subsection{Auxiliary predicates and functions}
\label{secAuxiliary}
These predicates and functions are used throughout the proof. Predicate $Chosen(s, v)$ is true iff there exists some ballot $b$ such that $ChosenIn(b, s, v)$ holds.
\begin{myeqn}\begin{aligned}
&Chosen(s \in \mathcal{S}, v \in \mathcal{V}) == \E b \in \mathcal{B} : ChosenIn(b, s, v)
\end{aligned}\end{myeqn}%
$ChosenIn(b, s, v)$ is true iff there exists some quorum of acceptors such that for each acceptor in the quorum, $VotedForIn(a, b, s, v)$ holds.
\begin{myeqn}\begin{aligned}
&ChosenIn(b \in \mathcal{B}, s \in \mathcal{S}, v \in \mathcal{V}) == \E Q \in \mathcal{Q} : \A a \in Q : VotedForIn(a, b, s, v)
\end{aligned}\end{myeqn}%
$VotedForIn(a, b, s, v)$ is true iff acceptor $a$ has voted value $v$ for slot $s$ in ballot $b$. This is realized in the algorithm by sending a $\textcolor{blue}{\textsf{2b}}$ message with ballot $b$ and with pair $\langle s, v\rangle$ in the message's $propSV$ in $Phase2b$. Putting everything together, the algorithm chooses value $v$ for slot $s$ if there exist some quorum $Q$ and ballot $b$ such that every acceptor in $Q$ has voted value $v$ for slot $s$ in ballot $b$:
%annie: reverse order of the 3?
%sc: done. More sensible that way
{
\begin{myeqn}\label{chosenmacro}\begin{aligned}
&VotedForIn(a \in \mathcal{A}, b \in \mathcal{B}, s \in \mathcal{S}, v \in \mathcal{V}) == \E m \in msgs :\\
&\halftab m.type = \textcolor{blue}{"2b"} /\ m.from = a /\ m.bal = b /\ \E d \in m.propSV : d.slot = s /\ d.val = v
\end{aligned}\end{myeqn}%
}
%annie: add explanations:
%$VotedForIn(a,b,s, v)$ means that some $\textcolor{blue}{\textsf{2b}}$ message $m$ from acceptor $a$ with ballot $b$ voted for some decree d with slot s and value v.
%annie: i think we should add s before v, rather than at the end.  better even always use the order of bsv.
%ChosenIn...
%Chosen...
% sc - done

% I changed /\ to \land below, because /\ breaks in this context.
\newcommand{\defSafeAt}{Predicate $SafeAt(b \in \mathcal{B}, s \in \mathcal{S}, v \in \mathcal{V})$ means that no value except possibly $v$ has been or will be chosen in any ballot lower than $b$ for slot $s$. This is realized by asserting that for each ballot $b2 < b$, there exists some quorum $Q$ such that for every acceptor in $Q$, either $VotedForIn(a, b2, s, v)$ holds or $WontVoteIn(a, b2, s)$ holds.
\begin{myeqn}\label{safeatmacro}\begin{aligned}
&SafeAt(b \in \mathcal{B}, s \in \mathcal{S}, v \in \mathcal{V}) \mathrel{\smash{\triangleq}} \A b2 \in 0..(b-1) : \E Q \in \mathcal{Q} :\\
&\halftab \A a \in Q : VotedForIn(a, b2, s, v) \/ WontVoteIn(a, b2, s)
\end{aligned}\end{myeqn}%
$WontVoteIn(a, b, s)$ holds iff acceptor $a$ has seen a higher ballot than $b$, and did not and will not vote any value in $b$ for slot $s$
\begin{myeqn}\begin{aligned}
&WontVoteIn(a \in \mathcal{A}, b \in \mathcal{B}, s \in \mathcal{S}) \mathrel{\smash{\triangleq}} aBal[a] > b \land \A v \in \mathcal{V} : ~ VotedForIn(a, b, s, v)
\end{aligned}\end{myeqn}%
}%
\fullonly{\defSafeAt}

Function $MaxBalInSlot(T \subseteq [bal : \mathcal{B}, slot : \mathcal{S}], s \in \mathcal{S})$ selects among set of elements in $T$ with slot $s$, the highest ballot, or -1 if no element has slot $s$.
{
\begin{myeqn}\label{MVBIS}\begin{aligned}
&Max(T) == \CHOOSE e \in T : \A f \in T : e \geq f\\
&MaxBalInSlot(T \subseteq [bal : \mathcal{B}, slot : \mathcal{S}], s \in \mathcal{S}) ==\\
&\halftab \LET E == \{e \in T : e.slot = s\} \IN \IF{E = \emptyset} \THEN -1 \LSE Max(\{e.bal: e \in E\})\FI\NI \\
\end{aligned}\end{myeqn}
}

\fullonly{
To prove the $Safe$ property in (\ref{eqnsafe}) for the algorithm, we prove the following two lemmas:
\begin{enumerate} 
    \item Lemma $VotedInv$. If any acceptor votes any triple $\langle b, s, v \rangle$, then the predicate $SafeAt(b, s, v)$ holds.  That is, $\A a \in \mathcal{A}, b \in \mathcal{B}, s \in \mathcal{S}, v \in \mathcal{V}: VotedForIn(a, b, s, v) \Rightarrow SafeAt(b, s, v)$.

    \item Lemma $VotedOnce$. If acceptor $a1$ votes triple $\langle b, s, v1 \rangle$ and acceptor $a2$ votes triple $\langle b, s, v2 \rangle$, then $v1 = v2$.  That is, $\A a1, a2 \in \mathcal{A}, b \in \mathcal{B}, s \in \mathcal{S}, v1, v2 \in \mathcal{V}:$\\ $VotedForIn(a1, b, s, v1) \land VotedForIn(a2, b, s, v2) \Rightarrow v1 = v2$.
    %annie: is this part of SafeAt, or something in addition? - sc: This is addition. SafeAt + this stuff leads to safety. done
    
    %\item After an acceptor votes for triple $\langle b, s, v \rangle$, it can vote for triple $\langle b2, s, v2 \rangle$ only if $b2 > b$.
    %annie: need to make more precise by using the b,s,v I used.
    % sc: These can be different b, s, v. done
\end{enumerate}
In fact, for other consensus algorithms, either Paxos extensions like Fast Paxos~\cite{lamport2006fast} and Byzantine Paxos~\cite{lamport2011byzantizing}, or Paxos alternatives like Viewstamped Replication~\cite{liskov2010viewstamped} and Raft~\cite{ongaro2014search}, safety can also be proved by asserting these two properties.
%proving this should be the case. %annie: should be what case?
%"this property must be proved" or, sc: no, because one can write some other proof for their algorithm
%"must hold" or sc: yes
%"it is also important to prove this property. sc: yes

%To prove these properties,
%%annie: which part "above"?  all aux predicates? sc - done
%we write invariants about the algorithm. 
}

\subsection{Type invariants}
\label{secTypeInv}
Type invariants are captured by $TypeOK$.\fullonly{ They specify the sets of values that the variables of the system can hold. For example, $pBal \in [\mathcal{P} \to \mathcal{B}]$ states that $pBal$ is a function whose domain is $\mathcal{P}$ and whose range is $\mathcal{B}$, i.e., for any $p$ in $\mathcal{P}$, $pBal[p]$ is in $\mathcal{B}$.}
{
\begin{myeqn}\label{typeinv}\begin{aligned}
&Messages ==\\
&\halftab \cup [type : \{\textcolor{blue}{"1a"}\}, bal : \mathcal{B}, from : \mathcal{P}]\\
&\halftab \cup [type : \{\textcolor{blue}{"1b"}\}, bal : \mathcal{B}, voted : \SUBSET [bal : \mathcal{B}, slot : \mathcal{S}, val : \mathcal{V}], from : \mathcal{A}]\!\!\!\!\\
&\halftab \cup [type : \{\textcolor{blue}{"2a"}\}, bal : \mathcal{B}, propSV : \SUBSET [slot : \mathcal{S}, val : \mathcal{V}], from : \mathcal{P}]\\
&\halftab \cup [type : \{\textcolor{blue}{"2b"}\}, bal : \mathcal{B}, propSV : \SUBSET [slot : \mathcal{S}, val : \mathcal{V}], from : \mathcal{A}]\\
%&\halftab \cup [type : \{\textcolor{blue}{"preempt"}\}, bal : \mathcal{B}, to : \mathcal{P}, maxBal : \mathcal{B}]\\
&TypeOK ==\\
&\halftab /\ msgs \subseteq Messages /\ pBal \in [\mathcal{P} -> \mathcal{B}] /\ aBal \in [\mathcal{A} -> \mathcal{B} \cup \{-1\}]\\
&\halftab /\ aVoted \in [\mathcal{A} -> \SUBSET [bal : \mathcal{B}, slot : \mathcal{S}, val : \mathcal{V}]]
\end{aligned}\end{myeqn}}

\newcommand{\defAccInv}{
\subsection{Invariants about acceptors}
\label{secAccInv}
The following predicate specifies invariants about acceptor processes. For each acceptor $a$, the first conjunct establishes the initial condition.
The second conjunct says that $aBal[a]$ is higher than or equal to the ballot number of each triple in $aVoted[a]$ and $a$ has voted each triple in $aVoted[a]$.
The third conjunct states that if acceptor $a$ has voted any value $v$ for slot $s$ in ballot $b$, then there is some triple $t$ in $aVoted[a]$ such that $t.bal\geq b$ and $t.slot=s$.
%the triple $\langle b2, s, v \rangle$ must be in $aVoted$ of $a$ where $b2 \geq b$ and $v$ is any value. 
The last conjunct says that acceptor $a$ has not voted for a value in any ballot higher than the highest it has seen per slot. %: vote for what -- done: sc
\begin{myeqn}\label{accinv}\begin{aligned}
&AccInv \mathrel{\smash{\triangleq}} \A a \in \mathcal{A}:\\
&\halftab \land (aBal[a] = -1) \Rightarrow (aVoted[a] = \emptyset)\\
&\halftab \land \A t \in aVoted[a] : aBal[a] \geq t.bal \land VotedForIn(a, t.bal, t.slot, t.val)\\
&\halftab \land \A b \in \mathcal{B}, s \in \mathcal{S}, v \in \mathcal{V} : VotedForIn(a, b, s, v) \Rightarrow \E t \in aVoted[a] : t.bal \geq b \land t.slot = s\\
&\halftab \land \A b \in \mathcal{B}, s \in \mathcal{S}, v \in \mathcal{V} : b > MaxBalInSlot(aVoted[a], s) \Rightarrow ~ VotedForIn(a, b, s, v)
\end{aligned}\end{myeqn}
}

\defAccInv

\subsection{Invariants about messages}
\label{secMsgInv}
The following invariant is for a $\textcolor{blue}{\textsf{1b}}$ message $m$, sent from an acceptor $m.from$, where $m.bal$ is the ballot number of the \msg{1a} message that $m$ is a reply of, and $m.voted$ is the value of $aVoted[m.from]$ when $m$ was sent.
The first conjunct says that $m.bal$ is no higher than the highest ballot number seen by $m.from$.
%the maximum ballot possible with respect to acceptor. 
The second conjunct states that $m.from$ has voted every triple in the set $m.voted$.
%annie: each... voted not explained sc - attempted
The last conjunct asserts that for each slot $s$ and ballot $b$ higher than the highest ballot that $m.from$ has voted in for slot $s$, and lower than the ballot in $m$, $m.from$ has not voted any value $v$ for slot $s$ in ballot $b$.
\begin{myeqn}\label{msginv1b}\begin{aligned}
&MsgInv1b(m) ==\\ 
&\halftab /\ m.bal \leq aBal[m.from]\\
&\halftab /\ \A t \in m.voted : VotedForIn(m.from, t.bal, t.slot, t.val)\\
&\halftab /\ \A b \in \mathcal{B}, s \in \mathcal{S}, v \in \mathcal{V} : b \in MaxBalInSlot(m.voted, s)+1..m.bal-1\\
&\tab => ~ VotedForIn(m.from, b, s, v)
\end{aligned}\end{myeqn}

\shortonly{The invariants for \textcolor{blue}{\textsf{2a}} and \textcolor{blue}{\textsf{2b}} messages are defined in Appendix \ref{appendix:prop}.}
The following invariant is for a $\textcolor{blue}{\textsf{2a}}$ message $m$, where $m.bal$ is the ballot for which a quorum of \msg{1b} replies were received by $m.from$, and $m.propSV$ is the set of proposals that $m.from$ proposes for ballot $m.bal$ based on the set of $voted$ triples in these \msg{1b} replies, as shown in $Phase2a$ (\figref{fig-spec-p2a}).
The first conjunct establishes safety for each proposal $d$ in $m$.
The second conjunct says that two proposals in $m$ with the same slot must be the same proposal. That is, for each slot, there is at most one proposal in $m$.
The third conjunct says that there is at most one $\textcolor{blue}{\textsf{2a}}$ message for each ballot. %annie: not sure i.e., a ballot number uniquely identifies a 2a message. sc - yes, for each ballot, there is at most one 2a message. All <s, v> proposals are clubbed into one message. done?
\begin{myeqn}\label{msginv2a}\begin{aligned}
&MsgInv2a(m) ==\\ 
&\halftab \land \A d \in m.propSV : SafeAt(m.bal, d.slot, d.val)\\
&\halftab \land \A d1,d2 \in m.propSV : d1.slot = d2.slot \Rightarrow d1 = d2\\
&\halftab \land \A m2 \in msgs : (m2.type = \textcolor{blue}{"2a"}) \land (m2.bal = m.bal) \Rightarrow m2 = m\\
\end{aligned}\end{myeqn}

The following invariant is for a $\textcolor{blue}{\textsf{2b}}$ message $m$ sent by acceptor $m.from$, where $m.bal$ is the ballot number of the \msg{2a} message $m2$ that $m$ is a reply of, and $m.propSV$ is the same set of proposals as in $m2$. The first conjunct states that there exists a $\textcolor{blue}{\textsf{2a}}$ message with the same ballot and same set of proposals as $m$. The second conjunct asserts that the ballot of $m$ is no higher than the highest ballot seen by $m.from$.
\begin{myeqn}\label{msginv2b}\begin{aligned}
&MsgInv2b(m) == \\ 
&\halftab \land \E m2 \in msgs : m2.type = \textcolor{blue}{"2a"}
                             \land m2.bal  = m.bal
                             \land m2.propSV = m.propSV\\
&\halftab \land m.bal \leq aBal[m.from]
\end{aligned}\end{myeqn}

The complete message invariant is the conjunction of $MsgInv1b$, $MsgInv2a$, and $MsgInv2b$:
\begin{myeqn}\begin{aligned}
&MsgInv == \A m \in msgs: /\ \,(m.type = \textcolor{blue}{"\textsf{1b}"})=> MsgInv1b(m)\\
&\hspace{115.5pt} /\ (m.type = \textcolor{blue}{"\textsf{2a}"}) => MsgInv2a(m)\\
&\hspace{115.5pt} /\ (m.type = \textcolor{blue}{"\textsf{2b}"}) => MsgInv2b(m)\\
\end{aligned}\end{myeqn}%
The complete invariants, auxiliary operators, and the safety property to be proved can be found in Appendix~\ref{appendix:prop}.

\subsection{Overall proof strategy}
\label{sec-strategy}

The main theorem to prove is $Safety$ as defined in (\ref{cons}). For this, we define $Inv$ and first prove $Inv => Safe$. Then, we prove $Spec => []Inv$ which by temporal logic, concludes $Spec => []Safe$. Note that property $Safety$ is called $Consistent$, and invariant $Safe$ is called $Consistency$ by Lamport et al.~\cite{basicpaxos2014}. 
\begin{myeqn}\label{cons}\begin{aligned}
&Inv == TypeOK /\ AccInv /\ MsgInv\\
&Safe == \A v1, v2 \in \mathcal{V}, s \in \mathcal{S} : Chosen(s, v1) /\ Chosen(s, v2) => v1 = v2\\
&\THEOREM Safety == Spec => []Safe
\end{aligned}\end{myeqn}%
\shortonly{
where $AccInv$ is an invariant about acceptors, and $MsgInv2a$ and $MsgInv2b$ are invariants for \textcolor{blue}{\textsf{2a}} and \textcolor{blue}{\textsf{2b}} messages, respectively, and these three invariants are defined in Appendix \ref{appendix:prop}.}

%We first establish $Spec => []Inv$, and we prove it by induction: first prove that $Inv$ holds initially, i.e., $Init => Inv$, and then prove $Inv /\ [Next]_{vars} => Inv'$, i.e., given that $Inv$ holds, executing any disjunct in $Next$ would still yield that $Inv$ holds. 

%Having proved $Spec => []Inv$, to prove $\THEOREM Safety$, we prove $Inv => Safe$ at the end. %nothing interesting; half page at the end.

The proof is developed following a standard hierarchical structure and uses proof by induction and contradiction. To prove $Spec => []Inv$, we employ a systematic proof strategy that works well for algorithms described in an event-driven style~\cite{wiki2019event}. %~\cite{musser2013structured}\shortonly{, including message-passing distributed algorithms}.
\fullonly{ Distributed algorithms are prime examples of this style as they can be specified in blocks of code, each triggered upon the event of receiving a certain set of messages, and ending with sending a set of messages.} We demonstrate the strategy with a few invariants in $Inv$ as examples.

First, consider invariant $TypeOK$. The goal to prove is $Spec => []TypeOK$. Recall $Spec \mathrel{\smash{\triangleq}} Init /\ [][Next]_{vars}$:
\begin{itemize}
    \item The induction basis, $Init => TypeOK$, is trivial in this case because the set of sent messages is empty. To prove this step, we need to instruct TLAPS to unfold the definitions of $Init$ and $TypeOK$ and of all other predicates in $TypeOK$ till we have only statements about the variables. In most cases, TLAPS would then prove this step without a manually written proof.
    
    \item The induction step is to prove $TypeOK /\ [Next]_{vars} => TypeOK'$, where the left side is the induction hypothesis, and right side is the goal to be proved. $[Next]_{vars}$ is a disjunction of phases, as for any distributed algorithm, and $TypeOK'$ is a conjunction of smaller invariants, as for many invariants.
\end{itemize}
The hypothesis of the induction step can be stripped down to each disjunct of $Next$ separately, and each smaller goal needs to be proved for each disjunct. This process is mechanical, and TLAPS provides a feature for precisely this expansion into smaller proof obligations. This breakdown is the first step in our proof strategy. For $TypeOK$, this expands to 4 smaller assertions; with 4 phases in $Next$, we obtain 16 even smaller proof obligations.

\subsection{Proving complex invariants using invariance lemmas and increments}
\label{sec-invinc}

$AccInv$ and $MsgInv$ are more involved. We proceed as we did for $TypeOK$ and create a proof tree, each branch of which aims to prove an invariant for a disjunct in $Next$. To explain the rest of our strategy, we show a complex combination: $MsgInv$ and $Phase1b$. (\ref{msginvproof}) shows the skeleton of the proof. 
\definecolor{commentGreen}{rgb}{0,0.5,0}
\newcommand{\mytlacomment}{{\text{\textbackslash}\vspace{-1ex}*}\,}
\begin{myeqn}\label{msginvproof}\begin{aligned}
&\langle4\rangle2. \ASSUME \NEW a \in \mathcal{A}, Phase1b(a)\ \PROVE  MsgInv'\\
&\halftab \langle5\rangle1. \PICK m \in msgs: Phase1b(a)!(m) \ldots 
    \hfill\hspace{14ex} \mytlacomment{} m \mbox{\normalfont\, is a witness of\,} Phase1b(a)\\
&\halftab  \langle5\rangle. \DEFINE m1 == [type |-> \textcolor{blue}{"1b"}, from |-> a, bal |-> m.bal, voted |-> aVoted[a]]\\
&\halftab \langle5\rangle2. (m1.bal \leq aBal[m1.from])' \ldots\\
&\halftab \langle5\rangle4. (\A r \in m1.voted : VotedForIn(m1.from, r.bal, r.slot, r.val))' \ldots\\
&\halftab \langle5\rangle12. (\A b \in \mathcal{B}, s \in \mathcal{S}, v \in \mathcal{V} : b \in MaxBalInSlot(m1.voted, s)+1..m1.bal-1 =>\\
&\tab   			   ~ VotedForIn(m1.from, b, s, v))' \ldots\\
&\halftab \langle5\rangle13. \CASE m.bal > aBal[a]\\
&\tab \langle6\rangle \ASSUME \NEW m2 \in msgs'\ \PROVE MsgInv!(m2)'
    \hfill\hspace{3ex} \mytlacomment{} m2 \mbox{\normalfont\, is the instantiation in\,} MsgInv'\\
&\tab \langle6\rangle1. (m2.type = \textcolor{blue}{"1b"} => MsgInv1b(m2))'\\
&\tab\halftab \langle7\rangle1. \CASE m2 \in msgs\ \ldots\\
&\tab\halftab \langle7\rangle2. \CASE m2 \in msgs' \setminus msgs\ \ldots\\
&\tab  \langle6\rangle2. ((m2.type = \textcolor{blue}{"2a"} => MsgInv2a(m2)) /\ \\
&\tab\hspace{5.5ex}(m2.type = \textcolor{blue}{"2b"} => MsgInv2b(m2)))' \ldots\\
\end{aligned}\end{myeqn}
%annie: what do little right triangles above mean?
%sc - it's supposed to be the comment delimiter. I can change it to TLA+ 1-line comment symbol
\newline 
The goal is to prove $\langle4\rangle2$, which states that $MsgInv'$ holds if acceptor $a$ executes $Phase1b$. $\langle6\rangle1$ proves $MsgInv1b'$ and $\langle6\rangle2$ proves $MsgInv2a'$ and $MsgInv2b'$.

We describe the use of invariance lemmas to prove $\langle6\rangle2$ and the additional use of increments to prove $\langle6\rangle1$.

\mypar{Invariance lemmas}
$\langle6\rangle2$ seems easy, because $Phase1b$ sends a $\textcolor{blue}{\textsf{1b}}$ message, and $MsgInv2a$ and $MsgInv2b$ are not invariants about $\textcolor{blue}{\textsf{1b}}$ messages. However, this is not the case because $MsgInv2a$ uses predicate $SafeAt$, which\shortonly{expresses whether it is safe to accept a given value for a given ballot for a given slot (the formal definition is in Appendix \ref{appendix:prop}) and} is more complicated.  To succeed, we use what we call \textit{invariance lemmas}.

An invariance lemma is a lemma that asserts that a predicate continues to hold as the system goes from one state to the next in a single step. This happens when a step does not affect the part of system state asserted by the predicates, and requires more tedious proofs otherwise. For example, the invariance lemma for $SafeAt$ states that $SafeAt$ continues to hold for any disjunct in $Next$, which includes $Phase1b$. The characteristic property of such lemmas is their reuse. In our proof of Multi-Paxos, we defined 5 invariance lemmas which are used in 27 places.

\mypar{Increments} 
$\langle6\rangle1$ is more complicated, because $MsgInv1b$ is about $\textcolor{blue}{\textsf{1b}}$ messages, and $Phase1b$ generates a $\textcolor{blue}{\textsf{1b}}$ message.  The prover needs more manual intervention. To proceed, we split the set of messages in the new state into two sets: $\langle7\rangle1$ for the old messages, and $\langle7\rangle2$ for the \textit{increment}, the new message sent in this step. For the old messages, we use invariance lemmas. The most challenging case is for the increment.

An increment is a new message sent in a phase or generally a new element in a set. In the example in (\ref{msginvproof}), the increment is $m1$, and we focus on the cause of the increment---here from the definition of $Phase1b$---and prove each conjunct of the goal $MsgInv1b$ for $m1$ separately, in $\langle5\rangle2,4,12$.  For $\langle5\rangle2$, the prover just needs the definition of $Phase1b$. For $\langle5\rangle4$, the prover needs in addition invariance lemma for $VotedForIn$ for $Phase1b$. For $\langle5\rangle12$, the prover further needs case-specific manual intervention. %This intervention cannot be generalized and depends on the problem at hand. 
Specifically, we helped the prover understand the change in limits of the set $MaxBalInSlot(m1.voted, s)+1..m1.bal-1$. This proves $MsgInv1b$ for $m1$. The only thing left to prove now is that $m1$ is in fact the increment due to $Phase1b$. This is proved in $\prfstepnum{7}{2}$ by %asserting 
adding the assertion $m2 = m1$. $\prfstepnum{7}{1}$ uses invariance lemmas to prove that the invariant continues to hold for the old messages. This shows an example of proof with set increment.
%meaning? sc- others in the next subsection! done?

% This is the extra manual stuff needed beyond the proof strategy
% todo - change title? suggestion - Beyond Proof Strategy
% ss: another suggestion: Strategies for Sets and Tuples in Proofs
\subsection{Induction for properties over sets, and ways of accessing components of tuples}
\label{sec-setstuples}

With the strategy discussed so far, we were still faced with certain assertions that were difficult to prove. One of the main difficulties lay in proving properties about tuples and sets of tuples for each of a set of processes in Multi-Paxos, as opposed to one or two values for each of a set of processes in Basic Paxos.  It may appear that, in many places, this requires simply adding an extra parameter for the slot, but the proof became significantly more difficult: even in places where an explicit inductive proof is not needed, auxiliary facts had to be added to help TLAPS succeed or proceed sufficiently fast.

For example, to prove theorem $Safety$ by adding a slot component to the proof of theorem $Safety$ for Basic Paxos, the prover took about 90 seconds to check the proof.  To aid the proof, we added $\E a \in \mathcal{A} : VotedForIn(a, b1, s, v1) /\ VotedForIn(a, b1, s, v2)$ as an intermediate fact derivable from $b1 = b2 /\ ChosenIn(b1, s, v1) /\ ChosenIn(b2, s, v2)$, 
under the case $b1 = b2$ 
%step $\prfstepnum{3}{1}$ of step $\prfstepnum{2}{1}$ 
in the proof of theorem $Safety$. 
Following this, the prover asserted the conclusion $v1 = v2$ in a few milliseconds, a 10,000$\times$ speedup.

%These 2 paras have been explained in Proof Strategy.
%An important class of difficult proof obligations was satisfied by using induction and by explicitly directing the prover to focus on the increment in each $Next$ step of induction.  Specifically, to verify properties about a set in an inductive step, we split the set into two parts: the old value of the set before the step (unprimed version in \tlaplus{}), and the increment to the value of the set in a $Next$ step (difference between primed version and unprimed version).  To prove that a property $P$ holds always, i.e., $[]P$, the induction works by establishing that the old value of the set continues to respect property $P$ at the $Next$ step, and that the increment in the set respects $P$ too. This systematic approach works well, partly because in most cases, the increment set is empty, making the proofs simple. 

%For example, observe that invariant $AccInv$ is only affected by variables $aBal$ and $aVoted$, which are variables about acceptors. Thus, $Phase1a$, $Phase2a$, and $Preempt$ in $Next$ give empty increment sets in the proof of $AccInv$.  $Phase2b$ changes $aBal$ and $aVoted$. To prove $AccInv$ here, we establish simple facts about the tuples in $aVoted'[a] \setminus aVoted[a]$; this helps the prover check the property within milliseconds. This explication was not needed in the proof for Basic Paxos because it simply has scalars for each acceptor.

Tuples have only a fixed number of components and therefore do not require separate inductive proofs.  However, they often turn out to be tricky and require special care in choosing the ways to access and test their components, to sufficiently reduce TLAPS's proof-checking time and observe progress.
%We tried alternative ways and selected the ways that help reduce TLAPS's proof-checking time.

For example, consider the definition of $VotedForIn$ in (\ref{chosenmacro}). Originally, $[slot |-> s,val |-> v] \in m.propSV$ was written for the last conjunct with existential quantification, $\E d \in m.propSV : d.slot = s /\ d.val = v$, because it was natural, but it had to be changed to the latter for the prover to make more observable progress.  With the original version, the proof did not carry through after 1 or 2 minutes.  After the change, the proof proceeded quickly.  One minute of waiting for such simple, small tests felt very long, making it uncertain whether the proof would carry through.
%While these observations certainly apply for TLAPS, the same could not be said for other proof systems because we do not compare proof systems in this paper.

\subsection{Note on validity condition}
\label{sec-valid}

% SC: I could just add it in the TLAPS proof. But that would mean all the numbers would change
We discuss the validity condition C1 in Section~\ref{secDistCons}. 
In our specification, value $v$ is proposed for slot $s$ if the following predicate holds:
\begin{myeqn}
Proposed(s, v) == \E m \in sent: m.type = \msgtype{2a} /\ \E d \in m.propSV: d.slot = s /\ d.val = v
\end{myeqn}
Then, C1 can be formally stated as
\begin{myeqn}
\A s \in \mathcal{S}, v \in \mathcal{V}: Chosen(s, v) => Proposed(s, v)
\end{myeqn}

From the definitions of $Chosen$, $ChosenIn$, and $VotedForIn$, one can easily conclude the following:
\begin{myeqn}\begin{aligned}
&\A s \in \mathcal{S}, v \in \mathcal{V}: Chosen(s, v) =>\\
&\halftab \E m \in sent: m.type = \msgtype{2b} /\ \E d \in m.propSV: d.slot = s /\ d.val = v
\end{aligned}\end{myeqn}
Then, from the first conjunct of $MsgInv2b$ for \msg{2b} messages, one can directly deduce that C1 holds.

\section{Multi-Paxos with Preemption, and overall proof improvements}
\label{secPreemption}

Preemption is described informally in Lamport's description of Basic Paxos in \figref{fig-lapaxos-paper}, in the paragraph after the two phases. % about abandoning a proposal number.
It is about when and how to let a proposer abandon its current ballot.  Specifically, preemption makes an acceptor reply to a proposer, instead of doing nothing as in the specifications discussed so far, in both Phases 1b and 2b, if the proposer's ballot is preempted, i.e., the acceptor has seen a higher ballot than the one just received from the proposer.

This reply informs the proposer of the highest ballot the acceptor has seen, and the proposer can abandon its lower ballot number and later choose a higher ballot.   This is an important performance optimization because otherwise, as in the specifications discussed so far, it can happen that proposers arbitrarily choose new ballots. In particular, if a proposer chooses a ballot lower than previously proposed ballots, its messages with a lower ballot are wasteful, contribute to network traffic, and cause unnecessary work for the acceptors.
\notes{
\begin{itemize}
    \item Proposers arbitrarily abandon their ballots. Consider the simple system with a single proposer. If we allow the proposer to arbitrarily abandon its ballot, consensus may never be reached if it keeps proposing a new higher ballot before a quorum of acceptors votes on the current ballot.
    \item Proposers arbitrarily choose new ballots. In particular, if a proposer chooses a ballot lower than previously proposed ballots, its messages with a lower ballot are wasteful, contribute to network traffic, and cause unnecessary work for the acceptors.
\end{itemize}
}
\begin{figure}[ht!]
 \centering
%\begin{table}[h]\small
\begin{tabular}{|p{0.975\textwidth}|}
\hline
$NewBal(b2 \in \mathcal{B}) \mathrel{\smash{\triangleq}} \CHOOSE b \in \mathcal{B} : b > b2$\\
%\smallskip
$Preempt(p \in \mathcal{P}) \mathrel{\smash{\triangleq}}\,$
\begin{tabular}[t]{@{\!}ll@{~~}}
$\E m \in msgs :$\\
$\quad /\ \, m.type = \textcolor{blue}{"preempt"}$\\
$\quad /\ \, m.to = p$\\
$\quad /\ \, m.bal > pBal[p]$\\
$\quad /\ \, pBal' = [pBal \EXCEPT ![p] = NewBal(m.bal)]$\\
$\quad /\ \, \UNCHANGED <<msgs, aBal, aVoted>>$
\end{tabular}
\end{tabular}
\begin{tabular}{|p{0.475\textwidth}|p{0.475\textwidth}|}
    \hline
    Without preemption & With preemption \\
    \hline
    \begin{tabular}{@{}l@{}}
        $Phase1a(p \in \mathcal{P}) == \E b \in \mathcal{B}:$\\
        $\quad /\ \nexists\, m \in msgs : /\ m.type = \textcolor{blue}{"1a"}$\\
        $\hspace{76.7pt} /\ m.bal = b$\\
        $\quad /\ Send([type |-> \textcolor{blue}{"1a"}, from |-> p, bal |-> b])$\\
        $\quad /\ pBal' = [pBal \EXCEPT![p] = b]$\\
        $\quad /\ \UNCHANGED <<aBal, aVoted>>$
    \end{tabular}
 &
    \begin{tabular}{@{}l@{}}
        $Phase1a(p \in \mathcal{P}) == $\\
        \\
        \\
        $\quad /\ Send([type |-> \textcolor{blue}{"1a"}, from |-> p,$\\
        $\qquad bal |-> pBal[p]])$\\
        $\quad /\ \UNCHANGED <<pBal, aBal, aVoted>>$
    \end{tabular}
\end{tabular}
\begin{tabular}{|p{0.475\textwidth}|p{0.475\textwidth}|}
    %\hline
    %$Phase1b$ without preemption & $Phase1b$ with preemption \\
    %\hline
    \begin{tabular}{@{}l@{}}
        \\
        $Phase1b(a \in \mathcal{A}) ==$\\
        $\E m \in msgs :$\\
        $\quad /\ m.type = \textcolor{blue}{"1a"}$\\
        $\quad /\ m.bal > aBal[a]$\\
        $\quad /\ Send([type |-> \textcolor{blue}{"1b"},$\\
        $\qquad from |-> a,$\\
        $\qquad bal |-> m.bal,$\\
        $\qquad voted |-> aVoted[a]])$\\
        $\quad /\ aBal' = [aBal \EXCEPT ![a] = m.bal]$\\
        $\quad /\ \UNCHANGED <<pBal, aVoted>>$\\
        \\
        \\
        \\
        \\
    \end{tabular}
 &
    \begin{tabular}{@{}l@{}}
        \\
        $Phase1b(a\in \mathcal{A}) ==$\\
        $\E m \in msgs :$\\
        $\quad /\ m.type = \textcolor{blue}{"1a"}$\\
        $\quad /\ ${\textcolor{purple}{\sc if}}$\, m.bal > aBal[a]$ {\textcolor{purple}{\sc then}} \\
        $\qquad /\ Send([type |-> \textcolor{blue}{"1b"},$\\
        $\qquad \quad from |-> a,$\\
        $\qquad \quad bal |-> m.bal,$\\
        $\qquad \quad voted |-> aVoted[a]])$\\
        $\qquad /\ aBal' = [aBal \EXCEPT ![a] = m.bal]$\\
        $\qquad /\ \UNCHANGED <<pBal, aVoted>>$\\
        $\quad\,\,$ {\textcolor{purple}{\sc else}}\\
        $\qquad /\ Send([type |-> \textcolor{blue}{"preempt"},$\\
        $\qquad \quad to |-> m.from, bal |-> aBal[a]])$\\
        $\qquad /\ \UNCHANGED <<pBal, aBal, aVoted>>$\\
    \end{tabular}\\
    \hline
\end{tabular}

 \caption[.]{Extension of Multi-Paxos to Multi-Paxos with Preemption. We show changes in $Phase1a$ and $Phase1b$. Changes to $Phase2b$ are similar to $Phase1b$ and omitted for brevity.}
 \label{fig-spec-pmt}
%\end{table}
\end{figure}

To specify preemption, we 
(1) define predicate $Preempt$ that specifies when and how proposers update $pBal$ upon receiving a \textcolor{blue}{\textsf{preempt}} message, 
(2) modify $Phase1a$ to send \msg{1a} message with $pBal$, and 
(3) add a new case to $Phases1b$ and $Phase2b$ for when the acceptor $a$ receives a ballot lower than $aBal[a]$. In this new case, the acceptor replies with a newly added \msg{preempt} message. $Preempt$ specifies that proposer $p$ abandons its current ballot only after receiving a \msg{preempt} message with a ballot higher than $pBal[p]$ and that it updates $pBal[p]$ to be a ballot higher than the ballot of the \msg{preempt} message

\figref{fig-spec-pmt} shows $Preempt$, and $Phase1a$ and $Phase1b$ with and without the modifications to add preemption. Modifications to $Phase2b$ are similar and are in Appendix~\ref{appendix:spec}. We also extend the definition of $Messages$ in (\ref{typeinv}) to include the new \msg{preempt} message. 
%Thus, Preemption adds a new phase, Preempt, as an action in $Next$, modifies definitions of existing phases, and adds a new type of message.
Adding preemption increases the width of the proof tree. Except for $TypeOK$, this new branch of the proof was proved by asserting invariance lemmas established earlier. The entire task of adding the new parts of specification and proof, except for the proof of $TypeOK$, took less than an hour.

\subsection{Remedial proof for TypeOK}
\label{secMadeupProof}
%Name suggestions - Remedial proof for TypeOK, 
The invariance of $TypeOK$, i.e., $[] TypeOK$, was automatically checked previously~\cite{chand2016formal} by instructing TLAPS to use its PTL (Propositional Temporal Logic) backend prover. However, we later discovered that due to an undocumented bug in TLAPS~\cite{tlapsbug}, using PTL in this way causes incorrect obligations to be marked correct. This bug lets one \textit{prove} that a primed formula is true if its unprimed form is an assumption. A primed formula is one with only primed variables and constants. For example, $x' = 0$ is primed but $x' = x + 1$ and $x = 0$ are not. The unprimed form of $x' = 0$ is $x = 0$. With this bug, TLAPS incorrectly verifies $TypeOK'$ under the assumption $TypeOK /\ [Next]_{vars}$.

To remedy this, we provide here the missing proof of invariance of $TypeOK$. Using our proof strategy, writing this proof of invariance took a matter of minutes for Multi-Paxos. The remedial proof of $TypeOK$ is longer by 39 lines, a 72\% increase from 54 for the previous incomplete proof~\cite{chand2016formal} to 93
, and took 3 seconds more for TLAPS to check, a 30\% increase from 10 to 13 seconds.\iftmi The reason for such a small apparent increase in proof-checking time despite a large increase in proof size, is an obligation whose proof takes almost 20 seconds for TLAPS to check and the fact that proofs are checked in parallel on a multi-core machine. Because the proof of this obligation is correct, it is left untouched in the remedy. Ignoring this special obligation revealed the real impact of the remedial proof on proof-checking time --- the incorrect proof in~\cite{chand2016formal} is \textit{checked} within a second whereas the remedial proof takes 9 seconds to check.
\fi

%old: We need to make up the missing proof for $TypeOK'$. 
%annie: not neeeded without preemption?
%sc - done
For Multi-Paxos with Preemption, our strategy failed at first because the invariants were not strong enough.  We added the conjunct $pBal \in [\mathcal{P} -> \mathcal{B}]$ to $TypeOK$.  Upon preemption, proposer $p$ changes $pBal[p]$ to ballot $b$ such that (i) no \textcolor{blue}{\textsf{1a}} message has been sent yet with $b$ as ballot, and (ii) $b$ is higher  
%old: higher
%annie: check consistent: high, for all
%sc: Lamport uses `highest-numbered` for max and `greater` or `higher` when comparing ballots.
than the ballot of the preempting message.
To prove that 
%annie: box is new?  why missing before?
%sc - no specific reason. If a formula is named, I use either `invariance of f` or `[]f` but when its an unnamed formula, I go with the latter. I can change it to `invariance of`
$[] pBal \in [\mathcal{P} -> \mathcal{B}]$, %annie: removed "in our specification," not sure what it means.
we need to establish that such a $b$ indeed exists.

To this end, we strengthen our invariants by adding the fact that $msgs$ is always a finite set, i.e., $IsFiniteSet(msgs)$. We add this as a conjunct in $TypeOK$.
%annie: don't see that you added?
%sc: its only added for preemption. so its not in equation (9) but its in the proof in the appendix
Now, it can be proved that only a finite number of \textcolor{blue}{\textsf{1a}} messages exist. Thus, only a finite number of ballots can ever be used.  Because $\mathcal{B}$ is infinite, there will always be some ballot available to be chosen. We prove this constructively by providing the prover with a witness: a ballot that is 1 greater than the highest ballot in \textcolor{blue}{\textsf{1a}} messages. This remedial proof of $TypeOK$ is 63 lines longer, an 84\% increase from 75 for the previous incomplete proof~\cite{chand2016formal} to 138, and took 29 seconds more for TLAPS to check, a 242\% increase from 12 to 41 seconds.\iftmi Just like for Multi-Paxos, ignoring the special obligation revealed that the incorrect proof in~\cite{chand2016formal} is checked within a second whereas the remedial proof takes 30 seconds to check. The proof improvements described in Section~\ref{sec-improve} enhance these proof statistics for the remedial proof and are summarized in Table~\ref{tab-remedial}.\fi

Table~\ref{tab-remedial} summarizes the results, in the first two columns under Multi-Paxos and under Multi-Paxos with Preemption; the third columns under those are from overall proof improvement described in Section~\ref{sec-improve}.

\newcommand{\cnterr}{\textasciicircum}
\newcommand{\prferr}{\textasciicircum\!\textasciicircum\!}
\newcommand{\oldver}{**}
\newcommand{\cntapp}{*}
\newcommand{\centerincell}[1]{\multicolumn{1}{c|}{#1}}
\begin{table}[htbp]
  \centering
\begin{tabular}{@{}l@{\,}|r|r|r|r|r|r@{}}
    \Xhline{2\arrayrulewidth}
    \multirow{4}{*}{Metric} & \multicolumn{3}{c|}{Multi-Paxos} & \multicolumn{3}{c}{Multi-Paxos with Preemption}\\
    \cline{2-7}
    & \centerincell{Old~\cite{chand2016formal}} & \centerincell{Remedial} & \centerincell{Improved} & \centerincell{Old~\cite{chand2016formal}} & \centerincell{Remedial} & \multicolumn{1}{c}{Improved}\\
    
    & & \centerincell{Proof} & \centerincell{Remedial} & & \centerincell{Proof} & \multicolumn{1}{c}{Remedial}\\
    
    & & & \centerincell{Proof} & & & \multicolumn{1}{c}{Proof}\\
    \Xhline{2\arrayrulewidth}
    Proof size & 54\prferr & 93 & 54 & 75\prferr & 138 & 77\\
    \iftmi Net\fi CPU check time (s) & 10\prferr & 13 & 4 & 12\prferr & 41 & 12\\
    \iftmi Net\fi Elapsed check time (s) & 20\prferr & 22 & 8 & 21\prferr & 35 & 25\\
    \iftmi
    Real Check time (seconds) & 1 & 9 & 8 & 1 & 30 & 25\\
    \fi
    \Xhline{2\arrayrulewidth}
\end{tabular}\vspace{1ex}

 \caption{Proof statistics for remedial proof of TypeOK and improvements on it. Proof size is measured as non-empty lines excluding comments.\iftmi Net Time is the time taken by TLAPS (in seconds) to check the proof with the special obligation mentioned in Section~\ref{secMadeupProof}. Real Time is the time taken by TLAPS (in seconds) to check the proof without it. *The improved proof does not have this special obligation. Thus, Net and Real Time are same.\fi\ \\
 \prferr{} indicates an incorrect number from the incomplete proof due to the TLAPS bug~\cite{tlapsbug}.
 }
 \label{tab-remedial}
\end{table}

\subsection{Overall proof improvement} %annie: make all headings case consistent. sc - done.
\label{sec-improve}

In Section~\ref{secProve} we discussed the proof of Multi-Paxos from~\cite{chand2016formal} and outlined its differences with the proof for Basic Paxos. Then, in Section~\ref{secMadeupProof} we explained (1) why that proof was incorrect due to a bug in TLAPS and (2) the remedial proof. In this section, we describe three main kinds of improvements that we made to the complete proof (including the remedial proof) to make it easier to read and understand.
%old++: To make the proof presented earlier in~\cite{chand2016formal} , we made three main kinds of improvements.
These improvements apply to the proofs for both Multi-Paxos and Multi-Paxos with Preemption; the numbers described are for the proof of Multi-Paxos with Preemption.
\begin{enumerate}
%sc: check tense
    \item \textbf{Refined invariants.}
    %annie: i see them in FM. sc- they are part of the text in the FM paper but not in the proof.
    $MsgInv$ is now defined as a conjunction of three new predicates $MsgInv1b$, $MsgInv2a$, and $MsgInv2b$, in (\ref{msginv1b})-(\ref{msginv2b}), which were not separate predicates in the proof for~\cite{chand2016formal},
    %annie in paper?
    even though the names were used in the text for ease of presentation. This introduction of intermediate predicates reduced the proof size by 75 lines, because, at 12 places, the new names are used in place of their expanded definitions.
    This also reduced the proof-checking time by about 2 seconds, because now only the portions of $MsgInv$ needed for examination in the proof are instructed to be expanded.

    \item \textbf{Merged proof cases.} Cases with similar proofs are merged into a single proof. %annie: or a single case?
    %Parts of the proof are restructured to avoid repetition.
    %For example, for proving $AccInv$, when acceptor $a$ executes $Phase1b$ with preemption, $AccInv$ continues to hold for all acceptors in $\mathcal{A} \setminus \{a\},$ and also for $a$ if $a$ sent a $preempt$ message instead of $1b$ message. This is because in either case, acceptor variables, $aBal$ and $aVoted$, do not change. Hence, $AccInv$ continues to hold. The proof in~\cite{chand2016formal} had these as separate obligations. However after realizing that the proofs of these two cases are similar, they are now combined into one obligation. 
    %
    For example, to prove that $AccInv$ continues to hold when acceptor $a$ executes $Phase1b$ with preemption, there are three cases to consider for each acceptor $a2$ in $\mathcal{A}$: (1) if $a2$ is not $a$, (2) if $a2$ is $a$, and it sends a $\textcolor{blue}{\textsf{preempt}}$ message in this action, and (3) if $a2$ is $a$, and it sends a $\textcolor{blue}{\textsf{1b}}$ message in this action. The proofs for cases (1) and (2) are similar because in both cases, $a2$'s state does not change.
    We found 5 instances in the proof described in~\cite{chand2016formal} where such cases were handled separately, creating unnecessarily longer proof.
    %annie: this is just an example.  how many of this case? sc- done
    %
    Merging them resulted in an overall reduction of 27 lines of proof and slightly reduced (by less than 2 seconds) proof-checking time.
    %annie: for both w and w/o preemption? sc - yes

    \item \textbf{Removed unnecessary obligations.}  When TLAPS fails to prove an assertion, a proof must be manually written to be checked automatically by TLAPS.
    %old:Obligations are proven using assumptions or other already proven obligations. %annie: rewrite
    We discovered that parts of these manually written proofs described in~\cite{chand2016formal} are unnecessary because they can be found automatically by TLAPS. %annie: what does this mean? sc - done?
    %and some predicate definitions were unfolded unnecessarily. %annie: mean?
    %annie new: the two cases sound the same after all
    %sc: yeah, i removed the two points.
    About 25 instances of these were removed causing 110 lines of proof to be removed and the proof-checking times to reduce by about 20 seconds. An additional 22 instances were removed from the remedial proof for $TypeOK$\iftmi
    ~(including the special obligation mentioned in Section~\ref{secMadeupProof})\fi, reducing the proof size by 61 lines, from 138 to 77, and the proof-checking time by 29 seconds, from 41 to 12. %todo 5!=41-12. sc: i ran the numbers again and the trend seems to be there. the former takes 30 seconds, the latter takes 10 seconds. Numbers are different because the version on my current machine is different from the one in the paper.
    %annie new: by how much? sc: i can't count for 1.5.3 because that's the old version which doesn't run and 1.5.4 doesn't prove the old one. But what i can do is to check the delta on 1.5.4 on the obligations which it actually proves (there are only around 5 obligations which it doesn't prove and all of them are in AccInv and MsgInv for Phase 2b). Should i do that?
\end{enumerate}

\subsection{Overall proof summary}
\label{secProofSummary}

%The definitions of auxiliary predicates and invariants $TypeOK$, $AccInv$, and $MsgInv$ (1 page).

Overall, the proof for Multi-Paxos with Preemption (about 8.5 pages total) consists of the following:
\begin{enumerate}

\item Helper lemmas and their proofs (1.25 pages). These are used to prove helpful properties of operators $MaxBalInSlot$, $NewSV$, and $MaxSV$. This also includes proofs for lemmas $VotedInv$ and $VotedOnce$.

\item The invariance lemmas and their proofs (0.75 page), as explained in Section~\ref{sec-invinc}.

\item The proof of type invariant $TypeOK$ (1 page).  It uses only a 1-level proof for each action. 

\item The proof of acceptor invariant $AccInv$ (1.5 pages).  This uses a 5-level proof for action $Phase2b$ because only $Phase2b$ changes variable $aVoted$ which affects $AccInv$. 
The proofs for other actions are only 1-level because of their independence from $AccInv$.

\item The proof of message invariant $MsgInv$ (3.75 pages).  This uses 4-level proofs for actions $Phase1b$ and $Phase2b$ taking about 1 page each, and a 6-level proof for $Phase2a$ taking 2 pages, because $MsgInv$ is over messages sent in these actions. The proof is long and complex in particular because each message is a tuple and its contents may contain sets of tuples, as explained in Section~\ref{sec-setstuples}.

\item The proof of theorem $Safety$ (0.25 page) using $Spec => []Inv$ and $Inv => Safe$, which together, by temporal logic, prove $Safety$.

\end{enumerate}
The complete TLAPS-checked proof for Multi-Paxos with Preemption is given in Appendix~\ref{appendix:proof}.

\section{Results of TLAPS-checked proofs}
\label{secResults}

Table~\ref{tab-sum} summarizes the results from our specification and proof. Some mistakes in the counting logic of the result generating script caused incorrect numbers to be reported in~\cite{chand2016formal} for specification and proof sizes and number of proofs by contradiction. These errors have been corrected here.

\begin{table}[htbp]
  \centering
\begin{tabular}{@{}l@{\,}|l@{\,}r|l@{\,}r|r|l@{\,}r|r@{\,}r@{}}
    \Xhline{2\arrayrulewidth}
    \multirow{2}{*}{Metric} & \multicolumn{2}{c|}{Basic} & \multicolumn{3}{c|}{Multi-Paxos} & \multicolumn{4}{c@{}}{Multi- with Preemption}\\
    \cline{4-10}
    & \multicolumn{2}{c|}{Paxos} & \multicolumn{2}{c|}{Old~\cite{chand2016formal}} & $\!$New$\!$ & \multicolumn{2}{c|}{Old~\cite{chand2016formal}} & \multicolumn{2}{c}{New}\\
    \Xhline{2\arrayrulewidth}
    Spec size (lines, excl.\ comments)  & -, & 52 & -, & 56\phantom{\prferr} & 55 & -, & 75\phantom{\prferr} & 74, & 52\cntapp{} \\
    
    Proof size (lines, excl.\ comments) & -, & 310 & -, & 1010\prferr & 787 & -,& 1054\prferr & 831,& 528\cntapp{}\\
    
    Spec size incl.\ comments (lines)   & 115\cnterr, & 106 & 133\cnterr, & 123\phantom{\prferr}  & 115 & 158\cnterr,& 151\phantom{\prferr} & 144,& 76\cntapp{}\\
    
    Proof size incl.\ comments (lines)  & 423\cnterr, & 432 & 1106\cnterr, & 1096\prferr & 868 & 1136\cnterr,& 1143\prferr & 915,& 603\cntapp{}\\
    \hline
    
    Max level of proof tree nodes & 7 & & 11 & & 10 & 11 & & 10 &\\
    
    Max degree of proof tree nodes & 3 & & 17 & & 17  & 17 & & 17 &\\ \hline
    
    \# lemmas & 4 & & 11 & & 11  & 12 & & 12 &\\
    
    \# invariance lemmas & 1 & & 5 & & 5  & 6 & & 6 &\\
    
    \# uses of invariance lemmas & 8 & & 27 & & 32  & 29 & &35 &\\ \hline
    
    \# proofs with set increment & 0 & & 4 & & 3  &  4 & & 3 &\\
    
    \# proofs by contradiction & 1\cnterr, & 2 & 1\cnterr, & 3\phantom{\prferr} & 3  & 1\cnterr,&  3\phantom{\prferr} & 3 &\\
    
    \# obligations in TLAPS & 239 & & 918\prferr & & 779 & 959\prferr & & 825&\\ \hline
    
    TypeOK CPU check time (s)    & -, & 1 & -, &10\prferr & 4 & -,& 12\phantom{\prferr} & 12 &\\
    
    Total CPU check time (s)     & -, & 40 & -, & 228\prferr & 175 & -,& 222\prferr{} & 180&\\
    
    TypeOK elapsed check time (s) & -, & 1 & -, & 20\prferr & 7 & -,& 22\prferr & 24&\\
    
    Total elapsed check time (s)  & 24\oldver, & 14 & 128\oldver, & 63\prferr & 51 & 94\oldver,& 69\prferr{} & 90 &\\
    \Xhline{2\arrayrulewidth}
\end{tabular}\vspace{1ex}

 \caption[.]{Summary of results.
 %Spec size and proof size are measured in lines excluding comments and empty lines.
 %
 %Width of proof is the product of \textit{Inv} conjuncts and \textit{Next} disjuncts explicitly combined as a goal for the prover to prove; %sc -- product, cartesian product: Inv x Next
 % what is a proof step? sc -- reworded a bit
 %should be $Inv$ and $Next$? - mathmode doesn't compile in this caption - hence, I'm using \textit Haha!!
 %depth is the maximum level of nested steps.
 The check time is on a 4-core Intel i7-4720HQ 2.6 GHz CPU with 16 GB of memory, running 64-bit Ubuntu 17.10 and TLAPS 1.5.6.\\
 %\textasciicircum Numbers in these metrics are corrected from~\cite{chand2016formal} where an erroneous result generating script reported incorrect numbers. 
 %*TLAPS check time of x, y means x seconds were spent in checking the invariance proof of TypeOK and y for the rest of the proof.  
 %**This checking-time is irrelevant because the invariance proof of TypeOK in the old proofs is incorrect.
 %
 \mbox{\,}- denotes an entry not in~\cite{chand2016formal}, followed by the now added measurement.\\
 \cnterr{} indicates a number with a count oversight in~\cite{chand2016formal}, followed by the now correct count.\\
 \prferr{}\, indicates an incorrect number from the incomplete proof due to the TLAPS bug~\cite{tlapsbug}.\\
 %in~\cite{chand2016formal}.
 %
 \oldver{} indicates a number that used TLAPS 1.5.3 in~\cite{chand2016formal}, followed by the number using the new version 1.5.6.\\
 \cntapp{} indicates a number for the specification or proof in Appendix~\ref{appendix:proof}, after removing unnecessary line breaks from default latex generated by \tlaplus{} Tools.
 }
 \label{tab-sum}
\end{table}

\begin{itemize}
%annie new: why the number of lines changed by a few for basic paxos?
%sc: Now I remember. It's because Lamport had written 2 lemmas within his spec body. For journal version, I moved them to the verification part of the proof. It seems I only moved one of them, QuorumNonEmpty, when I made this. There's another one, NoneNotAValue. Moved it to verification part this time. So there's a 4 line move from spec to proof part excl comments since each lemma is 2 lines long. For the part including comments, which is what FM paper shows, the move from spec to proof is 9 lines - 4 lines of lemma, 3 lines comment and 2 line breaks. Check the first para of this section. The other error is in prf by contradiction counting. I've written both at one place.
    \item The specification size grew by only 3 lines (6\%), from 52 lines for Basic Paxos to 55 lines for Multi-Paxos; another 19 lines (35\%) are added for preemption.

    \item Overall, the proof size increased significantly by 477 lines (154\%), from 310 for Basic Paxos to 787 for Multi-Paxos; only 44 more lines (6\%) were added for preemption, thanks to the reuse of all lemmas, especially invariance lemmas. As mentioned in Section~\ref{secMadeupProof}, adding the remedial proof to the incomplete proof reported in~\cite{chand2016formal} initially increased the proof size by 39 lines (4\%) from 1010 to 1049 for Multi-Paxos and by 63 lines (6\%) from 1054 to 1117 for Multi-Paxos with Preemption. However, the proof improvements decreased these numbers by 262 lines (25\%) from 1049 to 787 for Multi-Paxos and by 286 lines (26\%) from 1117 to 831 for Multi-Paxos with Preemption.
    
    \item The maximum level of proof tree nodes increased from 7 to 10 going from Basic Paxos to Multi-Paxos but remained 10 after adding preemption; this contrast is even stronger for the maximum degree of proof tree nodes, consistent with the challenge of going to Multi-Paxos. The proof improvements reduced the maximum level of proof tree nodes by 1 for both Multi-Paxos and Multi-Paxos with Preemption compared with~\cite{chand2016formal}.
    %
    %The proof width tripled going to Multi-Paxos, similar to the increase in proof size.  Then it increased by nearly 20\% after adding Preemption, but the proof size increased by only 2.7\%, thanks to the reuse of most lemmas. 
    %The proof depth remained the same even after adding Preemption.
    
    \item The increase in number of lemmas is due to the change from $Max$ in Basic Paxos to $MaxBalInSlot$ in Multi-Paxos, defined in (\ref{MVBIS}).  Five lemmas were needed for this predicate alone to aid the prover, as we moved from two numbers to a set of 3-tuples for each acceptor.

    \item No proof with set increment is used for Basic Paxos. Three such proofs are used for Multi-Paxos and for Multi-Paxos with Preemption.
    
    \item Proof by contradiction is used twice in the proof of Basic Paxos, and we extended them with slots in the proofs of Multi-Paxos and Multi-Paxos with Preemption. We use proof by contradiction one more time in our proofs in lieu of longer proof by induction.
    %annie new: why the number of proof by contradiction changed for basic paxos? 
    %annie new: also, is the one more for multi and multi with preemp due to the new proof for TypeOK (since that's the same as for basic paxos in FM)?  if so, it will be better to say this.
    %sc: It was a bug in the counting script. For FM, the script was looking for occurrences of 'PROVE <space> FALSE', but in 1 place in BP proof and 2 places of MP proofs both w/ and w/o preemption, the string in the proof was 'PROVE <space> <space> FALSE'. Because my new script was checking for the keywords PROVE <space>* FALSE in the same line, it caught those missed guys.

    \item The number of proof obligations for the provers increased most significantly, by 540 (226\%), from 239 for Basic Paxos to 779 for Multi-Paxos. Only another 46 (6\%) proof obligations were generated for Multi-Paxos with Preemption. Also, with our proof improvements, the number of obligations decreased by 139 (15\%) from 918 in~\cite{chand2016formal} to 779 for Multi-Paxos and by 134 (14\%), from 959 in~\cite{chand2016formal} to 825 for Multi-Paxos with Preemption.
\iftmi
    \item The checking time of invariance proof of $TypeOK$ initially increased by 21 seconds (2100\%) from 1 for Basic Paxos to 22 for Multi-Paxos due to the more complicated structures involved in Multi-Paxos. It further increased by 13 seconds (59\%) from 22 to 35 when preemption was added. This is due to new elements involved in the remedial proof of $TypeOK$ as explained in Section~\ref{secMadeupProof}. The proof improvements reduced checking time by 14 seconds (64\%) from 22 to 8 for Multi-Paxos and by 10 seconds (29\%) from 35 to 25 for Multi-Paxos with Preemption.
\else
    \item The checking time of invariance proof of $TypeOK$ increased by 3 seconds (300\%) from 1 for Basic Paxos to 4 for Multi-Paxos due to the more complicated structures involved in Multi-Paxos. It further increased by 8 seconds (200\%) from 4 to 12 when preemption was added. This is due to new aspects in the remedial proof of $TypeOK$ as explained in Section~\ref{secMadeupProof}.
\fi
    \item The checking time of the total proof increased by 135 seconds (338\%), from 40 for Basic Paxos to 175 for Multi-Paxos, despite our continuous efforts to help the prover reduce it. This is because of the greatly increased size and complexity of inductions, leading to significantly more obligations for the prover. A small increase of 5 seconds (3\%) is observed when preemption is added. As a result of the proof improvements, we obtain a 53 second decrease (23\%, from 228 to 175) and a 42 second decrease (19\%, from 222 to 180) in the proof-checking time for Multi-Paxos and Multi-Paxos with Preemption.
    %\item The proof-checking time increased by 80 seconds (364\%), from 22 for Basic Paxos to 102 for Multi-Paxos, despite our continuous efforts to help the prover reduce it, because of the greatly increased size and complexity of the inductions used, leading to significantly more obligations to the prover. A further increase of 46 seconds (45\%) is observed when Preemption is added.
    % notes: for the last: was desceasing in FM paper. now appears increasing, but w/o fixing the missing proof, the descrease still exists.  we think it was/is becuase of heuristics used by the solver, where the needed invaraints are more readily avialble when preemption is used.
\end{itemize}

\section{Related work and conclusion}
\label{sec-related}

%\section{Related work and conclusion}
%\label{sec-related}

% this repeats sec:intro.
%Paxos is an important algorithm that was not appreciated or even understood
%when it was first introduced~\cite{lamport2016writings}, but it gained
%attention in the late 1990's~\cite{lamport1998part,lamport2001paxos}.  Especially in
%recent years, numerous efforts have been devoted to specifying,
%checking, and verifying Paxos, mostly Basic Paxos and its variants. %

We discuss closest related results on verification of Paxos and related methods using \tlaplus and TLA tools, including works using any model checking or theorem proving techniques.

\mypar{Model checking}
%
%Model checking automatically explores the state space of
%systems~\cite{clarke2018model}.
%and has been used to check models and implementations of Paxos. 
%, mostly Basic Paxos and its variants.
%, e.g.,~\cite{yabandeh2010predicting,yang2009modist, delzanno2014model} 
Lamport has written \tlaplus{} specifications for Basic Paxos and its variants, e.g., 
Fast Paxos~\cite{lamport2006fast}, and checked them using the \tlaplus{} model checker 
TLC~\cite{yu1999model,kuppe2017thesis,kuppe2019tlcgit}, but not for Multi-Paxos or its variants;
a number of M.S. students at our university have also done this in course
projects, including for Multi-Paxos.
TLC is part of the TLA Tools~\cite{kuppe2019tlcgit,kuppe2018rise} that also includes TLAPS, and has been used in finding bugs in other models~\cite{newcombe2014tla,kuppe2019hyperscale}.
Delzanno et al.~\cite{delzanno2014model} modeled Basic Paxos
in Promela and 
% tested it using a simulator and 
checked it using the Spin model checker~\cite{spin16}.  
To reduce the state space, they use counting guards
to track majority, reset local variables after state operations, 
and use sorted \textit{send} instead of FIFO \textit{send}
% see http://spinroot.com/spin/Man/send.html
(with random \textit{receive}, to model non-FIFO channels).
They checked Basic Paxos for pairs of numbers of proposers and acceptors
up to (2,8), (3,5), (4,4), (5,3), and (8,2).
%total number of ... between 4 and 10, with each between 2 and 8.

Yabandeh et al.~\cite{yabandeh2010predicting} checked a C++ implementation
of Basic Paxos using CrystalBall, a tool built on Mace~\cite{killian2007mace}, 
which includes a model checker.
%MC can start from current state, not having to start from init configuration
%
Yang et al.~\cite{yang2009modist} used their model checker MoDist to check
%MPS, an anonymous name of 
a Multi-Paxos-based service system developed by a
Microsoft product team~\cite{liu2008d3s}.  With dynamic partial-order
reduction~\cite{flanagan2005dynamic}, they found 13 bugs including 2 bugs in
the Paxos implementation, with as few as 3 replicas and a few slots.
%MoDist requires no modification to the implementation making it industry
%friendly.

In all cases, existing work in model checking
% can only check a small number processes for Basic and Multi-Paxos and a
% small number of slots for Multi-Paxos.
either does not check Multi-Paxos or can check it for only a very small number of processes and slots.

\mypar{Deductive verification}
%
%Proof systems %Theorem provers and SMT solvers
%have been used increasingly to verify %prove or check manually written proofs
%Paxos, especially in most recent years.
%
Kellom\"{a}ki~\cite{kellomaki2004annotated} formally specified and verified Basic
Paxos using PVS~\cite{owre1992pvs}. % more inv and much bigger than in tlaps
%Charron-Bost and Schiper~\cite{charron2009heard} expressed Basic Paxos in the
%Heard-Of model,
% as \textit{LastVoting}: interpretation of paxos in HOM
and Charron-Bost and Merz~\cite{charron2009formal} verified a %Heard-Of model 
version of Basic Paxos
using Isabelle/HOL~\cite{isabelle15}.
Dr\u{a}goi et al.~\cite{druagoi2016psync} specified and verified a version
of Basic Paxos % last voting
in PSync, %a partially synchronous language
which is based on the Heard-Of model, so the specification and proof 
are similar to~\cite{charron2009heard,charron2009formal}.
Lamport et al.~\cite{basicpaxos2014} give a formal
specification of Basic Paxos in \tlaplus{} and a TLAPS-checked proof of its
correctness.
Lamport~\cite{lamport2011byzantizing} wrote a \tlaplus{} specification of Byzantine Paxos,
a variant of Basic Paxos that tolerates arbitrary failures, and a TLAPS-checked
proof that it
% in history: "it" is the Castro-Liskov algorithm
refines Basic Paxos. K{\"u}fner et al.~\cite{kufner2012formal} exhibit a methodology to
develop machine-checkable
parameterized proofs of correctness of fault-tolerant round-based distributed algorithms
with Basic Paxos as a case study. Their proof is approximately 10,000 lines in Isabelle/HOL.
Merz, Lu, and Weidenbach~\cite{merz2011pastry}, and Azmy, Merz, and Weidenbach~\cite{azmy2018pastry} develop several versions of TLAPS proofs 
for Pastry, a distributed hash table algorithm, but do not discuss general proof strategies.

% in his paper:
% written a formal, machine-checked proof that the Byzantized algorithm
% implements the ordinary Paxos consensus algorithm under a suitable
% refinement mapping.
%
%The motive here is to prove correctness of an implementation of the
%protocol. 
%

In the IronFleet project, Hawblitzel et al.~\cite{hawblitzel2015ironfleet} verified a
state machine replication
system that uses Multi-Paxos at its core.  Their specification mimics \tlaplus{}
models but is written in Dafny~\cite{leino2010dafny}, which has no direct concurrency support but has more automated
proof support than TLAPS.  This work is superior to its peers by proving
not only safety but also liveness properties. However, it is a complex system, with 3 levels and many components of specifications,
over 1000 lines, and proofs, over 30,000 lines.
% to read: msg delay, fail rate probability, processor speed bound
%
Schiper et al.~\cite{schiper2014developing} 
% carried out a similar but less extensive effort.  They 
used EventML~\cite{eventml12} to specify Multi-Paxos and used
NuPRL~\cite{constable1986implementing} to verify safety.
Using the Verdi framework, Wilcox et al.~\cite{wilcox2015verdi} expressed
Raft~\cite{ongaro2014search},
an algorithm similar to Multi-Paxos, in OCAML and verified it using
Coq~\cite{coq2019coq}. The proof is over 50,000 lines and takes almost 30 minutes to verify.
Padon et al.~\cite{padon2017paxos} specify variants of Paxos including Basic and
Multi-Paxos in first-order logic. They present a methodology aiming at automatic
verification based on Effectively Propositional Logic (EPR).
%a framework for implementing and formally verifying distributed systems in Coq
% P?

All these works either do not handle Multi-Paxos or handle it using more
restricted or less direct language models than \tlaplus{},
% e.g., psync (psync did multi-paxos?) or dafny
some with reformulated algorithms and some mixed in large systems, making the exact algorithm indirect to see and the essence of the
proof harder to find and understand. %discern

\mypar{Conclusion}
%
%Our work is the first to specify the exact phases of Multi-Paxos in a most
%direct and general language model, \tlaplus{}, with a complete proof
%automatically checked using TLAPS.
%
%Building on Lamport et al.'s specification and proof for Basic
%Paxos~\cite{basicpaxos2014}, we aim to facilitate the understanding of
%Multi-Paxos and its proof by minimizing the difference from those for basic Paxos.
%
%We further show this as a general way for specifying and proving variants of Multi-Paxos,
%by doing so for Multi-Paxos extended with preemption.
%
%We also discuss the significantly more complex but necessary subproofs by induction.
%
This work specifies the exact phases of Multi-Paxos in a most direct way in a general language, \tlaplus{}, and develops an automatically checked proof of its safety property using TLAPS.
As a general method for verifying variants of Paxos, we show precisely how to extend Lamport et al.'s specification and proof for Basic Paxos to Multi-Paxos and then to Multi-Paxos with Preemption. 
We further present our general strategies for proving complex properties and improving the proofs.

We have also used our method and general proof strategies in specifying and verifying other variants of Paxos, including simpler variants of Basic Paxos and Multi-Paxos specified using history variables~\cite{chand2018simpler}, and complete executable programs for Multi-Paxos with Preemption as well as optimized programs after state reduction and failure detection~\cite{liu2019moderately}. The latter also allowed us to discover a safety violation, not discovered by extensive testing and model checking, in an earlier program.

Future work may provide more automated proof by induction and support the verification of more variants that improve and extend Multi-Paxos.
\begin{acks}
We thank Stephan Merz for his helpful comments on the proofs and explanations of TLAPS.
We thank anonymous reviewers for their helpful comments on this work.
We thank Leslie Lamport for his encouragement for this work.
This work was supported in part by %\grantsponsor{sponsorid}{name}{url}
\grantsponsor{sponsorid}{National Science Foundation}{https://www.nsf.gov} 
  grants \grantnum{sponsorid}{CCF-1248184}, %distalgo eager
  \grantnum{sponsorid}{CCF-1414078}, %distalgo
  \grantnum{sponsorid}{CNS-1421893}, % trustworthy policies
  and \grantnum{sponsorid}{IIS-1447549},
  and \grantsponsor{sponsorid}{Office of Naval Research}{https://www.onr.navy.mil} grant \grantnum{sponsorid}{N000141512208}. %resilience
  Any opinions, findings, and conclusions or recommendations expressed in
  this material are those of the authors and do not necessarily reflect
  the views of these agencies.
\end{acks}
%}%ACM acknowledgment

\forACM{ %ACM bibstyle
\bibliographystyle{ACM-Reference-Format}
} %ACM bibstyle
\forThesis{
\bibliographystyle{plain}
}
\notforThesis{
\bibliography{mybib.bib}


\begin{thebibliography}{63}

%%% the Web address.

\ifx \showCODEN    \undefined \def \showCODEN     #1{\unskip}     \fi
\ifx \showDOI      \undefined \def \showDOI       #1{#1}\fi
\ifx \showISBNx    \undefined \def \showISBNx     #1{\unskip}     \fi
\ifx \showISBNxiii \undefined \def \showISBNxiii  #1{\unskip}     \fi
\ifx \showISSN     \undefined \def \showISSN      #1{\unskip}     \fi
\ifx \showLCCN     \undefined \def \showLCCN      #1{\unskip}     \fi
\ifx \shownote     \undefined \def \shownote      #1{#1}          \fi
\ifx \showarticletitle \undefined \def \showarticletitle #1{#1}   \fi
\ifx \showURL      \undefined \def \showURL       {\relax}        \fi
% invisible to TeX
\providecommand\bibfield[2]{#2}
\providecommand\bibinfo[2]{#2}
\providecommand\natexlab[1]{#1}
\providecommand\showeprint[2][]{arXiv:#2}

\bibitem[\protect\citeauthoryear{Agrawal and El~Abbadi}{Agrawal and
  El~Abbadi}{1992}]%
        {agrawal1992generalized}
\bibfield{author}{\bibinfo{person}{D. Agrawal} {and} \bibinfo{person}{A.
  El~Abbadi}.} \bibinfo{year}{1992}\natexlab{}.
\newblock \showarticletitle{The Generalized Tree Quorum Protocol: An Efficient
  Approach for Managing Replicated Data}.
\newblock \bibinfo{journal}{\emph{ACM Trans. Database Syst.}}
  \bibinfo{volume}{17}, \bibinfo{number}{4} (\bibinfo{date}{Dec.}
  \bibinfo{year}{1992}), \bibinfo{pages}{689--717}.
\newblock
\showISSN{0362-5915}
\urldef\tempurl%
\url{https://doi.org/10.1145/146931.146935}
\showDOI{\tempurl}


\bibitem[\protect\citeauthoryear{Azmy, Merz, and Weidenbach}{Azmy
  et~al\mbox{.}}{2018}]%
        {azmy2018pastry}
\bibfield{author}{\bibinfo{person}{Noran Azmy}, \bibinfo{person}{Stephan Merz},
  {and} \bibinfo{person}{Christoph Weidenbach}.}
  \bibinfo{year}{2018}\natexlab{}.
\newblock \showarticletitle{A machine-checked correctness proof for Pastry}.
\newblock \bibinfo{journal}{\emph{Science of Computer Programming}}
  \bibinfo{volume}{158} (\bibinfo{date}{June} \bibinfo{year}{2018}),
  \bibinfo{pages}{64--80}.
\newblock
\showISSN{0167-6423}
\urldef\tempurl%
\url{https://doi.org/10.1016/j.scico.2017.08.003}
\showDOI{\tempurl}
\newblock
\shownote{Abstract State Machines, Alloy, B, TLA, VDM and Z (ABZ 2016).}


\bibitem[\protect\citeauthoryear{Bickford, Constable, Eaton, Guaspari, and
  Rahli}{Bickford et~al\mbox{.}}{2012}]%
        {eventml12}
\bibfield{author}{\bibinfo{person}{Mark Bickford}, \bibinfo{person}{Robert~L.
  Constable}, \bibinfo{person}{Richard Eaton}, \bibinfo{person}{David
  Guaspari}, {and} \bibinfo{person}{Vincent Rahli}.}
  \bibinfo{year}{2012}\natexlab{}.
\newblock \bibinfo{title}{Introduction to EventML}.
\newblock
\newblock
\urldef\tempurl%
\url{http://www.nuprl.org/software/eventml/IntroductionToEventML.pdf}
\showURL{%
Retrieved November 6, 2019 from \tempurl}


\bibitem[\protect\citeauthoryear{Burrows}{Burrows}{2006}]%
        {burrows2006chubby}
\bibfield{author}{\bibinfo{person}{Mike Burrows}.}
  \bibinfo{year}{2006}\natexlab{}.
\newblock \showarticletitle{The Chubby Lock Service for Loosely-coupled
  Distributed Systems}. In \bibinfo{booktitle}{\emph{Proceedings of the 7th
  Symposium on Operating Systems Design and Implementation}}
  \emph{(\bibinfo{series}{OSDI '06})}. \bibinfo{publisher}{USENIX Association},
  \bibinfo{address}{Berkeley, CA, USA}, \bibinfo{pages}{335--350}.
\newblock
\showISBNx{1-931971-47-1}
\urldef\tempurl%
\url{http://dl.acm.org/citation.cfm?id=1298455.1298487}
\showURL{%
\tempurl}


\bibitem[\protect\citeauthoryear{Center}{Center}{2017}]%
        {tlaps}
\bibfield{author}{\bibinfo{person}{Microsoft Research-Inria~Joint Center}.}
  \bibinfo{year}{2017}\natexlab{}.
\newblock \bibinfo{title}{{TLA\textsuperscript{+} Proof System (TLAPS)}}.
\newblock
\newblock
\urldef\tempurl%
\url{http://tla.msr-inria.inria.fr/tlaps}
\showURL{%
Retrieved September 9, 2019 from \tempurl}


\bibitem[\protect\citeauthoryear{Chand and Liu}{Chand and Liu}{2018}]%
        {chand2018simpler}
\bibfield{author}{\bibinfo{person}{Saksham Chand} {and}
  \bibinfo{person}{Yanhong~A. Liu}.} \bibinfo{year}{2018}\natexlab{}.
\newblock \showarticletitle{Simpler Specifications and Easier Proofs of
  Distributed Algorithms Using History Variables}. In
  \bibinfo{booktitle}{\emph{NASA Formal Methods}} \emph{(\bibinfo{series}{NFM
  '18})}. \bibinfo{publisher}{Springer International Publishing},
  \bibinfo{address}{Cham, Switzerland}, \bibinfo{pages}{70--86}.
\newblock
\showISBNx{978-3-319-77935-5}
\urldef\tempurl%
\url{https://doi.org/10.1007/978-3-319-77935-5_5}
\showDOI{\tempurl}


\bibitem[\protect\citeauthoryear{Chand, Liu, and Stoller}{Chand
  et~al\mbox{.}}{2016}]%
        {chand2016formal}
\bibfield{author}{\bibinfo{person}{Saksham Chand}, \bibinfo{person}{Yanhong~A.
  Liu}, {and} \bibinfo{person}{Scott~D. Stoller}.}
  \bibinfo{year}{2016}\natexlab{}.
\newblock \showarticletitle{Formal Verification of Multi-Paxos for Distributed
  Consensus}. In \bibinfo{booktitle}{\emph{FM 2016: Formal Methods}}
  \emph{(\bibinfo{series}{FM '16})}. \bibinfo{publisher}{Springer International
  Publishing}, \bibinfo{address}{Cham, Switzerland}, \bibinfo{pages}{119--136}.
\newblock
\showISBNx{978-3-319-48989-6}
\urldef\tempurl%
\url{https://doi.org/10.1007/978-3-319-48989-6_8}
\showDOI{\tempurl}


\bibitem[\protect\citeauthoryear{Chandra, Griesemer, and Redstone}{Chandra
  et~al\mbox{.}}{2007}]%
        {chandra2007paxos}
\bibfield{author}{\bibinfo{person}{Tushar~D. Chandra}, \bibinfo{person}{Robert
  Griesemer}, {and} \bibinfo{person}{Joshua Redstone}.}
  \bibinfo{year}{2007}\natexlab{}.
\newblock \showarticletitle{Paxos Made Live: An Engineering Perspective}. In
  \bibinfo{booktitle}{\emph{Proceedings of the Twenty-sixth Annual ACM
  Symposium on Principles of Distributed Computing}}
  \emph{(\bibinfo{series}{PODC '07})}. \bibinfo{publisher}{ACM},
  \bibinfo{address}{New York, NY, USA}, \bibinfo{pages}{398--407}.
\newblock
\showISBNx{978-1-59593-616-5}
\urldef\tempurl%
\url{https://doi.org/10.1145/1281100.1281103}
\showDOI{\tempurl}


\bibitem[\protect\citeauthoryear{Charron-Bost and Merz}{Charron-Bost and
  Merz}{2009}]%
        {charron2009formal}
\bibfield{author}{\bibinfo{person}{Bernadette Charron-Bost} {and}
  \bibinfo{person}{Stephan Merz}.} \bibinfo{year}{2009}\natexlab{}.
\newblock \showarticletitle{{Formal Verification of a Consensus Algorithm in
  the Heard-Of Model}}.
\newblock \bibinfo{journal}{\emph{{International Journal of Software and
  Informatics (IJSI)}}} \bibinfo{volume}{3}, \bibinfo{number}{2-3}
  (\bibinfo{year}{2009}), \bibinfo{pages}{273--303}.
\newblock
\urldef\tempurl%
\url{https://hal.inria.fr/inria-00426388}
\showURL{%
\tempurl}


\bibitem[\protect\citeauthoryear{Charron-Bost and Schiper}{Charron-Bost and
  Schiper}{2009}]%
        {charron2009heard}
\bibfield{author}{\bibinfo{person}{Bernadette Charron-Bost} {and}
  \bibinfo{person}{Andr{\'e} Schiper}.} \bibinfo{year}{2009}\natexlab{}.
\newblock \showarticletitle{The Heard-Of Model: Computing in Distributed
  Systems with Benign Faults}.
\newblock \bibinfo{journal}{\emph{Distrib. Comput.}} \bibinfo{volume}{22},
  \bibinfo{number}{1} (\bibinfo{date}{April} \bibinfo{year}{2009}),
  \bibinfo{pages}{49--71}.
\newblock
\showISSN{0178-2770}
\urldef\tempurl%
\url{https://doi.org/10.1007/s00446-009-0084-6}
\showDOI{\tempurl}


\bibitem[\protect\citeauthoryear{Chaudhuri, Doligez, Lamport, and
  Merz}{Chaudhuri et~al\mbox{.}}{2008}]%
        {chaudhuri2008tlaps}
\bibfield{author}{\bibinfo{person}{Kaustuv Chaudhuri}, \bibinfo{person}{Damien
  Doligez}, \bibinfo{person}{Leslie Lamport}, {and} \bibinfo{person}{Stephan
  Merz}.} \bibinfo{year}{2008}\natexlab{}.
\newblock \showarticletitle{A TLA+ Proof System}. In
  \bibinfo{booktitle}{\emph{Proceedings of the LPAR Workshops, CEUR Workshop}},
  Vol.~\bibinfo{volume}{418}. \bibinfo{pages}{17--37}.
\newblock
\urldef\tempurl%
\url{http://ceur-ws.org/Vol-418/paper2.pdf}
\showURL{%
\tempurl}


\bibitem[\protect\citeauthoryear{Community}{Community}{2019a}]%
        {coq2019coq}
\bibfield{author}{\bibinfo{person}{Coq Community}.}
  \bibinfo{year}{2019}\natexlab{a}.
\newblock \bibinfo{title}{The {C}oq {P}roof {A}ssistant}.
\newblock
\newblock
\urldef\tempurl%
\url{http://coq.inria.fr/}
\showURL{%
Retrieved November 4, 2019 from \tempurl}


\bibitem[\protect\citeauthoryear{Community}{Community}{2019b}]%
        {isabelle15}
\bibfield{author}{\bibinfo{person}{Isabelle Community}.}
  \bibinfo{year}{2019}\natexlab{b}.
\newblock \bibinfo{title}{Isabelle (a generic proof assistant)}.
\newblock
\newblock
\urldef\tempurl%
\url{http://isabelle.in.tum.de}
\showURL{%
Retrieved November 1, 2019 from \tempurl}


\bibitem[\protect\citeauthoryear{Constable, Allen, Bromley, Cleaveland, Cremer,
  Harper, Howe, Knoblock, Mendler, Panangaden, Sasaki, and Smith}{Constable
  et~al\mbox{.}}{1986}]%
        {constable1986implementing}
\bibfield{author}{\bibinfo{person}{R.~L. Constable}, \bibinfo{person}{S.~F.
  Allen}, \bibinfo{person}{H.~M. Bromley}, \bibinfo{person}{W.~R. Cleaveland},
  \bibinfo{person}{J.~F. Cremer}, \bibinfo{person}{R.~W. Harper},
  \bibinfo{person}{D.~J. Howe}, \bibinfo{person}{T.~B. Knoblock},
  \bibinfo{person}{N.~P. Mendler}, \bibinfo{person}{P. Panangaden},
  \bibinfo{person}{J.~T. Sasaki}, {and} \bibinfo{person}{S.~F. Smith}.}
  \bibinfo{year}{1986}\natexlab{}.
\newblock \bibinfo{booktitle}{\emph{Implementing Mathematics with the Nuprl
  Proof Development System}}.
\newblock \bibinfo{publisher}{Prentice-Hall, Inc.}, \bibinfo{address}{Upper
  Saddle River, NJ, USA}.
\newblock
\showISBNx{0-13-451832-2}


\bibitem[\protect\citeauthoryear{contributors}{contributors}{2019}]%
        {wiki2019event}
\bibfield{author}{\bibinfo{person}{Wikipedia contributors}.}
  \bibinfo{year}{2019}\natexlab{}.
\newblock \bibinfo{title}{Event-driven programming --- {Wikipedia}{,} The Free
  Encyclopedia}.
\newblock
\newblock
\urldef\tempurl%
\url{https://en.wikipedia.org/w/index.php?title=Event-driven_programming&oldid=914440210}
\showURL{%
Retrieved September 9, 2019 from \tempurl}


\bibitem[\protect\citeauthoryear{Cousineau, Doligez, Lamport, Merz, Ricketts,
  and Vanzetto}{Cousineau et~al\mbox{.}}{2012}]%
        {cousineau2012tlaproofs}
\bibfield{author}{\bibinfo{person}{Denis Cousineau}, \bibinfo{person}{Damien
  Doligez}, \bibinfo{person}{Leslie Lamport}, \bibinfo{person}{Stephan Merz},
  \bibinfo{person}{Daniel Ricketts}, {and} \bibinfo{person}{Hern{\'a}n
  Vanzetto}.} \bibinfo{year}{2012}\natexlab{}.
\newblock \showarticletitle{TLA+ Proofs}. In \bibinfo{booktitle}{\emph{FM 2012:
  Formal Methods}}. \bibinfo{publisher}{Springer Berlin Heidelberg},
  \bibinfo{address}{Berlin, Heidelberg}, \bibinfo{pages}{147--154}.
\newblock
\showISBNx{978-3-642-32759-9}
\urldef\tempurl%
\url{https://doi.org/10.1007/978-3-642-32759-9_14}
\showDOI{\tempurl}


\bibitem[\protect\citeauthoryear{Delzanno, Tatarek, and Traverso}{Delzanno
  et~al\mbox{.}}{2014}]%
        {delzanno2014model}
\bibfield{author}{\bibinfo{person}{Giorgio Delzanno}, \bibinfo{person}{Michele
  Tatarek}, {and} \bibinfo{person}{Riccardo Traverso}.}
  \bibinfo{year}{2014}\natexlab{}.
\newblock \showarticletitle{Model Checking Paxos in Spin}. In
  \bibinfo{booktitle}{\emph{Proceedings of the 5th International Symposium on
  Games, Automata, Logics and Formal Verification}}
  \emph{(\bibinfo{series}{EPTCS})}. \bibinfo{pages}{131--146}.
\newblock
\urldef\tempurl%
\url{https://doi.org/10.4204/EPTCS.161.13}
\showDOI{\tempurl}


\bibitem[\protect\citeauthoryear{Dr\u{a}goi, Henzinger, and
  Zufferey}{Dr\u{a}goi et~al\mbox{.}}{2016}]%
        {druagoi2016psync}
\bibfield{author}{\bibinfo{person}{Cezara Dr\u{a}goi},
  \bibinfo{person}{Thomas~A. Henzinger}, {and} \bibinfo{person}{Damien
  Zufferey}.} \bibinfo{year}{2016}\natexlab{}.
\newblock \showarticletitle{PSync: A Partially Synchronous Language for
  Fault-tolerant Distributed Algorithms}.
\newblock \bibinfo{journal}{\emph{SIGPLAN Not.}} \bibinfo{volume}{51},
  \bibinfo{number}{1} (\bibinfo{date}{Jan.} \bibinfo{year}{2016}),
  \bibinfo{pages}{400--415}.
\newblock
\showISSN{0362-1340}
\urldef\tempurl%
\url{https://doi.org/10.1145/2914770.2837650}
\showDOI{\tempurl}


\bibitem[\protect\citeauthoryear{Flanagan and Godefroid}{Flanagan and
  Godefroid}{2005}]%
        {flanagan2005dynamic}
\bibfield{author}{\bibinfo{person}{Cormac Flanagan} {and}
  \bibinfo{person}{Patrice Godefroid}.} \bibinfo{year}{2005}\natexlab{}.
\newblock \showarticletitle{Dynamic Partial-order Reduction for Model Checking
  Software}.
\newblock \bibinfo{journal}{\emph{SIGPLAN Not.}} \bibinfo{volume}{40},
  \bibinfo{number}{1} (\bibinfo{date}{Jan.} \bibinfo{year}{2005}),
  \bibinfo{pages}{110--121}.
\newblock
\showISSN{0362-1340}
\urldef\tempurl%
\url{https://doi.org/10.1145/1047659.1040315}
\showDOI{\tempurl}


\bibitem[\protect\citeauthoryear{Ghemawat, Gobioff, and Leung}{Ghemawat
  et~al\mbox{.}}{2003}]%
        {ghemawat2003gfs}
\bibfield{author}{\bibinfo{person}{Sanjay Ghemawat}, \bibinfo{person}{Howard
  Gobioff}, {and} \bibinfo{person}{Shun-Tak Leung}.}
  \bibinfo{year}{2003}\natexlab{}.
\newblock \showarticletitle{The Google File System}.
\newblock \bibinfo{journal}{\emph{SIGOPS Oper. Syst. Rev.}}
  \bibinfo{volume}{37}, \bibinfo{number}{5} (\bibinfo{date}{Oct.}
  \bibinfo{year}{2003}), \bibinfo{pages}{29--43}.
\newblock
\showISSN{0163-5980}
\urldef\tempurl%
\url{https://doi.org/10.1145/1165389.945450}
\showDOI{\tempurl}


\bibitem[\protect\citeauthoryear{Hawblitzel, Howell, Kapritsos, Lorch, Parno,
  Roberts, Setty, and Zill}{Hawblitzel et~al\mbox{.}}{2015}]%
        {hawblitzel2015ironfleet}
\bibfield{author}{\bibinfo{person}{Chris Hawblitzel}, \bibinfo{person}{Jon
  Howell}, \bibinfo{person}{Manos Kapritsos}, \bibinfo{person}{Jacob~R. Lorch},
  \bibinfo{person}{Bryan Parno}, \bibinfo{person}{Michael~L. Roberts},
  \bibinfo{person}{Srinath Setty}, {and} \bibinfo{person}{Brian Zill}.}
  \bibinfo{year}{2015}\natexlab{}.
\newblock \showarticletitle{IronFleet: Proving Practical Distributed Systems
  Correct}. In \bibinfo{booktitle}{\emph{Proceedings of the 25th Symposium on
  Operating Systems Principles}} \emph{(\bibinfo{series}{SOSP '15})}.
  \bibinfo{publisher}{ACM}, \bibinfo{address}{New York, NY, USA},
  \bibinfo{pages}{1--17}.
\newblock
\showISBNx{978-1-4503-3834-9}
\urldef\tempurl%
\url{https://doi.org/10.1145/2815400.2815428}
\showDOI{\tempurl}


\bibitem[\protect\citeauthoryear{Holzmann}{Holzmann}{2011}]%
        {spin16}
\bibfield{author}{\bibinfo{person}{Gerard Holzmann}.}
  \bibinfo{year}{2011}\natexlab{}.
\newblock \bibinfo{booktitle}{\emph{The SPIN Model Checker: Primer and
  Reference Manual} (\bibinfo{edition}{1st} ed.)}.
\newblock \bibinfo{publisher}{Addison-Wesley Professional}.
\newblock
\showISBNx{0321773713, 9780321773715}


\bibitem[\protect\citeauthoryear{Hunt, Konar, Junqueira, and Reed}{Hunt
  et~al\mbox{.}}{2010}]%
        {hunt2010zookeeper}
\bibfield{author}{\bibinfo{person}{Patrick Hunt}, \bibinfo{person}{Mahadev
  Konar}, \bibinfo{person}{Flavio~P. Junqueira}, {and}
  \bibinfo{person}{Benjamin Reed}.} \bibinfo{year}{2010}\natexlab{}.
\newblock \showarticletitle{ZooKeeper: Wait-free Coordination for
  Internet-scale Systems}. In \bibinfo{booktitle}{\emph{Proceedings of the 2010
  USENIX Conference on USENIX Annual Technical Conference}}
  \emph{(\bibinfo{series}{USENIXATC'10})}. \bibinfo{publisher}{USENIX
  Association}, \bibinfo{address}{Berkeley, CA, USA}, \bibinfo{pages}{11--11}.
\newblock
\urldef\tempurl%
\url{http://dl.acm.org/citation.cfm?id=1855840.1855851}
\showURL{%
\tempurl}


\bibitem[\protect\citeauthoryear{Inc.}{Inc.}{2019}]%
        {dynamodb2019}
\bibfield{author}{\bibinfo{person}{Amazon Web~Services Inc.}}
  \bibinfo{year}{2019}\natexlab{}.
\newblock \bibinfo{title}{Amazon DynamoDB Developer Guide}.
\newblock
\newblock
\urldef\tempurl%
\url{https://docs.aws.amazon.com/en_pv/amazondynamodb/latest/developerguide/Introduction.html}
\showURL{%
Retrieved November 4, 2019 from \tempurl}


\bibitem[\protect\citeauthoryear{Isard}{Isard}{2007}]%
        {isard2007autopilot}
\bibfield{author}{\bibinfo{person}{Michael Isard}.}
  \bibinfo{year}{2007}\natexlab{}.
\newblock \showarticletitle{Autopilot: Automatic Data Center Management}.
\newblock \bibinfo{journal}{\emph{SIGOPS Oper. Syst. Rev.}}
  \bibinfo{volume}{41}, \bibinfo{number}{2} (\bibinfo{date}{April}
  \bibinfo{year}{2007}), \bibinfo{pages}{60--67}.
\newblock
\showISSN{0163-5980}
\urldef\tempurl%
\url{https://doi.org/10.1145/1243418.1243426}
\showDOI{\tempurl}


\bibitem[\protect\citeauthoryear{Kellom{\"a}ki}{Kellom{\"a}ki}{2004}]%
        {kellomaki2004annotated}
\bibfield{author}{\bibinfo{person}{Pertti Kellom{\"a}ki}.}
  \bibinfo{year}{2004}\natexlab{}.
\newblock \bibinfo{booktitle}{\emph{An annotated specification of the consensus
  protocol of {P}axos using superposition in {PVS}}}.
\newblock \bibinfo{type}{{T}echnical {R}eport}. \bibinfo{address}{Tampere,
  Finland}.
\newblock
\urldef\tempurl%
\url{http://www.cs.tut.fi/ohj/laitosraportit/report36-paxos.pdf}
\showURL{%
\tempurl}


\bibitem[\protect\citeauthoryear{Killian, Anderson, Braud, Jhala, and
  Vahdat}{Killian et~al\mbox{.}}{2007}]%
        {killian2007mace}
\bibfield{author}{\bibinfo{person}{Charles~Edwin Killian},
  \bibinfo{person}{James~W. Anderson}, \bibinfo{person}{Ryan Braud},
  \bibinfo{person}{Ranjit Jhala}, {and} \bibinfo{person}{Amin~M. Vahdat}.}
  \bibinfo{year}{2007}\natexlab{}.
\newblock \showarticletitle{Mace: Language Support for Building Distributed
  Systems}.
\newblock \bibinfo{journal}{\emph{SIGPLAN Not.}} \bibinfo{volume}{42},
  \bibinfo{number}{6} (\bibinfo{date}{June} \bibinfo{year}{2007}),
  \bibinfo{pages}{179--188}.
\newblock
\showISSN{0362-1340}
\urldef\tempurl%
\url{https://doi.org/10.1145/1273442.1250755}
\showDOI{\tempurl}


\bibitem[\protect\citeauthoryear{K\"{u}fner, Nestmann, and Rickmann}{K\"{u}fner
  et~al\mbox{.}}{2012}]%
        {kufner2012formal}
\bibfield{author}{\bibinfo{person}{Philipp K\"{u}fner}, \bibinfo{person}{Uwe
  Nestmann}, {and} \bibinfo{person}{Christina Rickmann}.}
  \bibinfo{year}{2012}\natexlab{}.
\newblock \showarticletitle{Formal Verification of Distributed Algorithms: From
  Pseudo Code to Checked Proofs}. In \bibinfo{booktitle}{\emph{Proceedings of
  the 7th IFIP TC 1/WG 202 International Conference on Theoretical Computer
  Science}} \emph{(\bibinfo{series}{TCS '12})}.
  \bibinfo{publisher}{Springer-Verlag}, \bibinfo{address}{Berlin, Heidelberg},
  \bibinfo{pages}{209--224}.
\newblock
\showISBNx{978-3-642-33474-0}
\urldef\tempurl%
\url{https://doi.org/10.1007/978-3-642-33475-7_15}
\showDOI{\tempurl}


\bibitem[\protect\citeauthoryear{Kuppe}{Kuppe}{2017}]%
        {kuppe2017thesis}
\bibfield{author}{\bibinfo{person}{Markus~A. Kuppe}.}
  \bibinfo{year}{2017}\natexlab{}.
\newblock \emph{\bibinfo{title}{A Verified and Scalable Hash Table for the TLC
  Model Checker: Towards an Order of Magnitude Speedup}}.
\newblock \bibinfo{thesistype}{Master's\ thesis}. \bibinfo{school}{University
  of Hamburg}.
\newblock
\urldef\tempurl%
\url{http://www.lemmster.de/talks/MSc_MarkusAKuppe_1497363471.pdf}
\showURL{%
\tempurl}


\bibitem[\protect\citeauthoryear{Kuppe}{Kuppe}{2018}]%
        {kuppe2018rise}
\bibfield{author}{\bibinfo{person}{Markus~A. Kuppe}.}
  \bibinfo{year}{2018}\natexlab{}.
\newblock \bibinfo{title}{Let TLA+ RiSE}.
\newblock \bibinfo{howpublished}{RiSE group all-hands meeting}.
\newblock
\urldef\tempurl%
\url{http://www.lemmster.de/talks/Let_TLA_RiSE_-_Markus_Alexander_Kuppe.pdf}
\showURL{%
Retrieved November 8, 2019 from \tempurl}


\bibitem[\protect\citeauthoryear{Kuppe}{Kuppe}{2019}]%
        {kuppe2019hyperscale}
\bibfield{author}{\bibinfo{person}{Markus~A. Kuppe}.}
  \bibinfo{year}{2019}\natexlab{}.
\newblock \bibinfo{title}{Debuggable Design with TLA+}.
\newblock \bibinfo{howpublished}{Hyperscale Verification}.
\newblock
\urldef\tempurl%
\url{http://www.lemmster.de/talks/Hyperscale_Verification_-_Markus_Alexander_Kuppe.pdf}
\showURL{%
Retrieved November 8, 2019 from \tempurl}


\bibitem[\protect\citeauthoryear{Kuppe et~al\mbox{.}}{Kuppe
  et~al\mbox{.}}{2019}]%
        {kuppe2019tlcgit}
\bibfield{author}{\bibinfo{person}{Markus~A. Kuppe} {et~al\mbox{.}}}
  \bibinfo{year}{2019}\natexlab{}.
\newblock \bibinfo{title}{TLC Github}.
\newblock
\newblock
\urldef\tempurl%
\url{https://github.com/tlaplus/tlaplus}
\showURL{%
Retrieved November 8, 2019 from \tempurl}


\bibitem[\protect\citeauthoryear{Lamport}{Lamport}{1978}]%
        {lamport1978time}
\bibfield{author}{\bibinfo{person}{Leslie Lamport}.}
  \bibinfo{year}{1978}\natexlab{}.
\newblock \showarticletitle{Time, Clocks, and the Ordering of Events in a
  Distributed System}.
\newblock \bibinfo{journal}{\emph{Commun. ACM}} \bibinfo{volume}{21},
  \bibinfo{number}{7} (\bibinfo{date}{July} \bibinfo{year}{1978}),
  \bibinfo{pages}{558--565}.
\newblock
\showISSN{0001-0782}
\urldef\tempurl%
\url{https://doi.org/10.1145/359545.359563}
\showDOI{\tempurl}


\bibitem[\protect\citeauthoryear{Lamport}{Lamport}{1994}]%
        {lamport1994temporal}
\bibfield{author}{\bibinfo{person}{Leslie Lamport}.}
  \bibinfo{year}{1994}\natexlab{}.
\newblock \showarticletitle{The Temporal Logic of Actions}.
\newblock \bibinfo{journal}{\emph{ACM Trans. Program. Lang. Syst.}}
  \bibinfo{volume}{16}, \bibinfo{number}{3} (\bibinfo{date}{May}
  \bibinfo{year}{1994}), \bibinfo{pages}{872--923}.
\newblock
\showISSN{0164-0925}
\urldef\tempurl%
\url{https://doi.org/10.1145/177492.177726}
\showDOI{\tempurl}


\bibitem[\protect\citeauthoryear{Lamport}{Lamport}{1998}]%
        {lamport1998part}
\bibfield{author}{\bibinfo{person}{Leslie Lamport}.}
  \bibinfo{year}{1998}\natexlab{}.
\newblock \showarticletitle{The Part-time Parliament}.
\newblock \bibinfo{journal}{\emph{ACM Trans. Comput. Syst.}}
  \bibinfo{volume}{16}, \bibinfo{number}{2} (\bibinfo{date}{May}
  \bibinfo{year}{1998}), \bibinfo{pages}{133--169}.
\newblock
\showISSN{0734-2071}
\urldef\tempurl%
\url{https://doi.org/10.1145/279227.279229}
\showDOI{\tempurl}


\bibitem[\protect\citeauthoryear{Lamport}{Lamport}{2001}]%
        {lamport2001paxos}
\bibfield{author}{\bibinfo{person}{Leslie Lamport}.}
  \bibinfo{year}{2001}\natexlab{}.
\newblock \showarticletitle{Paxos made simple}.
\newblock \bibinfo{journal}{\emph{ACM SIGACT News}} \bibinfo{volume}{32},
  \bibinfo{number}{4} (\bibinfo{date}{Dec.} \bibinfo{year}{2001}),
  \bibinfo{pages}{51--58}.
\newblock
\showISSN{0163-5700}
\urldef\tempurl%
\url{https://doi.org/10.1145/568425.568433}
\showDOI{\tempurl}


\bibitem[\protect\citeauthoryear{Lamport}{Lamport}{2002}]%
        {lamport2002specifying}
\bibfield{author}{\bibinfo{person}{Leslie Lamport}.}
  \bibinfo{year}{2002}\natexlab{}.
\newblock \bibinfo{booktitle}{\emph{Specifying Systems: The TLA+ Language and
  Tools for Hardware and Software Engineers}}.
\newblock \bibinfo{publisher}{Addison-Wesley Longman Publishing Co., Inc.},
  \bibinfo{address}{Boston, MA, USA}.
\newblock
\showISBNx{032114306X}


\bibitem[\protect\citeauthoryear{Lamport}{Lamport}{2006}]%
        {lamport2006fast}
\bibfield{author}{\bibinfo{person}{Leslie Lamport}.}
  \bibinfo{year}{2006}\natexlab{}.
\newblock \showarticletitle{Fast Paxos}.
\newblock \bibinfo{journal}{\emph{Distributed Computing}} \bibinfo{volume}{19},
  \bibinfo{number}{2} (\bibinfo{date}{Oct.} \bibinfo{year}{2006}),
  \bibinfo{pages}{79--103}.
\newblock
\showISSN{1432-0452}
\urldef\tempurl%
\url{https://doi.org/10.1007/s00446-006-0005-x}
\showDOI{\tempurl}


\bibitem[\protect\citeauthoryear{Lamport}{Lamport}{2011}]%
        {lamport2011byzantizing}
\bibfield{author}{\bibinfo{person}{Leslie Lamport}.}
  \bibinfo{year}{2011}\natexlab{}.
\newblock \showarticletitle{Byzantizing Paxos by Refinement}. In
  \bibinfo{booktitle}{\emph{Proceedings of the 25th International Conference on
  Distributed Computing}} \emph{(\bibinfo{series}{DISC'11})}.
  \bibinfo{publisher}{Springer-Verlag}, \bibinfo{address}{Berlin, Heidelberg},
  \bibinfo{pages}{211--224}.
\newblock
\showISBNx{978-3-642-24099-7}
\urldef\tempurl%
\url{http://dl.acm.org/citation.cfm?id=2075029.2075058}
\showURL{%
\tempurl}


\bibitem[\protect\citeauthoryear{Lamport}{Lamport}{2012}]%
        {lamport2012write}
\bibfield{author}{\bibinfo{person}{Leslie Lamport}.}
  \bibinfo{year}{2012}\natexlab{}.
\newblock \showarticletitle{How to write a 21\textsuperscript{st} century
  proof}.
\newblock \bibinfo{journal}{\emph{Journal of Fixed Point Theory and
  Applications}} \bibinfo{volume}{11}, \bibinfo{number}{1}
  (\bibinfo{date}{March} \bibinfo{year}{2012}), \bibinfo{pages}{43--63}.
\newblock
\showISSN{1661-7746}
\urldef\tempurl%
\url{https://doi.org/10.1007/s11784-012-0071-6}
\showDOI{\tempurl}


\bibitem[\protect\citeauthoryear{Lamport, Merz, and Doligez}{Lamport
  et~al\mbox{.}}{2014}]%
        {basicpaxos2014}
\bibfield{author}{\bibinfo{person}{Leslie Lamport}, \bibinfo{person}{Stephan
  Merz}, {and} \bibinfo{person}{Damien Doligez}.}
  \bibinfo{year}{2014}\natexlab{}.
\newblock \bibinfo{title}{Paxos.tla}.
\newblock
\newblock
\urldef\tempurl%
\url{https://github.com/tlaplus/v1-tlapm/blob/master/examples/paxos/Paxos.tla}
\showURL{%
Retrieved February 6, 2018 from \tempurl}


\bibitem[\protect\citeauthoryear{Leino}{Leino}{2010}]%
        {leino2010dafny}
\bibfield{author}{\bibinfo{person}{K.~Rustan~M. Leino}.}
  \bibinfo{year}{2010}\natexlab{}.
\newblock \showarticletitle{Dafny: An Automatic Program Verifier for Functional
  Correctness}. In \bibinfo{booktitle}{\emph{Proceedings of the 16th
  International Conference on Logic for Programming, Artificial Intelligence,
  and Reasoning}} \emph{(\bibinfo{series}{LPAR'10})}.
  \bibinfo{publisher}{Springer-Verlag}, \bibinfo{address}{Berlin, Heidelberg},
  \bibinfo{pages}{348--370}.
\newblock
\showISBNx{3-642-17510-4, 978-3-642-17510-7}
\urldef\tempurl%
\url{http://dl.acm.org/citation.cfm?id=1939141.1939161}
\showURL{%
\tempurl}


\bibitem[\protect\citeauthoryear{Liskov}{Liskov}{2010}]%
        {liskov2010viewstamped}
\bibfield{author}{\bibinfo{person}{Barbara Liskov}.}
  \bibinfo{year}{2010}\natexlab{}.
\newblock \showarticletitle{Replication}.
\newblock \bibinfo{publisher}{Springer-Verlag}, \bibinfo{address}{Berlin,
  Heidelberg}, Chapter From Viewstamped Replication to Byzantine Fault
  Tolerance, \bibinfo{pages}{121--149}.
\newblock
\showISBNx{3-642-11293-5, 978-3-642-11293-5}
\urldef\tempurl%
\url{http://dl.acm.org/citation.cfm?id=2172338.2172345}
\showURL{%
\tempurl}


\bibitem[\protect\citeauthoryear{Liu, Guo, Wang, Chen, Lian, Tang, Wu,
  Kaashoek, and Zhang}{Liu et~al\mbox{.}}{2008}]%
        {liu2008d3s}
\bibfield{author}{\bibinfo{person}{Xuezheng Liu}, \bibinfo{person}{Zhenyu Guo},
  \bibinfo{person}{Xi Wang}, \bibinfo{person}{Feibo Chen},
  \bibinfo{person}{Xiaochen Lian}, \bibinfo{person}{Jian Tang},
  \bibinfo{person}{Ming Wu}, \bibinfo{person}{M.~Frans Kaashoek}, {and}
  \bibinfo{person}{Zheng Zhang}.} \bibinfo{year}{2008}\natexlab{}.
\newblock \showarticletitle{D3S: Debugging Deployed Distributed Systems}. In
  \bibinfo{booktitle}{\emph{Proceedings of the 5th USENIX Symposium on
  Networked Systems Design and Implementation}}
  \emph{(\bibinfo{series}{NSDI'08})}. \bibinfo{publisher}{USENIX Association},
  \bibinfo{address}{Berkeley, CA, USA}, \bibinfo{pages}{423--437}.
\newblock
\showISBNx{111-999-5555-22-1}
\urldef\tempurl%
\url{http://dl.acm.org/citation.cfm?id=1387589.1387619}
\showURL{%
\tempurl}


\bibitem[\protect\citeauthoryear{Liu, Chand, and Stoller}{Liu
  et~al\mbox{.}}{2019}]%
        {liu2019moderately}
\bibfield{author}{\bibinfo{person}{Yanhong~A. Liu}, \bibinfo{person}{Saksham
  Chand}, {and} \bibinfo{person}{Scott~D. Stoller}.}
  \bibinfo{year}{2019}\natexlab{}.
\newblock \showarticletitle{Moderately Complex Paxos Made Simple: High-Level
  Executable Specification of Distributed Algorithms}. In
  \bibinfo{booktitle}{\emph{Proceedings of the 21st International Symposium on
  Principles and Practice of Programming Languages 2019}}
  \emph{(\bibinfo{series}{PPDP '19})}. \bibinfo{publisher}{ACM},
  \bibinfo{address}{New York, NY, USA}, Article \bibinfo{articleno}{15},
  \bibinfo{numpages}{15}~pages.
\newblock
\showISBNx{978-1-4503-7249-7}
\urldef\tempurl%
\url{https://doi.org/10.1145/3354166.3354180}
\showDOI{\tempurl}


\bibitem[\protect\citeauthoryear{Maekawa}{Maekawa}{1985}]%
        {maekawa1985algorithm}
\bibfield{author}{\bibinfo{person}{Mamoru Maekawa}.}
  \bibinfo{year}{1985}\natexlab{}.
\newblock \showarticletitle{A $\sqrt{N}$ Algorithm for Mutual Exclusion in
  Decentralized Systems}.
\newblock \bibinfo{journal}{\emph{ACM Trans. Comput. Syst.}}
  \bibinfo{volume}{3}, \bibinfo{number}{2} (\bibinfo{date}{May}
  \bibinfo{year}{1985}), \bibinfo{pages}{145--159}.
\newblock
\showISSN{0734-2071}
\urldef\tempurl%
\url{https://doi.org/10.1145/214438.214445}
\showDOI{\tempurl}


\bibitem[\protect\citeauthoryear{Merz}{Merz}{2003}]%
        {merz2003logic}
\bibfield{author}{\bibinfo{person}{Stephan Merz}.}
  \bibinfo{year}{2003}\natexlab{}.
\newblock \showarticletitle{On the Logic of TLA+}.
\newblock \bibinfo{journal}{\emph{Computing and Informatics}}
  \bibinfo{volume}{22}, \bibinfo{number}{3-4} (\bibinfo{year}{2003}),
  \bibinfo{pages}{351--379}.
\newblock
\showISSN{2585-8807}
\urldef\tempurl%
\url{http://www.cai.sk/ojs/index.php/cai/article/view/460/367}
\showURL{%
\tempurl}


\bibitem[\protect\citeauthoryear{Merz}{Merz}{2008}]%
        {merz2008spec}
\bibfield{author}{\bibinfo{person}{Stephan Merz}.}
  \bibinfo{year}{2008}\natexlab{}.
\newblock \bibinfo{booktitle}{\emph{The Specification Language TLA+}}.
\newblock \bibinfo{publisher}{Springer Berlin Heidelberg},
  \bibinfo{address}{Berlin, Heidelberg}, \bibinfo{pages}{401--451}.
\newblock
\showISBNx{978-3-540-74107-7}
\urldef\tempurl%
\url{https://doi.org/10.1007/978-3-540-74107-7_8}
\showDOI{\tempurl}


\bibitem[\protect\citeauthoryear{Merz}{Merz}{2017}]%
        {tlapsbug}
\bibfield{author}{\bibinfo{person}{Stephan Merz}.}
  \bibinfo{year}{2017}\natexlab{}.
\newblock \bibinfo{title}{Coalescing bug in {TLAPS}}.
\newblock
\newblock
\urldef\tempurl%
\url{https://github.com/tlaplus/tlaplus/issues/40}
\showURL{%
Retrieved March 16, 2018 from \tempurl}


\bibitem[\protect\citeauthoryear{Merz, Lu, and Weidenbach}{Merz
  et~al\mbox{.}}{2011}]%
        {merz2011pastry}
\bibfield{author}{\bibinfo{person}{Stephan Merz}, \bibinfo{person}{Tianxiang
  Lu}, {and} \bibinfo{person}{Christoph Weidenbach}.}
  \bibinfo{year}{2011}\natexlab{}.
\newblock \showarticletitle{{Towards Verification of the Pastry Protocol using
  TLA+}}. In \bibinfo{booktitle}{\emph{{31st IFIP International Conference on
  Formal Techniques for Networked and Distributed Systems}}}
  \emph{(\bibinfo{series}{FMOODS/FORTE 2011})}, Vol.~\bibinfo{volume}{6722}.
  \bibinfo{pages}{244--258}.
\newblock
\urldef\tempurl%
\url{https://hal.inria.fr/inria-00593523}
\showURL{%
\tempurl}


\bibitem[\protect\citeauthoryear{Merz and Vanzetto}{Merz and Vanzetto}{2012a}]%
        {merz2012automatic}
\bibfield{author}{\bibinfo{person}{Stephan Merz} {and}
  \bibinfo{person}{Hern\'{a}n Vanzetto}.} \bibinfo{year}{2012}\natexlab{a}.
\newblock \showarticletitle{Automatic Verification of TLA+ Proof Obligations
  with SMT Solvers}. In \bibinfo{booktitle}{\emph{Proceedings of the 18th
  International Conference on Logic for Programming, Artificial Intelligence,
  and Reasoning}} \emph{(\bibinfo{series}{LPAR'12})}.
  \bibinfo{publisher}{Springer-Verlag}, \bibinfo{address}{Berlin, Heidelberg},
  \bibinfo{pages}{289--303}.
\newblock
\showISBNx{978-3-642-28716-9}
\urldef\tempurl%
\url{https://doi.org/10.1007/978-3-642-28717-6_23}
\showDOI{\tempurl}


\bibitem[\protect\citeauthoryear{Merz and Vanzetto}{Merz and Vanzetto}{2012b}]%
        {merz2012harnessing}
\bibfield{author}{\bibinfo{person}{Stephan Merz} {and}
  \bibinfo{person}{Hern{\'a}n Vanzetto}.} \bibinfo{year}{2012}\natexlab{b}.
\newblock \showarticletitle{Harnessing SMT Solvers for TLA+ Proofs}. In
  \bibinfo{booktitle}{\emph{Proceedings of the 12th International Workshop on
  Automated Verification of Critical Systems}} \emph{(\bibinfo{series}{AVoCS
  '12})}, Vol.~\bibinfo{volume}{53}. \bibinfo{publisher}{European Association
  of Software Science and Technology}, \bibinfo{pages}{1--15}.
\newblock
\showISSN{1863-2122}
\urldef\tempurl%
\url{https://doi.org/10.14279/tuj.eceasst.53.766}
\showDOI{\tempurl}


\bibitem[\protect\citeauthoryear{Newcombe}{Newcombe}{2014}]%
        {newcombe2014tla}
\bibfield{author}{\bibinfo{person}{Chris Newcombe}.}
  \bibinfo{year}{2014}\natexlab{}.
\newblock \showarticletitle{Why Amazon Chose TLA+}. In
  \bibinfo{booktitle}{\emph{Proceedings of the 4th International Conference on
  Abstract State Machines, Alloy, B, TLA, VDM, and Z - Volume 8477}}
  \emph{(\bibinfo{series}{ABZ 2014})}. \bibinfo{publisher}{Springer-Verlag New
  York, Inc.}, \bibinfo{address}{New York, NY, USA}, \bibinfo{pages}{25--39}.
\newblock
\showISBNx{978-3-662-43651-6}
\urldef\tempurl%
\url{https://doi.org/10.1007/978-3-662-43652-3_3}
\showDOI{\tempurl}


\bibitem[\protect\citeauthoryear{Ongaro and Ousterhout}{Ongaro and
  Ousterhout}{2014}]%
        {ongaro2014search}
\bibfield{author}{\bibinfo{person}{Diego Ongaro} {and} \bibinfo{person}{John
  Ousterhout}.} \bibinfo{year}{2014}\natexlab{}.
\newblock \showarticletitle{In Search of an Understandable Consensus
  Algorithm}. In \bibinfo{booktitle}{\emph{Proceedings of the 2014 USENIX
  Conference on USENIX Annual Technical Conference}}
  \emph{(\bibinfo{series}{USENIX ATC'14})}. \bibinfo{publisher}{USENIX
  Association}, \bibinfo{address}{Berkeley, CA, USA},
  \bibinfo{pages}{305--320}.
\newblock
\showISBNx{978-1-931971-10-2}
\urldef\tempurl%
\url{http://dl.acm.org/citation.cfm?id=2643634.2643666}
\showURL{%
\tempurl}


\bibitem[\protect\citeauthoryear{Owre, Rushby, and Shankar}{Owre
  et~al\mbox{.}}{1992}]%
        {owre1992pvs}
\bibfield{author}{\bibinfo{person}{Sam Owre}, \bibinfo{person}{John~M. Rushby},
  {and} \bibinfo{person}{Natarajan Shankar}.} \bibinfo{year}{1992}\natexlab{}.
\newblock \showarticletitle{PVS: A Prototype Verification System}. In
  \bibinfo{booktitle}{\emph{Proceedings of the 11th International Conference on
  Automated Deduction: Automated Deduction}}
  \emph{(\bibinfo{series}{CADE-11})}. \bibinfo{publisher}{Springer-Verlag},
  \bibinfo{address}{London, UK, UK}, \bibinfo{pages}{748--752}.
\newblock
\showISBNx{3-540-55602-8}
\urldef\tempurl%
\url{http://dl.acm.org/citation.cfm?id=648230.752639}
\showURL{%
\tempurl}


\bibitem[\protect\citeauthoryear{Padon, Losa, Sagiv, and Shoham}{Padon
  et~al\mbox{.}}{2017}]%
        {padon2017paxos}
\bibfield{author}{\bibinfo{person}{Oded Padon}, \bibinfo{person}{Giuliano
  Losa}, \bibinfo{person}{Mooly Sagiv}, {and} \bibinfo{person}{Sharon Shoham}.}
  \bibinfo{year}{2017}\natexlab{}.
\newblock \showarticletitle{Paxos Made EPR: Decidable Reasoning About
  Distributed Protocols}.
\newblock \bibinfo{journal}{\emph{Proc. ACM Program. Lang.}}
  \bibinfo{volume}{1}, \bibinfo{number}{OOPSLA}, Article
  \bibinfo{articleno}{108} (\bibinfo{date}{Oct.} \bibinfo{year}{2017}),
  \bibinfo{numpages}{31}~pages.
\newblock
\showISSN{2475-1421}
\urldef\tempurl%
\url{https://doi.org/10.1145/3140568}
\showDOI{\tempurl}


\bibitem[\protect\citeauthoryear{research INRIA Joint~Centre}{research INRIA
  Joint~Centre}{2014}]%
        {tlaps2019provers}
\bibfield{author}{\bibinfo{person}{Microsoft research INRIA Joint~Centre}.}
  \bibinfo{year}{2014}\natexlab{}.
\newblock \bibinfo{title}{TLA+ Proof System, Tactics}.
\newblock
\newblock
\urldef\tempurl%
\url{https://tla.msr-inria.inria.fr/tlaps/content/Documentation/Tutorial/Tactics.html}
\showURL{%
Retrieved November 1, 2019 from \tempurl}


\bibitem[\protect\citeauthoryear{Schiper, Rahli, Renesse, Bickford, and
  Constable}{Schiper et~al\mbox{.}}{2014}]%
        {schiper2014developing}
\bibfield{author}{\bibinfo{person}{Nicolas Schiper}, \bibinfo{person}{Vincent
  Rahli}, \bibinfo{person}{Robbert~Van Renesse}, \bibinfo{person}{Marck
  Bickford}, {and} \bibinfo{person}{Robert~L. Constable}.}
  \bibinfo{year}{2014}\natexlab{}.
\newblock \showarticletitle{Developing Correctly Replicated Databases Using
  Formal Tools}. In \bibinfo{booktitle}{\emph{Proceedings of the 2014 44th
  Annual IEEE/IFIP International Conference on Dependable Systems and
  Networks}} \emph{(\bibinfo{series}{DSN '14})}. \bibinfo{publisher}{IEEE
  Computer Society}, \bibinfo{address}{Washington, DC, USA},
  \bibinfo{pages}{395--406}.
\newblock
\showISBNx{978-1-4799-2233-8}
\urldef\tempurl%
\url{https://doi.org/10.1109/DSN.2014.45}
\showDOI{\tempurl}


\bibitem[\protect\citeauthoryear{Van~Renesse and Altinbuken}{Van~Renesse and
  Altinbuken}{2015}]%
        {van2015paxos}
\bibfield{author}{\bibinfo{person}{Robbert Van~Renesse} {and}
  \bibinfo{person}{Deniz Altinbuken}.} \bibinfo{year}{2015}\natexlab{}.
\newblock \showarticletitle{Paxos Made Moderately Complex}.
\newblock \bibinfo{journal}{\emph{ACM Comput. Surv.}} \bibinfo{volume}{47},
  \bibinfo{number}{3}, Article \bibinfo{articleno}{42} (\bibinfo{date}{Feb.}
  \bibinfo{year}{2015}), \bibinfo{numpages}{36}~pages.
\newblock
\showISSN{0360-0300}
\urldef\tempurl%
\url{https://doi.org/10.1145/2673577}
\showDOI{\tempurl}


\bibitem[\protect\citeauthoryear{Wilcox, Woos, Panchekha, Tatlock, Wang, Ernst,
  and Anderson}{Wilcox et~al\mbox{.}}{2015}]%
        {wilcox2015verdi}
\bibfield{author}{\bibinfo{person}{James~R. Wilcox}, \bibinfo{person}{Doug
  Woos}, \bibinfo{person}{Pavel Panchekha}, \bibinfo{person}{Zachary Tatlock},
  \bibinfo{person}{Xi Wang}, \bibinfo{person}{Michael~D. Ernst}, {and}
  \bibinfo{person}{Thomas Anderson}.} \bibinfo{year}{2015}\natexlab{}.
\newblock \showarticletitle{Verdi: A Framework for Implementing and Formally
  Verifying Distributed Systems}.
\newblock \bibinfo{journal}{\emph{SIGPLAN Not.}} \bibinfo{volume}{50},
  \bibinfo{number}{6} (\bibinfo{date}{June} \bibinfo{year}{2015}),
  \bibinfo{pages}{357--368}.
\newblock
\showISSN{0362-1340}
\urldef\tempurl%
\url{https://doi.org/10.1145/2813885.2737958}
\showDOI{\tempurl}


\bibitem[\protect\citeauthoryear{Yabandeh, Kne\v{z}evi\'{c}, Kosti\'{c}, and
  Kuncak}{Yabandeh et~al\mbox{.}}{2010}]%
        {yabandeh2010predicting}
\bibfield{author}{\bibinfo{person}{Maysam Yabandeh}, \bibinfo{person}{Nikola
  Kne\v{z}evi\'{c}}, \bibinfo{person}{Dejan Kosti\'{c}}, {and}
  \bibinfo{person}{Viktor Kuncak}.} \bibinfo{year}{2010}\natexlab{}.
\newblock \showarticletitle{Predicting and Preventing Inconsistencies in
  Deployed Distributed Systems}.
\newblock \bibinfo{journal}{\emph{ACM Trans. Comput. Syst.}}
  \bibinfo{volume}{28}, \bibinfo{number}{1}, Article \bibinfo{articleno}{2}
  (\bibinfo{date}{Aug.} \bibinfo{year}{2010}), \bibinfo{numpages}{49}~pages.
\newblock
\showISSN{0734-2071}
\urldef\tempurl%
\url{https://doi.org/10.1145/1731060.1731062}
\showDOI{\tempurl}


\bibitem[\protect\citeauthoryear{Yang, Chen, Wu, Xu, Liu, Lin, Yang, Long,
  Zhang, and Zhou}{Yang et~al\mbox{.}}{2009}]%
        {yang2009modist}
\bibfield{author}{\bibinfo{person}{Junfeng Yang}, \bibinfo{person}{Tisheng
  Chen}, \bibinfo{person}{Ming Wu}, \bibinfo{person}{Zhilei Xu},
  \bibinfo{person}{Xuezheng Liu}, \bibinfo{person}{Haoxiang Lin},
  \bibinfo{person}{Mao Yang}, \bibinfo{person}{Fan Long},
  \bibinfo{person}{Lintao Zhang}, {and} \bibinfo{person}{Lidong Zhou}.}
  \bibinfo{year}{2009}\natexlab{}.
\newblock \showarticletitle{MODIST: Transparent Model Checking of Unmodified
  Distributed Systems}. In \bibinfo{booktitle}{\emph{Proceedings of the 6th
  USENIX Symposium on Networked Systems Design and Implementation}}
  \emph{(\bibinfo{series}{NSDI'09})}. \bibinfo{publisher}{USENIX Association},
  \bibinfo{address}{Berkeley, CA, USA}, \bibinfo{pages}{213--228}.
\newblock
\urldef\tempurl%
\url{http://dl.acm.org/citation.cfm?id=1558977.1558992}
\showURL{%
\tempurl}


\bibitem[\protect\citeauthoryear{Yu, Manolios, and Lamport}{Yu
  et~al\mbox{.}}{1999}]%
        {yu1999model}
\bibfield{author}{\bibinfo{person}{Yuan Yu}, \bibinfo{person}{Panagiotis
  Manolios}, {and} \bibinfo{person}{Leslie Lamport}.}
  \bibinfo{year}{1999}\natexlab{}.
\newblock \showarticletitle{Model Checking TLA+ Specifications}. In
  \bibinfo{booktitle}{\emph{Proceedings of the 10th IFIP WG 10.5 Advanced
  Research Working Conference on Correct Hardware Design and Verification
  Methods}} \emph{(\bibinfo{series}{CHARME '99})}.
  \bibinfo{publisher}{Springer-Verlag}, \bibinfo{address}{London, UK},
  \bibinfo{pages}{54--66}.
\newblock
\showISBNx{3-540-66559-5}
\urldef\tempurl%
\url{http://dl.acm.org/citation.cfm?id=646704.702012}
\showURL{%
\tempurl}


\end{thebibliography}
}
%\end{document}

\shortonly{
\appendix
\section{Additional predicates and invariants}
\label{appendix:defs}

This appendix defines predicates and invariants not defined above due to space limits.

\mypar{Predicates}
\defSafeAt
\defAccInv

\mypar{Invariants about messages}
\defMsgInvTwo
}

\notforThesis{
\newpage
\appendix
\setlength{\baselineskip}{1.98ex}
\input{tlatex.sty}
\section{\texorpdfstring{\tlaplus{}}{TLA+} specification of Multi-Paxos with Preemption}
\label{appendix:spec}
\tlatex
\@x{}\moduleLeftDash\@xx{ {\MODULE} MultiPaxosSpec}\moduleRightDash\@xx{}%
\begin{lcom}{0}%
\begin{cpar}{0}{F}{F}{0}{0}{}%
This is a specification in TLA+ and machine checked proof in TLAPS of
Multi-Paxos with Preemption.
\end{cpar}%
\end{lcom}%

%P, A, Q, V  Sets of proposers, acceptors, quorums of acceptors, values to propose
%msgs, pBal, aBal, aVoted
%vote v for s in/with b; vote <b,s,v>
\@x{ {\EXTENDS} Integers ,\, TLAPS ,\, FiniteSets ,\, FiniteSetTheorems}%
\@x{ {\CONSTANTS} {\mathcal{P}} ,\, {\mathcal{A}} ,\, {\mathcal{Q}} ,\, {\mathcal{V}} \@s{17.75}}%
\@y{\@s{0.0}%
 Sets of proposers, acceptors, quorums of acceptors, and values to propose
}%
\@pvspace{8.0pt}%

\@x{ {\VARIABLES} msgs ,\,\@s{10.93}}%
\@y{\@s{0.0}%
 Set of sent messages
}%
\@xx{}%
\@x{\@s{41.14} pBal ,\,\@s{11.67}}%
\@y{\@s{0.0}%
 For each proposer, the current ballot of the proposer
}%
\@xx{}%
\@x{\@s{41.14} aBal ,\,\@s{11.67}}%
\@y{\@s{0.0}%
 For each acceptor, the highest ballot seen by the acceptor
}%
\@xx{}%
\@x{\@s{41.14} aVoted \,\@s{7.35}}%
\@y{\@s{0.0}%
 For each acceptor, a subset of \ensuremath{{\langle}ballot,\, slot,\, value{\rangle}} triples that the acceptor has voted %\ensuremath{VotedForIn} %annie: not use as it is not yet defined
}
\@pvspace{0.0pt}%
%\@x{\@s{75.75}}%
%\@y{\@s{0.0}%
%For every slot voted in, the acceptor only stores the triple with the highest ballot
%}%
\@xx{}%
\@x{ {\ASSUME} QuorumAssumption \.{\defeq} {\mathcal{Q}} \.{\subseteq} {\SUBSET}
 {\mathcal{A}} \.{\land} \A\, Q1 ,\, Q2 \.{\in} {\mathcal{Q}} \.{:} Q1 \.{\cap} Q2
 \.{\neq} {\emptyset}}%
\@pvspace{8.0pt}%
\@x{ {\mathcal{B}} \.{\defeq} {\mathds{N}}\@s{9.5}}%
\@y{\@s{0.0}%
 Set of ballots
}%
\@xx{}%
\@x{ {\mathcal{S}} \.{\defeq} {\mathds{N}}\@s{10}}%
\@y{\@s{0.0}%
 Set of slots
}%
\@xx{}%
\@x{ vars \.{\defeq} {\langle} msgs ,\, pBal ,\, aBal ,\, aVoted {\rangle}}%
\@x{ Send ( m ) \.{\defeq} msgs \.{'} \.{=} msgs \.{\cup} \{ m \}}%
\@pvspace{8.0pt}%
%\begin{lcom}{0}%
%\begin{cpar}{0}{F}{F}{0}{0}{}%
%Following is the specification of the phases of \textit{Multi-Paxos}. These are described in Section~\ref{secSpec}.
%\end{cpar}%
%\end{lcom}%

\begin{lcom}{0}%
\begin{cpar}{0}{F}{F}{0}{0}{}%
 Phase \ensuremath{1a}: For a proposer \ensuremath{p}, this phase selects some ballot number \ensuremath{pBal[p]} with which
 a \textsf{1a} message has not been sent,
 %This number is sent to any set of acceptors which contains at least one quorum from \ensuremath{{\mathcal{Q}}}.
 %Trivially it can be broadcast 
 and sends it (to all processes). 
 %For safety, any subset of
 %\ensuremath{{\mathcal{A}}} would suffice. For liveness, a subset containing at least one
 %quorum is needed.
\end{cpar}%
\end{lcom}%
 \@x{ Phase1a ( p ) \.{\defeq} }%
 \@x{\@s{4.0} \.{\land} Send ( [ type \.{\mapsto}\@w{1a} ,\, from
 \.{\mapsto} p ,\, bal \.{\mapsto} pBal [ p ] ] )}%
 \@x{\@s{4.0} \.{\land} {\UNCHANGED} {\langle} pBal ,\, aBal ,\, aVoted
 {\rangle}}%
\@pvspace{8.0pt}%
\begin{lcom}{0}%
\begin{cpar}{0}{F}{F}{0}{0}{}%
 Phase \ensuremath{1b}: For an acceptor \ensuremath{a}, if there is
 a \textsf{1a} message \ensuremath{m} with
 ballot \ensuremath{m.bal} that is
 %than that of any \textsf{1a} message to which it has already responded,
 higher than the highest it has seen, % \ensuremath{aBal[a]}, then
 \ensuremath{a} sends a \textsf{1b} message with \ensuremath{m.bal} and with the set of highest-numbered triples it has voted for each slot, and it updates the highest ballot it has seen to be $m.bal$;
 %a promise, that is a \textsf{1b} message, not to accept any more proposals
 %for ballots numbered less than \ensuremath{m.bal} by updating its \ensuremath{aBal} to \ensuremath{m.bal}; 
 otherwise it sends a \textsf{preempt} message
back with the highest ballot it has seen.
 %In case of a \textsf{1b} reply, the acceptor writes a mapping in
 %\ensuremath{{\mathcal{S}} \.{\rightarrow}
 %{\mathcal{B}} \.{\times} {\mathcal{V}}}. This mapping reveals for each slot, the value
 %that the acceptor most recently (that is, highest ballot) voted on, if
 %any. This is maintained in \ensuremath{aVoted}.
 %annie: I think voted are now triples, not S->B x V any more. ?  same to fix in paper body.
\end{cpar}%
\end{lcom}%
 \@x{ Phase1b ( a ) \.{\defeq} \E\, m \.{\in} msgs \.{:} m . type \.{=}\@w{1a}
 \.{\land}}%
\@x{\@s{4.0} {\IF} m . bal \.{>} aBal [ a ] \.{\THEN}}%
 \@x{\@s{13.72} \.{\land} Send ( [ type \.{\mapsto}\@w{1b} ,\, from
 \.{\mapsto} a ,\, bal \.{\mapsto} m . bal ,\, voted \.{\mapsto} aVoted [ a ]
 ] )}%
 \@x{\@s{13.72} \.{\land} aBal \.{'} \.{=} [ aBal {\EXCEPT} {\bang} [ a ]
 \.{=} m . bal ]}%
\@x{\@s{13.72} \.{\land} {\UNCHANGED} {\langle} pBal ,\, aVoted {\rangle}}%
\@x{\@s{2.0} \.{\ELSE}}%
 \@x{\@s{10.0} \.{\land} Send ( [ type \.{\mapsto}\@w{preempt} ,\, to
 \.{\mapsto} m . from ,\, bal \.{\mapsto} aBal [ a ] ] )}%
 \@x{\@s{10.0} \.{\land} {\UNCHANGED} {\langle} pBal ,\, aBal ,\, aVoted
 {\rangle}}%
\@pvspace{8.0pt}%
\begin{lcom}{0}%
\begin{cpar}{0}{F}{F}{0}{0}{}%
 Phase \ensuremath{2a}: For a proposer \ensuremath{p}, 
 if there is no \textsf{2a} message with current ballot \ensuremath{pBal[p]}, and 
 %there is a set \ensuremath{S} of %it receives a response to its \textsf{1a} message (for
 %\textsf{1b} messages with \ensuremath{pBal[p]} such that every acceptor in some quorum has sent some message in \ensuremath{S}, 
 a quorum of acceptors has sent a set \ensuremath{S} of \textsf{1b} messages with \ensuremath{pBal[p]},
 \ensuremath{p} sends a
 \textsf{2a} message %to all acceptors 
 with \ensuremath{pBal[p]} and a set of proposals \ensuremath{PropSV(T)}, where \ensuremath{T} is the union of all voted triples in messages in \ensuremath{S}. \ensuremath{PropSV(T)} includes \ensuremath{MaxSV(T)}, the set of slot-value pairs with the highest ballot for each slot in \ensuremath{T}, and \ensuremath{NewSV(T)}, a set of new slot-value pairs for slots not in \ensuremath{T}.
\end{cpar}%
\end{lcom}%
%notes:
%MaxSV: bmax in FM16, Pmax in vrapaxos, pmax in paxos in da
%MaxBSV: partialMaxSV in proof, max_prop in paxos in da
%UnusedS: Free slots
%NewSV: new props
%PropSV: proposals; for field propSV
 \@x{ MaxBSV ( T )\@s{8.96} \.{\defeq} \{ t \.{\in} T \.{:} \A\, t2
 \.{\in} T \.{:} t2 . slot \.{=} t . slot \.{\implies} t2 . bal \.{\leq} t .
 bal \}}%
 \@x{ MaxSV ( T )\@s{15} \.{\defeq} \{ [ slot \.{\mapsto} t . slot ,\, val
 \.{\mapsto} t . val ] \.{:} t \.{\in} MaxBSV ( T ) \}}%
 \@x{ UnusedS ( T )\@s{12} \.{\defeq} \{ s \.{\in} {\mathcal{S}} \.{:} {\nexists}
 \, t \.{\in} T \.{:} t . slot \.{=} s \}}%
 \@x{ NewSV ( T )\@s{15} \.{\defeq} {\CHOOSE} D \,{\subseteq} [
 slot \,{:} UnusedS ( T ) ,\, val \,{:} {\mathcal{V}} ]
 \,{:} \A\, d1 ,\, d2 \,{\in} D \,{:} d1 . slot \,{=} d2 . slot
 \,{\implies} d1 \,{=} d2}%
\@x{ PropSV ( T )\@s{13} \.{\defeq} MaxSV ( T ) \.{\cup} NewSV ( T )}%
\@pvspace{8.0pt}%
\@x{ Phase2a ( p ) \.{\defeq}}%
 \@x{\@s{4.0} \.{\land} {\nexists}\, m \.{\in} msgs \.{:} ( m . type
 \.{=}\@w{2a} ) \.{\land} ( m . bal \.{=} pBal [ p ] )}%
 \@x{\@s{4.0} \.{\land} \E\, Q \.{\in} {\mathcal{Q}} ,\, S \.{\subseteq} \{ m
 \.{\in} msgs \.{:}  m . type \.{=}\@w{1b}  \.{\land}  m . bal \.{=} pBal
 [ p ]  \} \.{:}}%
 \@x{\@s{16.89} \.{\land} \A\, a \.{\in} Q \.{:} \E\, m \.{\in} S \.{:} m .
 from \.{=} a}%
 \@x{\@s{16.89} \.{\land} Send ( [ type \.{\mapsto}\@w{2a} ,\, from
 \.{\mapsto} p ,\, bal \.{\mapsto} pBal [ p ] ,\, propSV \.{\mapsto}
 PropSV ( {\UNION} \{ m . voted \.{:} m \.{\in} S \} ) ] )}%
 \@x{\@s{4.0} \.{\land} {\UNCHANGED} {\langle} pBal ,\, aBal ,\, aVoted
 {\rangle}}%
\@pvspace{8.0pt}%
\begin{lcom}{0}%
\begin{cpar}{0}{F}{F}{0}{0}{}%
 Phase \ensuremath{2b}: For an acceptor \ensuremath{a}, if there is a \textsf{2a} message \ensuremath{m} with ballot \ensuremath{m.bal}
 that is higher than or equal to the highest it has seen, \ensuremath{a} 
 %votes for all the message\mbox{\footnotesize'}s values
 %in ballot \ensuremath{m.bal}; 
 sends a \textsf{2b} message with \ensuremath{m.bal} and \ensuremath{m.propSV}, 
 updates the highest ballot it has seen to \ensuremath{m.bal}, and 
 updates set of voted triples using \ensuremath{m.propSV};
 otherwise it sends a \textsf{preempt} message back with the highest ballot it has seen.
\end{cpar}%
\end{lcom}%
 \@x{ Phase2b ( a ) \.{\defeq} \E\, m \.{\in} msgs \.{:} m . type \.{=}\@w{2a}
 \.{\land}}%
\@x{\@s{4.0} {\IF} m . bal \.{\geq} aBal [ a ] \.{\THEN}}%
 \@x{\@s{13.72} \.{\land} Send ( [ type \.{\mapsto}\@w{2b} ,\, from
 \.{\mapsto} a ,\, bal \.{\mapsto} m . bal ,\, propSV \.{\mapsto} m .
 propSV ] )}%
 \@x{\@s{13.72} \.{\land} aBal \.{'} \.{=} [ aBal {\EXCEPT} {\bang} [ a ]
 \.{=} m . bal ]}%
 \@x{\@s{13.72} \.{\land} aVoted \.{'} \.{=} [ aVoted {\EXCEPT} {\bang} [ a ]
 \.{=} \{ [ bal \.{\mapsto} m . bal ,\, slot \.{\mapsto} d . slot ,\,
 val \.{\mapsto} d . val ] \.{:} d \.{\in} m . propSV \} \.{\cup}}%
 \@x{\@s{145.1} \{ e \.{\in} aVoted [ a ] \.{:} {\nexists}\, r \.{\in} m . propSV
 \.{:} e . slot \.{=} r . slot \} ]}%
\@x{\@s{13.72} \.{\land} {\UNCHANGED} {\langle} pBal {\rangle}}%
\@x{\@s{2.0} \.{\ELSE}}%
 \@x{\@s{10.0} \.{\land} Send ( [ type \.{\mapsto}\@w{preempt} ,\, to
 \.{\mapsto} m . from ,\, bal \.{\mapsto} aBal [ a ] ] )}%
 \@x{\@s{10.0} \.{\land} {\UNCHANGED} {\langle} pBal ,\, aBal ,\, aVoted
 {\rangle}}%
\@pvspace{8.0pt}%
\begin{lcom}{0}%
\begin{cpar}{0}{F}{F}{0}{0}{}%
Preempt: For a proposer \ensuremath{p}, if there is a \textsf{preempt} message \ensuremath{m} with ballot \ensuremath{m.bal} that is higher than \ensuremath{p}'s current ballot, \ensuremath{p} updates its current ballot to a new ballot that is higher than \ensuremath{m.bal}
 and with which no \textsf{1a} message has been sent.
\end{cpar}%
\end{lcom}%
 \@x{ NewBal ( b2 ) \.{\defeq} {\CHOOSE} b \.{\in} {\mathcal{B}} \.{:} b \.{>} b2
 %\.{\land} {\nexists}\, m \.{\in} msgs \.{:} m . type \.{=}\@w{1a} \.{\land}
 %m . bal \.{=} b
 }%
\@x{ Preempt ( p ) \.{\defeq} \E\, m \.{\in} msgs \.{:}}%
 \@x{\@s{4.0} \.{\land} m . type \.{=}\@w{preempt} \.{\land} m . to \.{=} p
 \.{\land} m . bal \.{>} pBal [ p ]}%
 \@x{\@s{4.0} \.{\land} pBal \.{'} \.{=} [ pBal {\EXCEPT} {\bang} [ p ] \.{=}
 NewBal ( m . bal ) ]}%
 \@x{\@s{4.0} \.{\land} {\UNCHANGED} {\langle} msgs ,\, aBal ,\, aVoted
 {\rangle}}%
\@pvspace{8.0pt}%
 \@x{ Init\@s{3.3} \.{\defeq} msgs \.{=} {\emptyset}\@s{2.0} \.{\land} pBal \.{=} [ p \.{\in}
 {\mathcal{P}} \.{\mapsto} 0 ]\@s{2.0} \.{\land} aBal \.{=} [ a \.{\in}
 {\mathcal{A}} \.{\mapsto} \.{-} 1 ]\@s{2.0} \.{\land} aVoted \.{=} [ a \.{\in} {\mathcal{A}} \.{\mapsto}
 {\emptyset} ]}%
 \@x{ Next \.{\defeq} \.{\lor} \E\, p\@s{0.1} \.{\in} {\mathcal{P}} \.{:} Phase1a
 ( p ) \.{\lor} Phase2a ( p ) \.{\lor} Preempt ( p )}%
 \@x{\@s{31.86} \.{\lor} \E\, a \.{\in} {\mathcal{A}}\@s{0.6} \.{:} Phase1b ( a
 )\@s{0.41} \.{\lor} Phase2b ( a )}%
\@x{ Spec\@s{1.17} \.{\defeq} Init \.{\land} {\Box} [ Next ]_{ vars}}%
\@x{}\bottombar\@xx{}%

\section{Safety property to prove for Multi-Paxos with Preemption and invariants used in proof}
\label{appendix:prop}
\tlatex
\@x{}\moduleLeftDash\@xx{ {\MODULE} MultiPaxosProp}\moduleRightDash\@xx{}%
\iffalse
\begin{lcom}{0}%
\begin{cpar}{0}{F}{F}{0}{0}{}%
How a value is chosen:
%\end{cpar}%
%\vshade{5.0}%
%\begin{cpar}{0}{F}{F}{0}{0}{}%
This spec does not contain any actions in which a value is explicitly
 chosen (or a chosen value learned). What it means for a value to be
 chosen is defined by the operator \ensuremath{Chosen}, where
 \ensuremath{Chosen(s,\, v)} means that
 value \ensuremath{v} has been chosen for slot \ensuremath{s}. These are described in detail in Section~\ref{secAuxiliary}.
\end{cpar}%
\end{lcom}%
\fi
\begin{lcom}{0}%
\begin{cpar}{0}{F}{F}{0}{0}{}%
\ensuremath{VotedForIn(a, b, s, v)} means that acceptor \ensuremath{a} has sent some \textsf{2b} message \ensuremath{m} with \ensuremath{m.bal} equal to \ensuremath{b} and some proposal in \ensuremath{m.propSV} with \ensuremath{slot} equal to \ensuremath{s} and \ensuremath{val} equal to \ensuremath{v}. This specifies that acceptor \ensuremath{a} has voted the triple \ensuremath{\langle b, s, v\rangle}.
\end{cpar}%
\end{lcom}%
\@x{ VotedForIn ( a ,\, b ,\, s ,\, v ) \.{\defeq} \E\, m \.{\in} msgs \.{:}}%
 \@x{\@s{4.0} m . type \.{=}\@w{2b} \.{\land} m . from\@s{0.36} \.{=} a
 \.{\land} m . bal \.{=} b \.{\land} \E\, d \.{\in} m . propSV \.{:} d .
 slot \.{=} s \.{\land} d . val \.{=} v}%

\@pvspace{8.0pt}%
\begin{lcom}{0}%
\begin{cpar}{0}{F}{F}{0}{0}{}%
\ensuremath{ChosenIn(b, s, v)} means that every acceptor in some quorum \ensuremath{Q} has voted the triple \ensuremath{\langle b, s, v\rangle}.
\end{cpar}%
\end{lcom}%
 \@x{ ChosenIn ( b ,\, s ,\, v ) \.{\defeq} \E\, Q \.{\in} {\mathcal{Q}} \.{:} \A\,
 a \.{\in} Q \.{:} VotedForIn ( a ,\, b ,\, s ,\, v )}%
 
\@pvspace{8.0pt}%
\begin{lcom}{0}%
\begin{cpar}{0}{F}{F}{0}{0}{}%
\ensuremath{Chosen(s, v)} means that for some ballot \ensuremath{b}, \ensuremath{ChosenIn(b, s, v)} holds. 
%We define \ensuremath{Safe} property in terms of this operator.
\end{cpar}%
\end{lcom}%
 \@x{ Chosen ( s ,\, v ) \.{\defeq} \E\, b \.{\in} {\mathcal{B}} \.{:} ChosenIn ( b
 ,\, s ,\, v )}%
\@pvspace{8.0pt}%
\begin{lcom}{0}%
\begin{cpar}{0}{F}{F}{0}{0}{}%
 \ensuremath{WontVoteIn(a,\, b,\, s)} means that acceptor \ensuremath{a} has seen a higher ballot than \ensuremath{b}, and did not and will not vote 
 any value with \ensuremath{b} for slot \ensuremath{s}.
\end{cpar}%
\end{lcom}%
 \@x{ WontVoteIn ( a ,\, b ,\, s ) \.{\defeq} aBal [ a ] \.{>} b \.{\land}
 \A\, v \.{\in} {\mathcal{V}} \.{:} {\lnot} VotedForIn ( a ,\, b ,\, s ,\, v )}%
\@pvspace{8.0pt}%
\begin{lcom}{0}%
\begin{cpar}{0}{F}{F}{0}{0}{}%
 \ensuremath{SafeAt(b,\, s,\, v)} means that no value except perhaps \ensuremath{v} has been or will be chosen in any ballot lower
 than \ensuremath{b} for slot \ensuremath{s}.
\end{cpar}%
\end{lcom}%
 \@x{ SafeAt ( b ,\, s ,\, v ) \.{\defeq} \A\, b2 \.{\in} 0 \.{\dotdot} ( b
 \.{-} 1 ) \.{:} \E\, Q \.{\in} {\mathcal{Q}} \.{:} \A\, a \.{\in} Q \.{:}
 VotedForIn ( a ,\, b2 ,\, s ,\, v ) \.{\lor} WontVoteIn ( a ,\, b2 ,\, s )}%
\@pvspace{8.0pt}%
\begin{lcom}{0}%
\begin{cpar}{0}{F}{F}{0}{0}{}%
\ensuremath{Safe} states that at most one value can be chosen for each slot.
\end{cpar}%
\end{lcom}%
 \@x{ Safe \.{\defeq} \A\, v1 ,\, v2 \.{\in} {\mathcal{V}} ,\, s \.{\in} {\mathcal{S}}
 \.{:} Chosen ( s ,\, v1 ) \.{\land} Chosen ( s ,\, v2 ) \.{\implies} v1
 \.{=} v2 }%
\@pvspace{8.0pt}%

\@x{}\midbar\@xx{}%
\begin{lcom}{0}%
\begin{cpar}{0}{F}{F}{0}{0}{}%
\ensuremath{Messages} defines the set of valid messages. % that can be sent in this specification.
\ensuremath{TypeOK} defines invariants for the types of the variables. 
%These are described in detail in Section~\ref{secTypeInv}.
\end{cpar}%
\end{lcom}%
 \@x{ Messages \.{\defeq} [ type \.{:} \{\@w{1a} \} ,\, from\@s{0.29} \.{:}
 {\mathcal{P}} ,\, bal \.{:} {\mathcal{B}} ] \.{\cup}}%
 \@x{\@s{49.3} [ type \.{:} \{\@w{1b} \} ,\, from \.{:} {\mathcal{A}} ,\,
 bal\@s{0.6} \.{:} {\mathcal{B}} ,\, voted \.{:} {\SUBSET} [ bal \.{:} {\mathcal{B}} ,\,
 slot \.{:} {\mathcal{S}} ,\, val \.{:} {\mathcal{V}} ] ] \.{\cup}}%
 \@x{\@s{49.3} [ type \.{:} \{\@w{2a} \} ,\, from\@s{0.29} \.{:} {\mathcal{P}}
 ,\, bal \.{:} {\mathcal{B}} ,\, propSV \.{:} {\SUBSET} [ slot \.{:} {\mathcal{S}} ,\,
 val\@s{0.99} \.{:} {\mathcal{V}} ] ] \.{\cup}}%
 \@x{\@s{49.3} [ type \.{:} \{\@w{2b} \} ,\, from \.{:} {\mathcal{A}} ,\,
 bal\@s{0.6} \.{:} {\mathcal{B}} ,\, propSV \.{:} {\SUBSET} [ slot \.{:} {\mathcal{S}}
 ,\, val\@s{0.99} \.{:} {\mathcal{V}} ] ] \.{\cup}}%
 \@x{\@s{49.3} [ type \.{:} \{\@w{preempt} \} ,\, to \.{:} {\mathcal{P}} ,\, bal
 \.{:} {\mathcal{B}} ]}%
 \@x{ TypeOK \.{\defeq} \.{\land} msgs \.{\subseteq} Messages \.{\land} IsFiniteSet ( msgs ) \.{\land} pBal \.{\in} [ {\mathcal{P}}
 \.{\rightarrow} {\mathcal{B}} ]}%
 \@x{\@s{44.5} \.{\land} aBal\@s{0.74} \.{\in} [ {\mathcal{A}} \.{\rightarrow}
 {\mathcal{B}} \.{\cup} \{ \.{-} 1 \} ] \.{\land} aVoted \.{\in} [ {\mathcal{A}} \.{\rightarrow} {\SUBSET} [ bal \.{:} {\mathcal{B}} ,\,
 slot \.{:} {\mathcal{S}} ,\, val \.{:} {\mathcal{V}} ] ]}%
\@pvspace{8.0pt}%
\begin{lcom}{0}%
\begin{cpar}{0}{F}{F}{0}{0}{}%
\ensuremath{Max(T)} selects the largest element in nonempty set \ensuremath{T}.
\end{cpar}%
\end{lcom}%
 \@x{ Max ( T ) \.{\defeq} {\CHOOSE} e \.{\in} T \.{:} \A\, f \.{\in} T \.{:}
 e \.{\geq} f}%
\@pvspace{8.0pt}%
\begin{lcom}{0}%
\begin{cpar}{0}{F}{F}{0}{0}{}%
% Given a set of records \ensuremath{S} and a slot \ensuremath{s},
\ensuremath{MaxBalInSlot(T,\, s)} 
%selects the highest ballot \ensuremath{b} such that \ensuremath{[bal \mapsto b, slot \mapsto s] \in S}
selects, among set of elements in \ensuremath{T} with slot \ensuremath{s}, the highest ballot, or -1 if no element has slot \ensuremath{s}.
\end{cpar}%
\end{lcom}%
 \@x{ MaxBalInSlot ( T ,\, s ) \.{\defeq} \.{\LET} E \.{\defeq} \{ e \.{\in} T \.{:} e . slot \.{=} s \} \@s{2.0} \.{\IN} \.{\IF} E \.{=} {\emptyset}
 \.{\THEN} \.{-} 1 \.{\ELSE} Max ( \{e.bal \.{:} e \.{\in} E\} )}%
\@pvspace{8.0pt}%
\begin{lcom}{0}%
\begin{cpar}{0}{F}{F}{0}{0}{}%
\ensuremath{MsgInv} defines properties satisfied by the contents of messages, for 1b, 2a, and 2b messages.
%specifies the message invariant, as described in Section~\ref{secMsgInv}. These invariants specify properties satisfied by the content of the messages classified by message types. In this specification we have three types, \ensuremath{1b}, \ensuremath{2a}, and \ensuremath{2b} for which invariants have been specified by predicates \ensuremath{MsgInv1b}, \ensuremath{MsgInv2a}, and \ensuremath{MsgInv2b}.
\end{cpar}%
\end{lcom}%
 \@x{ MsgInv1b ( m )\@s{0.51} \.{\defeq} \.{\land} m . bal \.{\leq} aBal [ m .
 from ]}%
 \@x{\@s{63.75} \.{\land} \A\, r \.{\in} m . voted \.{:} VotedForIn ( m . from
 ,\, r . bal ,\, r . slot ,\, r . val )}%
 \@x{\@s{63.75} \.{\land} \A\, b \.{\in} {\mathcal{B}} ,\, s \.{\in} {\mathcal{S}} ,\, v
 \.{\in} {\mathcal{V}} \.{:} b \.{\in} MaxBalInSlot ( m . voted ,\, s ) \.{+} 1
 \.{\dotdot} m . bal \.{-} 1 \.{\implies}}%
\@x{\@s{80.1} {\lnot} VotedForIn ( m . from ,\, b ,\, s ,\, v )}%
%\@pvspace{8.0pt}%
 \@x{ MsgInv2a ( m ) \.{\defeq} \.{\land} \A\, d \.{\in} m . propSV \.{:}
 SafeAt ( m . bal ,\, d . slot ,\, d . val )}%
 \@x{\@s{63.75} \.{\land} \A\, d1 ,\, d2 \.{\in} m . propSV \.{:} d1 . slot
 \.{=} d2 . slot \.{\implies} d1 \.{=} d2}%
 \@x{\@s{63.75} \.{\land} \A\, m2 \.{\in} msgs \.{:} ( m2 . type \.{=}\@w{2a}
 \.{\land} m2 . bal \.{=} m . bal ) \.{\implies} m2 \.{=} m}%
%\@pvspace{8.0pt}%
 \@x{ MsgInv2b ( m )\@s{0.51} \.{\defeq} \.{\land} \E\, m2 \.{\in} msgs \.{:}
 m2 . type \.{=}\@w{2a} \.{\land} m2 . bal \.{=} m . bal \.{\land} m2 .
 propSV \.{=} m . propSV}%
\@x{\@s{63.75} \.{\land} m . bal \.{\leq} aBal [ m . from ]}%
%\@pvspace{8.0pt}%
 \@x{ MsgInv \.{\defeq} \A\, m \.{\in} msgs \.{:} \.{\land} ( m . type
 \.{=}\@w{1b} ) \.{\implies} MsgInv1b ( m )}%
 \@x{\@s{90.0} \.{\land} ( m . type \.{=}\@w{2a} )\@s{0.29} \.{\implies}
 MsgInv2a ( m )}%
 \@x{\@s{90.0} \.{\land} ( m . type \.{=}\@w{2b} )\@s{0.29} \.{\implies}
 MsgInv2b ( m )}%
\@pvspace{8.0pt}%
\begin{lcom}{0}%
\begin{cpar}{0}{F}{F}{0}{0}{}%
\ensuremath{AccInv} defines 
%as described in Section~\ref{secAccInv}. These invariants specify 
properties satisfied by the data maintained by the acceptors. % processes throughout execution.
\end{cpar}%
\end{lcom}%
\@x{ AccInv\@s{1.63} \.{\defeq} \A\, a\@s{2.45} \.{\in} {\mathcal{A}} \.{:}}%
 \@x{\@s{4.0} \.{\land} aBal [ a ] \.{=} \.{-} 1 \.{\implies} aVoted [ a ]
 \.{=} {\emptyset}}%
 \@x{\@s{4.0} \.{\land} \A\, r \.{\in} aVoted [ a ] \.{:} aBal [ a ] \.{\geq}
 r . bal \.{\land} VotedForIn ( a ,\, r . bal ,\, r . slot ,\, r . val )}%
 \@x{\@s{4.0} \.{\land} \A\, b\@s{0.05} \.{\in} {\mathcal{B}} ,\, s \.{\in} {\mathcal{S}}
 ,\, v \.{\in} {\mathcal{V}} \.{:} VotedForIn ( a ,\, b ,\, s ,\, v ) \.{\implies}
 \E\, r \.{\in} aVoted [ a ] \.{:} r . bal \.{\geq} b \.{\land} r . slot
 \.{=} s}%
 \@x{\@s{4.0} \.{\land} \A\, b\@s{0.05} \.{\in} {\mathcal{B}} ,\, s \.{\in} {\mathcal{S}}
 ,\, v \.{\in} {\mathcal{V}} \.{:} b \.{>} MaxBalInSlot ( aVoted [ a ] ,\, s )
 \.{\implies} {\lnot} VotedForIn ( a ,\, b ,\, s ,\, v )}%
\@pvspace{8.0pt}%
\begin{lcom}{0}%
\begin{cpar}{0}{F}{F}{0}{0}{}%
\ensuremath{Inv} is the complete inductive invariant.
\end{cpar}%
\end{lcom}%
\@x{ Inv \.{\defeq} TypeOK \.{\land} AccInv \.{\land} MsgInv}%
\@x{}\bottombar\@xx{}%

%annie: should we move consistency to before WontVoteIn?

\section{TLAPS checked proof of Multi-Paxos with Preemption}
\label{appendix:proof}
\tlatex
\@x{}\moduleLeftDash\@xx{ {\MODULE} MultiPaxosProof}\moduleRightDash\@xx{}%
\begin{lcom}{0}%
\begin{cpar}{0}{F}{F}{0}{0}{}%
The following 2 axioms and 10 lemmas are straightforward consequences of the predicates defined above. 
\end{cpar}%
\end{lcom}%
 \@x{ {\AXIOM} MaxInSet\@s{4.9} \.{\defeq} \A\, S \.{\in}  {\SUBSET} {\mathds{N}} 
  \.{:} Max ( S ) \.{\in} S}%
 \@x{ {\AXIOM} MaxOnNat \.{\defeq} \A\, S \.{\in} {\SUBSET} {\mathds{N}} \.{:} {\nexists}
 \, s \.{\in} S \.{:} Max ( S ) \.{<} s}%
\@pvspace{8.0pt}%
 \@x{ {\LEMMA} MaxOnNatS \.{\defeq} \A\, S1 ,\, S2 \.{\in}  {\SUBSET} {\mathds{N}} 
  \.{:} S1 \.{\subseteq} S2 \.{\implies} Max ( S1 )
 \.{\leq} Max ( S2 )\@s{2.0} {\BY} MaxInSet}%
\@pvspace{8.0pt}%
 \@x{ {\LEMMA} MaxBinSType \.{\defeq} \A\, S \.{\in} {\SUBSET} [ bal \.{:}
 {\mathcal{B}} ,\, slot \.{:} {\mathcal{S}} ,\, val \.{:} {\mathcal{V}} ] ,\, s \.{\in} {\mathcal{S}}
 \.{:} MaxBalInSlot ( S ,\, s ) \.{\in} {\mathcal{B}} \.{\cup} \{ \.{-} 1 \}}%
\@x{ {\BY} MaxInSet {\DEF} MaxBalInSlot}%
\@pvspace{8.0pt}%
 \@x{ {\LEMMA} MaxBinSSubsets \.{\defeq} \A\, S1 ,\, S2 \.{\in} {\SUBSET} [ bal
 \.{:} {\mathcal{B}} ,\, slot \.{:} {\mathcal{S}} ,\, val \.{:} {\mathcal{V}} ] ,\, s \.{\in}
 {\mathcal{S}} \.{:} S1 \.{\subseteq} S2 \.{\implies}}%
 \@x{\@s{95.72} MaxBalInSlot ( S1 ,\, s ) \.{\leq} MaxBalInSlot ( S2
 ,\, s )}%
 \@x{\@s{4.0}\@pfstepnum{1}{}\  {\SUFFICES} {\ASSUME} {\NEW} S1 \.{\in}
 {\SUBSET} [ bal \.{:} {\mathcal{B}} ,\, slot \.{:} {\mathcal{S}} ,\, val \.{:} {\mathcal{V}} ]
 ,\, {\NEW} s \.{\in} {\mathcal{S}} ,\,}%
 \@x{\@s{83.25} {\NEW} S2 \.{\in} {\SUBSET} [ bal \.{:} {\mathcal{B}} ,\, slot
 \.{:} {\mathcal{S}} ,\, val \.{:} {\mathcal{V}} ] ,\, S1 \.{\subseteq} S2}%
 \@x{\@s{52.31} {\PROVE} MaxBalInSlot ( S1 ,\, s ) \.{\leq}
 MaxBalInSlot ( S2 ,\, s ) \@s{4.0} {\OBVIOUS}}%
 \@x{\@s{4.0}\@pfstepnum{1}{1.} {\CASE} {\nexists}\, d \.{\in} S1 \.{:} d .
 slot \.{=} s}%
 \@x{\@s{8.0}\@pfstepnum{2}{1.}\  MaxBalInSlot ( S1 ,\, s ) \.{=} \.{-}
 1\@s{2.0} {\BY}\@pfstepnum{1}{1}\  {\DEF} MaxBalInSlot}%
 \@x{\@s{8.0}\@pfstepnum{2}{}\  {\QED} {\BY}\@pfstepnum{2}{1} ,\, MaxBinSType
 {\DEF} {\mathcal{B}}}%
 \@x{\@s{4.0}\@pfstepnum{1}{2.} {\CASE} \E\, d \.{\in} S1 \.{:} d . slot \.{=}
 s}%
 \@x{\@s{8.0}\@pfstepnum{2}{1.} {\CASE} {\nexists}\, d \.{\in} S2
 \.{\,\backslash\,} S1 \.{:} d . slot \.{=} s}%
 \@x{\@s{12.0}\@pfstepnum{3}{1.}\  MaxBalInSlot ( S1 ,\, s ) \.{=}
 MaxBalInSlot ( S2 ,\, s )\@s{2.0} {\BY}\@pfstepnum{2}{1}
 ,\,\@pfstepnum{1}{2}\  {\DEF} MaxBalInSlot}%
 \@x{\@s{12.0}\@pfstepnum{3}{}\  {\QED} {\BY}\@pfstepnum{3}{1} ,\, MaxBinSType
 {\DEF} {\mathcal{B}}}%
 \@x{\@s{8.0}\@pfstepnum{2}{2.} {\CASE} \E\, d \.{\in} S2 \.{\,\backslash\,}
 S1 \.{:} d . slot \.{=} s\@s{2.0} {\BY}\@pfstepnum{2}{2}
 ,\,\@pfstepnum{1}{2} ,\, MaxBinSType ,\, MaxOnNatS {\DEF} {\mathcal{B}} ,\,
 MaxBalInSlot}%
 \@x{\@s{8.0}\@pfstepnum{2}{}\  {\QED} {\BY}\@pfstepnum{2}{1}
 ,\,\@pfstepnum{2}{2}\ }%
 \@x{\@s{4.0}\@pfstepnum{1}{}\  {\QED} {\BY}\@pfstepnum{1}{1}
 ,\,\@pfstepnum{1}{2}\ }%
\@pvspace{8.0pt}%
 \@x{ {\LEMMA} MaxBinSNoSlot \.{\defeq} \A\, S \.{\in} {\SUBSET} [ bal \.{:}
 {\mathcal{B}} ,\, slot \.{:} {\mathcal{S}} ,\, val \.{:} {\mathcal{V}} ] ,\, s \.{\in} {\mathcal{S}}
 \.{:}}%
 \@x{\@s{93.8} ( {\nexists}\, d \.{\in} S \.{:} d . slot \.{=} s )
 \.{\equiv} MaxBalInSlot ( S ,\, s ) \.{=} \.{-} 1}%
\@x{ {\BY} MaxInSet {\DEF} MaxBalInSlot ,\, {\mathcal{B}}}%
\@pvspace{8.0pt}%
 \@x{ {\LEMMA} MaxBinSExists \.{\defeq} \A\, S \.{\in} {\SUBSET} [ bal \.{:}
 {\mathcal{B}} ,\, slot \.{:} {\mathcal{S}} ,\, val \.{:} {\mathcal{V}} ] ,\, s \.{\in} {\mathcal{S}}
 \.{:} MaxBalInSlot ( S ,\, s ) \.{\in} {\mathcal{B}} \.{\implies}}%
 \@x{\@s{91.16} \E\, d \.{\in} S \.{:} d . slot \.{=} s \.{\land} d . bal
 \.{=} MaxBalInSlot ( S ,\, s )}%
 \@x{\@s{4.0}\@pfstepnum{1}{}\  {\SUFFICES} {\ASSUME} {\NEW} S \.{\subseteq}
 [ bal \.{:} {\mathcal{B}} ,\, slot \.{:} {\mathcal{S}} ,\, val \.{:} {\mathcal{V}} ]
 , {\NEW} s\@s{1.52} \.{\in} {\mathcal{S}} ,\, MaxBalInSlot ( S ,\, s
 ) \.{\in} {\mathcal{B}}}%
 \@x{\@s{52.31} {\PROVE} \E\, d \.{\in} S \.{:} d . bal \.{=}
 MaxBalInSlot ( S ,\, s ) \.{\land} d . slot \.{=} s \@s{4.0} {\OBVIOUS}}%
 \@x{\@s{8.0}\@pfstepnum{1}{1.}\  \E\, d \.{\in} S \.{:} d . slot \.{=}
 s\@s{2.0} {\BY} {\DEF} MaxBalInSlot ,\, {\mathcal{B}}}%
 \@x{\@s{8.0}\@pfstepnum{1}{2.}\  MaxBalInSlot ( S ,\, s ) \.{=} Max ( \{
 d . bal \.{:} d \.{\in} \{ d \.{\in} S \.{:} d . slot \.{=} s \} \}
 )\@s{2.0} {\BY}\@pfstepnum{1}{1}\  {\DEF} MaxBalInSlot}%
 \@x{\@s{4.0}\@pfstepnum{1}{}\  {\QED} {\BY}\@pfstepnum{1}{1}
 ,\,\@pfstepnum{1}{2} ,\, MaxInSet}%
\@pvspace{8.0pt}%
 \@x{ {\LEMMA} MaxBinSNoMore \.{\defeq} \A\, S \.{\in} {\SUBSET} [ bal \.{:}
 {\mathcal{B}} ,\, slot \.{:} {\mathcal{S}} ,\, val \.{:} {\mathcal{V}} ] ,\, s \.{\in} {\mathcal{S}}
 \.{:}}%
 \@x{\@s{98.26} {\nexists}\, d \.{\in} S \.{:} d . bal \.{>} MaxBalInSlot
 ( S ,\, s ) \.{\land} d . slot \.{=} s}%
 \@x{\@s{4.0}\@pfstepnum{1}{}\  {\SUFFICES} {\ASSUME} {\NEW} S \.{\in}
 {\SUBSET} [ bal \.{:} {\mathcal{B}} ,\, slot \.{:} {\mathcal{S}} ,\, val \.{:} {\mathcal{V}} ]
 ,\, {\NEW} s \.{\in} {\mathcal{S}}}%
 \@x{\@s{52.31} {\PROVE} {\nexists}\, d \.{\in} S \.{:} d . bal \.{>}
 MaxBalInSlot ( S ,\, s ) \.{\land} d . slot \.{=} s \@s{4.0} {\OBVIOUS}}%
 \@x{\@s{8.0}\@pfstepnum{1}{1.} {\CASE} {\nexists}\, d \.{\in} S \.{:} d .
 slot \.{=} s\@s{2.0} {\BY}\@pfstepnum{1}{1}\ }%
 \@x{\@s{8.0}\@pfstepnum{1}{2.} {\CASE} \E\, d \.{\in} S \.{:} d . slot \.{=}
 s}%
 \@x{\@s{12.0}\@pfstepnum{2}{1.}\  {\nexists}\, b \.{\in} \{ d . bal \.{:} d
 \.{\in} \{ d \.{\in} S \.{:} d . slot \.{=} s \} \} \.{:} b \.{>}
 MaxBalInSlot ( S ,\, s )}%
 \@x{\@s{31.1} {\BY}\@pfstepnum{1}{2} ,\, MaxOnNat {\DEF} MaxBalInSlot ,\,
 {\mathcal{B}} ,\, {\mathcal{S}}}%
 \@x{\@s{12.0}\@pfstepnum{2}{2.}\  {\nexists}\, d \.{\in} S \.{:} ( d . slot
 \.{=} s \.{\land} {\lnot} ( d . bal \.{\leq} MaxBalInSlot ( S ,\, s ) )
 )\@s{2.0} {\BY}\@pfstepnum{2}{1}\ }%
\@x{\@s{12.0}\@pfstepnum{2}{}\  {\QED} {\BY}\@pfstepnum{2}{2}\ }%
 \@x{\@s{4.0}\@pfstepnum{1}{}\  {\QED} {\BY}\@pfstepnum{1}{1}
 ,\,\@pfstepnum{1}{2}\ }%
\@pvspace{8.0pt}%
 \@x{ {\LEMMA} Misc \.{\defeq} \A\, S \.{:} \@s{1.25} \.{\land} NewSV ( S )
 \.{\in} ( {\SUBSET} [ slot \.{:} UnusedS ( S ) ,\, val \.{:} {\mathcal{V}} ] )
 }%
 \@x{\@s{79.71} \.{\land} \A\, t1 ,\, t2 \.{\in} NewSV ( S ) \.{:} t1 .
 slot \.{=} t2 . slot \.{\implies} t1 \.{=} t2}%
 \@x{\@s{79.71} \.{\land} {\nexists}\, t1 \.{\in} MaxSV ( S ) ,\, t2 \.{\in}
 NewSV ( S ) \.{:} t1 . slot \.{=} t2 . slot}%
\@x{\@s{4.0}\@pfstepnum{1}{}\  {\SUFFICES} {\ASSUME} {\NEW} S}%
 \@x{\@s{52.31} {\PROVE}\@s{-1.5} \.{\land} NewSV ( S ) \.{\in} (
 {\SUBSET} [ slot \.{:} UnusedS ( S ) ,\, val \.{:} {\mathcal{V}} ] )
 }%
 \@x{\@s{81.63} \.{\land} \A\, t1 ,\, t2 \.{\in} NewSV ( S ) \.{:} t1
 . slot \.{=} t2 . slot \.{\implies} t1 \.{=} t2}%
 \@x{\@s{81.63} \.{\land} {\nexists}\, t1 \.{\in} MaxSV ( S ) ,\, t2 \.{\in}
 NewSV ( S ) \.{:} t1 . slot \.{=} t2 . slot \@s{4.0} {\OBVIOUS}}%
 \@x{\@s{4.0}\@pfstepnum{1}{1.}\  \E\, T \.{\in}  {\SUBSET} [ slot \.{:}
 UnusedS ( S ) ,\, val \.{:} {\mathcal{V}} ]   \.{:} \A\,
 t1 ,\, t2 \.{\in} T \.{:} t1 . slot \.{=} t2 . slot \.{\implies} t1 \.{=}
 t2\@s{2.0} {\BY} {\DEF} UnusedS}%
%\@x{\@s{23.1} {\BY} {\DEF} UnusedS}%
 \@x{\@s{4.0}\@pfstepnum{1}{2.}\  NewSV ( S ) \.{\in} ( {\SUBSET} [
 slot \.{:} UnusedS ( S ) ,\, val \.{:} {\mathcal{V}} ] ) 
 \@s{2.0} {\BY}\@pfstepnum{1}{1}\  {\DEF} NewSV}%
 \@x{\@s{4.0}\@pfstepnum{1}{3.}\  \A\, t1 ,\, t2 \.{\in} NewSV ( S )
 \.{:} t1 . slot \.{=} t2 . slot \.{\implies} t1 \.{=} t2\@s{2.0}
 {\BY}\@pfstepnum{1}{1}\  {\DEF} NewSV}%
 \@x{\@s{4.0}\@pfstepnum{1}{4.}\  {\nexists}\, t1 \.{\in} MaxSV ( S ) ,\, t2
 \.{\in}  [ slot {:} UnusedS ( S ) ,\, val {:} {\mathcal{V}} ]
   \.{:} t1 . slot \.{=} t2 . slot \@s{2.0} {\BY} {\DEF} MaxSV ,\, MaxBSV ,\, UnusedS}%
 \@x{\@s{4.0}\@pfstepnum{1}{}\  {\QED} {\BY}\@pfstepnum{1}{2}
 ,\,\@pfstepnum{1}{3} ,\,\@pfstepnum{1}{4}\ }%
%\@pvspace{8.0pt}%
\begin{lcom}{0}%
\begin{cpar}{0}{F}{F}{0}{0}{}%
\ensuremath{VotedInv} asserts that if any acceptor \ensuremath{a} voted any triple \ensuremath{\langle b, s, v \rangle}, then that triple is safe.
\end{cpar}%
\end{lcom}%
 \@x{ {\LEMMA} VotedInv \.{\defeq}\@s{2.0} MsgInv \.{\land} TypeOK
 \.{\implies} \A\, a \.{\in} {\mathcal{A}} ,\, b \.{\in} {\mathcal{B}} ,\, s \.{\in}
 {\mathcal{S}} ,\, v \.{\in} {\mathcal{V}} \.{:}}%
 \@x{\@s{81.11} VotedForIn ( a ,\, b ,\, s ,\, v ) \.{\implies} SafeAt ( b
 ,\, s ,\, v ) \.{\land} b \.{\leq} aBal [ a ]}%
 \@x{ {\BY} {\DEF} VotedForIn ,\, MsgInv ,\, Messages ,\, TypeOK ,\, MsgInv2a
 ,\, MsgInv1b ,\, MsgInv2b}%
%\@pvspace{8.0pt}%
\begin{lcom}{0}%
\begin{cpar}{0}{F}{F}{0}{0}{}%
\ensuremath{VotedOnce} asserts that if any acceptor \ensuremath{a1} voted triple \ensuremath{\langle b, s, v1 \rangle} and acceptor \ensuremath{a2} voted triple \ensuremath{\langle b, s, v2 \rangle}, then \ensuremath{v1 = v2}.
%That is, for any given ballot and slot, only one value can be voted by any acceptor.
\end{cpar}%
\end{lcom}%
 \@x{ {\LEMMA} VotedOnce \.{\defeq} MsgInv \.{\implies} \A\, a1 ,\, a2 \.{\in}
 {\mathcal{A}} ,\, b \.{\in} {\mathcal{B}}  ,\, s \.{\in} {\mathcal{S}} ,\, v1 ,\, v2 \.{\in} {\mathcal{V}} \.{:}}%
 \@x{\@s{85.18} VotedForIn ( a1 ,\, b ,\, s ,\, v1 ) \.{\land} VotedForIn (
 a2 ,\, b ,\, s ,\, v2 ) \.{\implies} ( v1 \.{=} v2 )}%
 \@x{ {\BY} {\DEF} MsgInv ,\, VotedForIn ,\, MsgInv2a ,\, MsgInv1b ,\,
 MsgInv2b}%
%\@pvspace{8.0pt}%
\begin{lcom}{0}%
\begin{cpar}{0}{F}{F}{0}{0}{}%
\ensuremath{VotedUnion} asserts that, 
%for any \ensuremath{1b} messages \ensuremath{m1} and \ensuremath{m2}, for any \ensuremath{d1} and \ensuremath{d2} in their respective \ensuremath{voted} field, if \ensuremath{d1} and \ensuremath{d2} have the same ballot and slot, then they have the same value.
in any two \ensuremath{1b} messages' \ensuremath{voted} field, triples with the same ballot and slot have the same value.
\end{cpar}%
\end{lcom}%
 \@x{ {\LEMMA} VotedUnion \.{\defeq} MsgInv \.{\land} TypeOK \.{\implies} \A\,
 m1 ,\, m2 \.{\in} msgs \.{:} m1 . type \.{=}\@w{1b} \.{\land} m2 . type
 \.{=}\@w{1b} \.{\implies}}%
 \@x{\@s{89.09} \A\, d1 \.{\in} m1 . voted ,\, d2 \.{\in} m2 . voted \.{:} (
 d1 . bal \.{=} d2 . bal \.{\land} d1 . slot \.{=} d2 . slot ) \.{\implies}}%
\@x{\@s{93.09} d1 . val \.{=} d2 . val}%
 \@x{\@s{4.0}\@pfstepnum{1}{}\  {\SUFFICES} {\ASSUME} MsgInv ,\, TypeOK ,\,
 {\NEW} m1 \.{\in} msgs ,\, {\NEW} m2 \.{\in} msgs ,\, m1 . type \.{=}\@w{1b}
 ,\, m2 . type \.{=}\@w{1b} ,\,}%
 \@x{\@s{83.25} {\NEW} d1 \.{\in} m1 . voted ,\, {\NEW} d2 \.{\in} m2 . voted
 ,\, d1 . bal \.{=} d2 . bal ,\, d1 . slot \.{=} d2 . slot}%
\@x{\@s{52.31} {\PROVE} d1 . val \.{=} d2 . val \@s{4.0} {\OBVIOUS}}%
 \@x{\@s{8.0}\@pfstepnum{1}{1.}\  VotedForIn ( m1 . from ,\, d1 . bal ,\, d1
 . slot ,\, d1 . val )\@s{2.0} {\BY} {\DEF} MsgInv ,\, MsgInv1b}%
 \@x{\@s{8.0}\@pfstepnum{1}{2.}\  VotedForIn ( m2 . from ,\, d2 . bal ,\, d2
 . slot ,\, d2 . val )\@s{2.0} {\BY} {\DEF} MsgInv ,\, MsgInv1b}%
 \@x{\@s{4.0}\@pfstepnum{1}{}\  {\QED} {\BY}\@pfstepnum{1}{1}
 ,\,\@pfstepnum{1}{2} ,\, VotedOnce {\DEF} TypeOK ,\, Messages}%
%\@pvspace{8.0pt}%

\begin{lcom}{0}%
\begin{cpar}{0}{F}{F}{0}{0}{}%
The following 5 invariance lemmas 
assert that, for acceptor \ensuremath{a}, ballot \ensuremath{b}, slot \ensuremath{s}, and value \ensuremath{v}, and for all phases and preempt, if \ensuremath{VotedForIn(a, b, s, v)} holds then \ensuremath{VotedForIn(a, b, s, v)'} holds;
for all except \ensuremath{Phase2b}, the inverse also holds.
\end{cpar}%
\end{lcom}%

 \@x{ {\LEMMA} Phase1aVotedForInv \.{\defeq} TypeOK \.{\implies} \A\, p
 \.{\in} {\mathcal{P}} \.{:} Phase1a ( p ) \.{\implies} \A\, a \.{\in} {\mathcal{A}}
 ,\, b \.{\in} {\mathcal{B}} ,\, s \.{\in} {\mathcal{S}} ,\, v \.{\in} {\mathcal{V}} \.{:}}%
 \@x{\@s{120.43} VotedForIn ( a ,\, b ,\, s ,\, v ) \.{\equiv} VotedForIn ( a
 ,\, b ,\, s ,\, v ) \.{'}}%
\@x{ {\BY} {\DEF} VotedForIn ,\, Send ,\, TypeOK ,\, Messages ,\, Phase1a}%
\@pvspace{8.0pt}%
 \@x{ {\LEMMA} Phase1bVotedForInv \.{\defeq} TypeOK \.{\implies} \A\, a
 \.{\in} {\mathcal{A}} \.{:} Phase1b ( a ) \.{\implies} \A\, a2 \.{\in} {\mathcal{A}}
 ,\, b \.{\in} {\mathcal{B}} ,\, s \.{\in} {\mathcal{S}} ,\, v \.{\in} {\mathcal{V}} \.{:}}%
 \@x{\@s{120.02} VotedForIn ( a2 ,\, b ,\, s ,\, v ) \.{\equiv} VotedForIn (
 a2 ,\, b ,\, s ,\, v ) \.{'}}%
\@x{ {\BY} {\DEF} VotedForIn ,\, Send ,\, TypeOK ,\, Messages ,\, Phase1b}%
\@pvspace{8.0pt}%
 \@x{ {\LEMMA} Phase2aVotedForInv \.{\defeq} TypeOK \.{\implies} \A\, p
 \.{\in} {\mathcal{P}} \.{:} Phase2a ( p ) \.{\implies} \A\, a \.{\in} {\mathcal{A}}
 ,\, b \.{\in} {\mathcal{B}} ,\, s \.{\in} {\mathcal{S}} ,\, v \.{\in} {\mathcal{V}} \.{:}}%
 \@x{\@s{120.43} VotedForIn ( a ,\, b ,\, s ,\, v ) \.{\equiv} VotedForIn ( a
 ,\, b ,\, s ,\, v ) \.{'}}%
 \@x{ {\BY} \A\, p \.{\in} {\mathcal{P}} \.{:} Phase2a ( p ) \.{\implies} \A\, m
 \.{\in} msgs \.{'} \.{\,\backslash\,} msgs \.{:} m . type \.{=}\@w{2a}
 {\DEF} VotedForIn ,\,}%
\@x{\@s{12.0} Send ,\, TypeOK ,\, Messages ,\, Phase2a}%
\@pvspace{8.0pt}%
 \@x{ {\LEMMA} Phase2bVotedForInv \.{\defeq} TypeOK \.{\implies} \A\, a
 \.{\in} {\mathcal{A}} \.{:} Phase2b ( a ) \.{\implies} \A\, a2 \.{\in} {\mathcal{A}}
 ,\, b \.{\in} {\mathcal{B}} ,\, s \.{\in} {\mathcal{S}} ,\, v \.{\in} {\mathcal{V}} \.{:}}%
 \@x{\@s{120.02} VotedForIn ( a2 ,\, b ,\, s ,\, v ) \.{\implies} VotedForIn
 ( a2 ,\, b ,\, s ,\, v ) \.{'}}%
\@x{ {\BY} {\DEF} VotedForIn ,\, Send ,\, TypeOK ,\, Messages ,\, Phase2b}%
\@pvspace{8.0pt}%
 \@x{ {\LEMMA} PreemptVotedForInv \.{\defeq} TypeOK \.{\implies} \A\, p
 \.{\in} {\mathcal{P}} \.{:} Preempt ( p ) \.{\implies} \A\, a \.{\in} {\mathcal{A}}
 ,\, b \.{\in} {\mathcal{B}} ,\, s \.{\in} {\mathcal{S}} ,\, v \.{\in} {\mathcal{V}} \.{:}}%
 \@x{\@s{119.82} VotedForIn ( a ,\, b ,\, s ,\, v ) \.{\equiv} VotedForIn ( a
 ,\, b ,\, s ,\, v ) \.{'}}%
\@x{ {\BY} {\DEF} VotedForIn ,\, Send ,\, TypeOK ,\, Messages ,\, Preempt}%
%\@pvspace{8.0pt}%
%\@x{}\midbar\@xx{}%
\begin{lcom}{0}%
\begin{cpar}{0}{F}{F}{0}{0}{}%
Invariance lemma \ensuremath{SafeAtStable} asserts that if \ensuremath{SafeAt(b,\,s,\,v)} holds then \ensuremath{SafeAt(b,\,s,\,v)'} holds in the next state.
\end{cpar}%
\end{lcom}%
 \@x{ {\LEMMA} SafeAtStable \.{\defeq} Inv \.{\land} Next \.{\land} TypeOK
 \.{'} \.{\implies} \A\, b \.{\in} {\mathcal{B}} ,\, s \.{\in} {\mathcal{S}} ,\, v \.{\in}
 {\mathcal{V}} \.{:} SafeAt ( b ,\, s ,\, v ) \.{\implies} SafeAt ( b ,\, s ,\, v
 ) \.{'}}%
 \@x{\@pfstepnum{1}{}\  {\SUFFICES} {\ASSUME} Inv ,\, Next ,\, TypeOK \.{'}
 ,\, {\NEW} b \.{\in} {\mathcal{B}} ,\, {\NEW} s \.{\in} {\mathcal{S}} ,\, {\NEW} v \.{\in}
 {\mathcal{V}} ,\, SafeAt ( b ,\, s ,\, v )}%
\@x{\@s{48.31} {\PROVE} SafeAt ( b ,\, s ,\, v ) \.{'} \@s{4.0} {\OBVIOUS}}%
\@x{\@pfstepnum{1}{}\  {\USE} {\DEF} Send ,\, Inv ,\, {\mathcal{B}}}%
 \@x{\@pfstepnum{1}{1.} {\CASE} \E\, p\@s{0.1} \.{\in} {\mathcal{P}} \.{:}
 Phase1a ( p )\@s{2.0} {\BY}\@pfstepnum{1}{1}\  {\DEF} SafeAt ,\, Phase1a ,\,
 VotedForIn ,\, WontVoteIn}%
 \@x{\@pfstepnum{1}{2.} {\CASE} \E\, a \.{\in} {\mathcal{A}}\@s{0.6} \.{:}
 Phase1b ( a )}%
 \@x{\@s{19.0} {\BY}\@pfstepnum{1}{2} ,\, QuorumAssumption {\DEF} TypeOK ,\,
 SafeAt ,\, WontVoteIn ,\, VotedForIn ,\, Phase1b}%
 \@x{\@pfstepnum{1}{3.}\  {\ASSUME} {\NEW} p \.{\in} {\mathcal{P}} ,\, Phase2a ( p
 )\@s{2.0} {\PROVE}\@s{-4.1} SafeAt ( b ,\, s ,\, v ) \.{'}}%
 \@x{\@s{4.0}\@pfstepnum{2}{1.}\  \A\, a \.{\in} {\mathcal{A}} ,\, b2 \.{\in}
 {\mathcal{B}} ,\, s2 \.{\in} {\mathcal{S}} \.{:} WontVoteIn ( a ,\, b2 ,\, s2 ) \.{\equiv}
 WontVoteIn ( a ,\, b2 ,\, s2 ) \.{'}}%
 \@x{\@s{23.1} {\BY}\@pfstepnum{1}{3} ,\, Phase2aVotedForInv {\DEF}
 WontVoteIn ,\, Send ,\, Phase2a}%
 \@x{\@s{4.0}\@pfstepnum{2}{}\  {\QED} {\BY}\@pfstepnum{2}{1} ,\,
 QuorumAssumption ,\, Phase2aVotedForInv ,\,\@pfstepnum{1}{3}\  {\DEF}
 SafeAt}%
 \@x{\@pfstepnum{1}{4.}\  {\ASSUME} {\NEW} a \.{\in} {\mathcal{A}} ,\, Phase2b ( a
 )\@s{2.0} {\PROVE} SafeAt ( b ,\, s ,\, v ) \.{'}}%
 \@x{\@s{4.0}\@pfstepnum{2}{1.}\  {\PICK} m \.{\in} msgs \.{:} Phase2b ( a )
 {\bang} ( m )\@s{2.0} {\BY}\@pfstepnum{1}{4}\  {\DEF} Phase2b}%
 \@x{\@s{4.0}\@pfstepnum{2}{2.}\  \A\, a2 \.{\in} {\mathcal{A}} ,\, b2 \.{\in}
 {\mathcal{B}} \.{:} aBal [ a2 ] \.{>} b2 \.{\implies} aBal \.{'} [ a2 ] \.{>}
 b2\@s{2.0} {\BY}\@pfstepnum{2}{1}\  {\DEF} TypeOK}%
 \@x{\@s{4.0}\@pfstepnum{2}{3.}\  {\ASSUME} {\NEW} a2 \.{\in} {\mathcal{A}} ,\,
 {\NEW} b2 \.{\in} {\mathcal{B}} ,\, {\NEW} s2 \.{\in} {\mathcal{S}} ,\, {\NEW} v2 \.{\in}
 {\mathcal{V}} ,\, WontVoteIn ( a2 ,\, b2 ,\, s2 ) ,\,}%
 \@x{\@s{53.7} VotedForIn ( a2 ,\, b2 ,\, s2 ,\, v2 ) \.{'} ,\, {\NEW} S
 \.{\subseteq} [ slot \.{:} {\mathcal{S}} \.{\,\backslash\,} \{ s2 \} ,\, val
 \.{:} {\mathcal{V}} ] \@s{4.0} {\PROVE} {\FALSE}}%
 \@x{\@s{8.0}\@pfstepnum{3}{1.}\  \E\, m1 \.{\in} msgs \.{'}
 \.{\,\backslash\,} msgs \.{:} \.{\land} m1 . type \.{=}\@w{2b} \.{\land} m1
 . bal \.{=} b2 \.{\land} m1 . from \.{=} a2}%
 \@x{\@s{105.13} \.{\land} \E\, d \.{\in} m1 . propSV \.{:} d . slot \.{=} s2
 \.{\land} d . val \.{=} v2}%
\@x{\@s{27.1} {\BY}\@pfstepnum{2}{3}\  {\DEF} VotedForIn ,\, WontVoteIn}%
 \@x{\@s{8.0}\@pfstepnum{3}{2.}\  a2 \.{=} a \.{\land} m . bal \.{=}
 b2\@s{2.0} {\BY}\@pfstepnum{2}{1} ,\,\@pfstepnum{3}{1}\  {\DEF} TypeOK}%
 \@x{\@s{8.0}\@pfstepnum{3}{}\  {\QED}\@s{2.0} {\BY}\@pfstepnum{2}{1}
 ,\,\@pfstepnum{2}{3} ,\,\@pfstepnum{3}{2} ,\,\@pfstepnum{3}{1}\  {\DEF}
 Phase2b ,\, WontVoteIn ,\, TypeOK}%
 \@x{\@s{4.0}\@pfstepnum{2}{4.}\  \A\, a2 \.{\in} {\mathcal{A}} ,\, b2 \.{\in}
 {\mathcal{B}} ,\, s2 \.{\in} {\mathcal{S}} \.{:} WontVoteIn ( a2 ,\, b2 ,\, s2 )
 \.{\implies} WontVoteIn ( a2 ,\, b2 ,\, s2 ) \.{'}}%
 \@x{\@s{23.1} {\BY}\@pfstepnum{2}{2} ,\,\@pfstepnum{2}{3}\  {\DEF}
 WontVoteIn}%
 \@x{\@s{4.0}\@pfstepnum{2}{}\  {\QED} {\BY} Phase2bVotedForInv
 ,\,\@pfstepnum{2}{4} ,\, QuorumAssumption ,\,\@pfstepnum{1}{4}\  {\DEF}
 SafeAt}%
 \@x{\@pfstepnum{1}{5.} {\CASE} \E\, p \.{\in} {\mathcal{P}} \.{:} Preempt ( p
 )\@s{2.0} {\BY}\@pfstepnum{1}{5}\  {\DEF} SafeAt ,\, Preempt ,\, VotedForIn
 ,\, WontVoteIn}%
 \@x{\@pfstepnum{1}{}\  {\QED} {\BY}\@pfstepnum{1}{1} ,\,\@pfstepnum{1}{2}
 ,\,\@pfstepnum{1}{3} ,\,\@pfstepnum{1}{4} ,\,\@pfstepnum{1}{5}\  {\DEF}
 Next}%
\@pvspace{8.0pt}%
\begin{lcom}{0}%
\begin{cpar}{0}{F}{F}{0}{0}{}%
\ensuremath{Invariant} asserts the temporal formula that if \ensuremath{Spec} holds then \ensuremath{Inv} always holds. 
%Details about the proof structure and strategy can be found in Section~\ref{secProveStruct}.
\end{cpar}%
\end{lcom}%
\@x{ {\THEOREM} Invariant \.{\defeq} Spec \.{\implies} {\Box} Inv}%
\@x{\@pfstepnum{1}{}\  {\USE} {\DEF} {\mathcal{B}} ,\, {\mathcal{S}}}%
 \@x{\@pfstepnum{1}{1.}\  Init \.{\implies} Inv\@s{2.0} {\BY} FS\_EmptySet
 {\DEF} Init ,\, Inv ,\, TypeOK ,\, AccInv ,\, MsgInv ,\, VotedForIn}%
 \@x{\@pfstepnum{1}{2.}\  Inv \.{\land} [ Next ]_{ vars} \.{\implies} Inv
 \.{'}}%
 \@x{\@s{4.0}\@pfstepnum{2}{}\  {\SUFFICES} {\ASSUME} Inv ,\, [ Next ]_{ vars}
 {\PROVE} Inv \.{'} \@s{4.0} {\OBVIOUS}}%
\@x{\@s{4.0}\@pfstepnum{2}{}\  {\USE} {\DEF} Inv}%
\@x{\@s{4.0}\@pfstepnum{2}{1.} {\CASE} Next}%
\begin{lcom}{6.5}%
\begin{cpar}{0}{F}{F}{0}{0}{}%
\@pfstepnum{3}{1} proves \ensuremath{TypeOK'} for \ensuremath{Next}. 
%As described in Section~\ref{secProveStruct}, a new step is created for each action in \ensuremath{Next} and proved separately. In this case, 
Each of \@pfstepnum{4}{1-4} assumes the action of a phase and proves \ensuremath{TypeOK'} for that case.
\end{cpar}%
\end{lcom}%

\@x{\@s{4.0}\@pfstepnum{3}{1.}\  TypeOK \.{'}}%
 \@x{\@s{8.0}\@pfstepnum{4}{1.}\  {\ASSUME} {\NEW} p \.{\in} {\mathcal{P}} ,\,
 Phase1a ( p )\@s{2.0} {\PROVE} TypeOK \.{'}}%
 \@x{\@s{27.1} {\BY}\@pfstepnum{4}{1} ,\, msgs \.{'} \.{\,\backslash\,} msgs
 \.{\subseteq} Messages ,\, FS\_AddElement {\DEF} Phase1a ,\, TypeOK ,\, Send
 ,\, Messages}%
 \@x{\@s{8.0}\@pfstepnum{4}{2.}\  {\ASSUME} {\NEW} p \.{\in} {\mathcal{P}} ,\,
 Phase2a ( p )\@s{2.0} {\PROVE} TypeOK \.{'}}%
 \@x{\@s{12.0}\@pfstepnum{5}{1.}\  {\PICK} Q \.{\in} {\mathcal{Q}} ,\, S \.{\in}
 {\SUBSET} \{ m \.{\in} msgs \.{:} ( m . type \.{=}\@w{1b} ) \.{\land} ( m .
 bal \.{=} pBal [ p ] ) \} \.{:}}%
 \@x{\@s{39.1} \.{\land} \A\, a \.{\in} Q \.{:} \E\, m \.{\in} S \.{:} m .
 from \.{=} a}%
 \@x{\@s{39.1} \.{\land} Send ( [ type \.{\mapsto}\@w{2a} ,\, from
 \.{\mapsto} p ,\, bal \.{\mapsto} pBal [ p ] ,\,}%
 \@x{\@s{51.99} propSV \.{\mapsto} PropSV ( {\UNION} \{ m . voted
 \.{:} m \.{\in} S \} ) ] )\@s{2.0} {\BY}\@pfstepnum{4}{2}\  {\DEF} Phase2a}%
 \@x{\@s{12.0}\@pfstepnum{5}{2.}\  UnusedS ( {\UNION} \{ m . voted \.{:} m
 \.{\in} S \} ) \.{\subseteq} {\mathcal{S}}\@s{2.0} {\BY}\@pfstepnum{5}{1}\  {\DEF}
 UnusedS ,\, TypeOK ,\, Messages}%
 \@x{\@s{12.0}\@pfstepnum{5}{3.}\  MaxBSV ( {\UNION} \{ m . voted \.{:} m
 \.{\in} S \} ) \.{\subseteq} [ bal \.{:} {\mathcal{B}} ,\, slot \.{:} {\mathcal{S}} ,\,
 val \.{:} {\mathcal{V}} ]}%
\@x{\@s{31.1} {\BY} {\DEF} MaxBSV ,\, TypeOK ,\, Messages}%
 \@x{\@s{12.0}\@pfstepnum{5}{4.} \.{\land} NewSV ( {\UNION} \{ m .
 voted \.{:} m \.{\in} S \} ) \.{\subseteq} [ slot \.{:} {\mathcal{S}} ,\, val \.{:}
 {\mathcal{V}} ]}%
 \@x{\@s{28.44} \.{\land} MaxSV ( {\UNION} \{ m . voted \.{:} m \.{\in} S \} )
 \.{\subseteq} [ slot \.{:} {\mathcal{S}} ,\, val \.{:} {\mathcal{V}} ]\@s{2.0}
 {\BY}\@pfstepnum{5}{3} ,\,\@pfstepnum{5}{2} ,\, Misc {\DEF} MaxSV}%
 \@x{\@s{12.0}\@pfstepnum{5}{5.}\  PropSV ( {\UNION} \{ m . voted
 \.{:} m \.{\in} S \} ) \.{\subseteq} [ slot \.{:} {\mathcal{S}} ,\, val \.{:} {\mathcal{V}}
 ]\@s{2.0} {\BY}\@pfstepnum{5}{4}\  {\DEF} PropSV}%
 \@x{\@s{12.0}\@pfstepnum{5}{6.}\  \A\, m2 \.{\in} msgs \.{'}
 \.{\,\backslash\,} msgs \.{:} \.{\land} m2 . type \.{=}\@w{2a} \.{\land} m2
 . from \.{=} p \.{\land} m2 . bal \.{=} pBal [ p ]}%
 \@x{\@s{109.13} \.{\land} m2 . propSV \.{=} PropSV ( {\UNION} \{ m .
 voted \.{:} m \.{\in} S \} )}%
 \@x{\@s{31.1} {\BY}\@pfstepnum{5}{1} ,\,\@pfstepnum{5}{5}\  {\DEF} Send ,\,
 TypeOK ,\, Messages}%
 \@x{\@s{12.0}\@pfstepnum{5}{7.}\  ( msgs \.{\subseteq} Messages )
 \.{'}\@s{2.0} {\BY}\@pfstepnum{4}{2} ,\,\@pfstepnum{5}{6}
 ,\,\@pfstepnum{5}{1} ,\,\@pfstepnum{5}{5} ,\, msgs \.{'} \.{\,\backslash\,}
 msgs \.{\subseteq} Messages {\DEF} Phase2a ,\,}%
\@x{\@s{31.1} TypeOK ,\, Send ,\, Messages}%
 \@x{\@s{12.0}\@pfstepnum{5}{}\  {\QED} {\BY}\@pfstepnum{5}{7}
 ,\,\@pfstepnum{4}{2} ,\, FS\_AddElement {\DEF} Phase2a ,\, TypeOK ,\, Send}%
 \@x{\@s{8.0}\@pfstepnum{4}{3.}\  {\ASSUME} {\NEW} a \.{\in} {\mathcal{A}} ,\,
 Phase1b ( a )\@s{2.0} {\PROVE} TypeOK \.{'}}%
 \@x{\@s{12.0}\@pfstepnum{5}{1.}\  {\PICK} m \.{\in} msgs \.{:} Phase1b ( a )
 {\bang} ( m )\@s{2.0} {\BY}\@pfstepnum{4}{3}\  {\DEF} Phase1b}%
 \@x{\@s{12.0}\@pfstepnum{5}{2.} \.{\lor} msgs \.{'} \.{=} msgs \.{\cup} \{ [
 type \.{\mapsto}\@w{1b} ,\, from \.{\mapsto} a ,\, bal \.{\mapsto} m . bal
 ,\, voted \.{\mapsto} aVoted [ a ] ] \}}%
 \@x{\@s{28.44} \.{\lor} msgs \.{'} \.{=} msgs \.{\cup} \{ [ type
 \.{\mapsto}\@w{preempt} ,\, to \.{\mapsto} m . from ,\, bal \.{\mapsto} aBal
 [ a ] ] \}\@s{2.0} {\BY}\@pfstepnum{5}{1}\  {\DEF} Send}%
 \@x{\@s{12.0}\@pfstepnum{5}{3.}\  \A\, m2 \.{\in} msgs \.{'}
 \.{\,\backslash\,} msgs \.{:} \.{\lor} m2 . type \.{=}\@w{1b} \.{\land} m2 .
 from \.{=} a \.{\land} m2 . bal \.{=} m . bal \.{\land} m2 . voted \.{=}
 aVoted [ a ]}%
 \@x{\@s{109.13} \.{\lor} m2 . type \.{=}\@w{preempt} \.{\land} m2 . to \.{=}
 m . from \.{\land} m2 . bal \.{=} aBal [ a ]}%
\@x{\@s{31.1} {\BY}\@pfstepnum{5}{1}\  {\DEF} Send}%
 \@x{\@s{12.0}\@pfstepnum{5}{4.}\  ( msgs \.{\subseteq} Messages )
 \.{'}\@s{2.0} {\BY}\@pfstepnum{5}{1} ,\,\@pfstepnum{5}{3}\  {\DEF} TypeOK
 ,\, Messages ,\, Send}%
 \@x{\@s{12.0}\@pfstepnum{5}{5.}\  IsFiniteSet ( msgs ) \.{'}\@s{2.0}
 {\BY}\@pfstepnum{5}{2} ,\, FS\_AddElement {\DEF} TypeOK}%
 \@x{\@s{12.0}\@pfstepnum{5}{}\  {\QED} {\BY}\@pfstepnum{5}{4}
 ,\,\@pfstepnum{5}{5} ,\,\@pfstepnum{4}{3}\  {\DEF} Phase1b ,\, TypeOK}%
 \@x{\@s{8.0}\@pfstepnum{4}{4.}\  {\ASSUME} {\NEW} a \.{\in} {\mathcal{A}} ,\,
 Phase2b ( a )\@s{2.0} {\PROVE} TypeOK \.{'}}%
 \@x{\@s{12.0}\@pfstepnum{5}{1.}\  {\PICK} m \.{\in} msgs \.{:} Phase2b ( a )
 {\bang} ( m )\@s{2.0} {\BY}\@pfstepnum{4}{4}\  {\DEF} Phase2b}%
 \@x{\@s{12.0}\@pfstepnum{5}{2.} \.{\lor} msgs \.{'} \.{=} msgs \.{\cup} \{ [
 type \.{\mapsto}\@w{2b} ,\, bal \.{\mapsto} m . bal ,\, from \.{\mapsto} a
 ,\, propSV \.{\mapsto} m . propSV ] \}}%
 \@x{\@s{28.44} \.{\lor} msgs \.{'} \.{=} msgs \.{\cup} \{ [ type
 \.{\mapsto}\@w{preempt} ,\, to \.{\mapsto} m . from ,\, bal \.{\mapsto} aBal
 [ a ] ] \}\@s{2.0} {\BY}\@pfstepnum{5}{1}\  {\DEF} Send}%
 \@x{\@s{12.0}\@pfstepnum{5}{3.}\  IsFiniteSet ( msgs ) \.{'}\@s{2.0}
 {\BY}\@pfstepnum{5}{2} ,\, FS\_AddElement {\DEF} TypeOK}%
 \@x{\@s{12.0}\@pfstepnum{5}{4.}\  \A\, m2 \.{\in} msgs \.{'}
 \.{\,\backslash\,} msgs \.{:} \.{\lor} m2 . type \.{=}\@w{2b} \.{\land} m2 .
 from \.{=} a \.{\land} m2 . bal \.{=} m . bal \.{\land} m2 . propSV \.{=} m
 . propSV}%
 \@x{\@s{109.13} \.{\lor} m2 . type \.{=}\@w{preempt} \.{\land} m2 . to \.{=}
 m . from \.{\land} m2 . bal \.{=} aBal [ a ]}%
\@x{\@s{31.1} {\BY}\@pfstepnum{5}{1}\  {\DEF} Send}%
 \@x{\@s{12.0}\@pfstepnum{5}{5.}\  ( msgs \.{\subseteq} Messages )
 \.{'}\@s{2.0} {\BY}\@pfstepnum{5}{1} ,\,\@pfstepnum{5}{4}\  {\DEF} TypeOK
 ,\, Send ,\, Messages}%
\@x{\@s{12.0}\@pfstepnum{5}{6.} \.{\lor} aVoted \.{=} aVoted \.{'}}%
 \@x{\@s{28.44} \.{\lor} \.{\land} {\DOMAIN} aVoted \.{=} {\DOMAIN} aVoted
 \.{'}}%
 \@x{\@s{37.33} \.{\land} aVoted \.{'} [ a ] \.{=} \{ d \.{\in} aVoted [ a ]
 \.{:} {\nexists}\, d2 \.{\in} m . propSV \.{:} d . slot \.{=} d2 . slot \}
 \.{\cup}}%
 \@x{\@s{94.38} \{ [ bal \.{\mapsto} m . bal ,\, slot \.{\mapsto} d . slot
 ,\, val \.{\mapsto} d . val ] \.{:} d \.{\in} m . propSV \}}%
 \@x{\@s{37.33} \.{\land} \A\, a2 \.{\in} {\mathcal{A}} \.{\,\backslash\,} \{ a \}
 \.{:} aVoted [ a2 ] \.{=} aVoted \.{'} [ a2 ]\@s{2.0}
 {\BY}\@pfstepnum{5}{1}\  {\DEF} TypeOK}%
 \@x{\@s{12.0}\@pfstepnum{5}{7.}\  ( aVoted \.{\in} [ {\mathcal{A}}
 \.{\rightarrow} {\SUBSET} [ bal \.{:} {\mathcal{B}} ,\, slot \.{:} {\mathcal{S}} ,\, val
 \.{:} {\mathcal{V}} ] ] ) \.{'}\@s{2.0} {\BY}\@pfstepnum{5}{1} ,\,\@pfstepnum{5}{6}
 ,\,}%
 \@x{\@s{31.1} \{ [ bal \.{\mapsto} m . bal ,\, slot \.{\mapsto} d . slot ,\,
 val \.{\mapsto} d . val ] \.{:} d \.{\in} m . propSV \} \.{\subseteq} [ bal
 \.{:} {\mathcal{B}} ,\, slot \.{:} {\mathcal{S}} ,\, val \.{:} {\mathcal{V}} ] ,\,}%
 \@x{\@s{31.1} \{ d \.{\in} aVoted [ a ] \.{:} {\nexists}\, d2 \.{\in} m .
 propSV \.{:} d . slot \.{=} d2 . slot \} \.{\subseteq} [ bal \.{:} {\mathcal{B}}
 ,\, slot \.{:} {\mathcal{S}} ,\, val \.{:} {\mathcal{V}} ] ,\,}%
 \@x{\@s{31.1} aVoted \.{'} [ a ] \.{\subseteq} [ bal \.{:} {\mathcal{B}} ,\, slot
 \.{:} {\mathcal{S}} ,\, val \.{:} {\mathcal{V}} ] {\DEF} TypeOK ,\, Messages}%
 \@x{\@s{12.0}\@pfstepnum{5}{}\  {\QED} {\BY}\@pfstepnum{5}{5}
 ,\,\@pfstepnum{5}{7} ,\,\@pfstepnum{4}{4} ,\,\@pfstepnum{5}{3}\  {\DEF}
 Phase2b ,\, TypeOK}%
\begin{lcom}{10.5}%
\begin{cpar}{0}{F}{F}{0}{0}{}%
\@pfstepnum{4}{5} proves \ensuremath{TypeOK'} for \ensuremath{Preempt}. This is the new part of the remedial proof described in Section~\ref{secMadeupProof}.
\end{cpar}%
\end{lcom}%

 \@x{\@s{8.0}\@pfstepnum{4}{5.}\  {\ASSUME} {\NEW} p \.{\in} {\mathcal{P}} ,\,
 Preempt ( p )\@s{2.0} {\PROVE} TypeOK \.{'}}%
 \@x{\@s{12.0}\@pfstepnum{5}{}\  {\DEFINE} S \.{\defeq} \{ m1 \.{\in} msgs
 \.{:} m1 . type \.{=}\@w{1a} \}\@s{2.0} T \.{\defeq} \{ s . bal \.{:} s
 \.{\in} S \}\@s{2.0} f \.{\defeq} [ s \.{\in} S \.{\mapsto} s . bal ]}%
\@x{\@s{12.0}\@pfstepnum{5}{}\  {\HIDE} {\DEF} S ,\, T ,\, f}%
 \@x{\@s{12.0}\@pfstepnum{5}{1.}\  {\PICK} m \.{\in} msgs \.{:} Preempt ( p )
 {\bang} ( m )\@s{2.0} {\BY}\@pfstepnum{4}{5}\  {\DEF} Preempt}%
 \@x{\@s{12.0}\@pfstepnum{5}{2.}\  T \.{\subseteq} {\mathcal{B}}\@s{2.0} {\BY}
 {\DEF} T ,\, S ,\, TypeOK ,\, Messages}%
 \@x{\@s{12.0}\@pfstepnum{5}{3.}\  \E\, b \.{\in} {\mathcal{B}} \.{:} b \.{>} m .
 bal \.{\land} b \.{\notin} T}%
 \@x{\@s{31.1} {\BY}\@pfstepnum{5}{2} ,\, MaxInSet ,\, Max ( T \.{\cup} \{ m
 . bal \} ) \.{+} 1 \.{>} m . bal ,\,\@pfstepnum{5}{1}\  {\DEF} Max ,\,
 TypeOK ,\, Messages}%
 \@x{\@s{12.0}\@pfstepnum{5}{4.}\  NewBal ( m . bal ) \.{\in}
 {\mathcal{B}}\@s{2.0} {\BY}\@pfstepnum{5}{3}\  {\DEF} NewBal ,\, TypeOK ,\,
 Messages ,\, T ,\, S}%
 \@x{\@s{12.0}\@pfstepnum{5}{5.}\  ( pBal \.{\in} [ {\mathcal{P}} \.{\rightarrow}
 {\mathcal{B}} ] ) \.{'}\@s{2.0} {\BY}\@pfstepnum{5}{1} ,\,\@pfstepnum{5}{4}\ 
 {\DEF} TypeOK ,\, Messages}%
 \@x{\@s{12.0}\@pfstepnum{5}{}\  {\QED} {\BY}\@pfstepnum{4}{5}
 ,\,\@pfstepnum{5}{5}\  {\DEF} Preempt ,\, TypeOK}%
 \@x{\@s{8.0}\@pfstepnum{4}{}\  {\QED} {\BY}\@pfstepnum{2}{1}
 ,\,\@pfstepnum{4}{1} ,\,\@pfstepnum{4}{2} ,\,\@pfstepnum{4}{3}
 ,\,\@pfstepnum{4}{4} ,\,\@pfstepnum{4}{5}\  {\DEF} Next}%
%\@pvspace{8.0pt}%
\begin{lcom}{6.5}%
\begin{cpar}{0}{F}{F}{0}{0}{}%
\@pfstepnum{3}{2} proves \ensuremath{AccInv'} for \ensuremath{Next}.  
%It uses the same strategy as for proving \ensuremath{TypeOK'} in \@pfstepnum{3}{1}.
Each of \@pfstepnum{4}{1-4} assumes the action of a phase and proves \ensuremath{AccInv'} for that case.\\
Only \ensuremath{Phase2b} is challenging because only it updates acceptor variables.
\end{cpar}\end{lcom}
\@x{\@s{4.0}\@pfstepnum{3}{2.}\  AccInv \.{'}}%
 \@x{\@s{8.0}\@pfstepnum{4}{1.} {\CASE} \E\, p \.{\in} {\mathcal{P}} \.{:}
 Phase1a ( p )\@s{2.0} {\BY}\@pfstepnum{4}{1} ,\,\@pfstepnum{3}{1} ,\,
 Phase1aVotedForInv {\DEF} AccInv ,\, TypeOK ,\, Phase1a ,\, Send}%
 \@x{\@s{8.0}\@pfstepnum{4}{2.}\  {\ASSUME} {\NEW} p \.{\in} {\mathcal{P}} ,\,
 Phase2a ( p )\@s{2.0} {\PROVE} AccInv \.{'}}%
 \@x{\@s{12.0}\@pfstepnum{5}{1.}\  \A\, a \.{\in} {\mathcal{A}} ,\, b \.{\in}
 {\mathcal{B}} ,\, s \.{\in} {\mathcal{S}} ,\, v \.{\in} {\mathcal{V}} \.{:} VotedForIn ( a ,\, b
 ,\, s ,\, v ) \.{\equiv} VotedForIn ( a ,\, b ,\, s ,\, v ) \.{'}}%
\@x{\@s{31.1} {\BY}\@pfstepnum{4}{2} ,\, Phase2aVotedForInv}%
 \@x{\@s{12.0}\@pfstepnum{5}{}\  {\QED} {\BY}\@pfstepnum{3}{1}
 ,\,\@pfstepnum{4}{2} ,\,\@pfstepnum{5}{1}\  {\DEF} AccInv ,\, TypeOK ,\,
 Phase2a ,\, Send ,\, Messages}%
 \@x{\@s{8.0}\@pfstepnum{4}{3.} {\CASE} \E\, a \.{\in} {\mathcal{A}} \.{:}
 Phase1b ( a )\@s{2.0} {\BY}\@pfstepnum{4}{3} ,\,\@pfstepnum{3}{1} ,\,
 Phase1bVotedForInv {\DEF} AccInv ,\, TypeOK ,\, Phase1b ,\, Send}%
%\begin{lcom}{10.5}%
%\begin{cpar}{0}{F}{F}{0}{0}{}%
% To prove \ensuremath{AccInv'}, only \ensuremath{Phase2b} is challenging because only it updates acceptor variables.
%\end{cpar}\end{lcom}
 \@x{\@s{8.0}\@pfstepnum{4}{4.}\  {\ASSUME} {\NEW} a \.{\in} {\mathcal{A}} ,\,
 Phase2b ( a )\@s{2.0} {\PROVE} AccInv \.{'}}%
 \@x{\@s{12.0}\@pfstepnum{5}{}\  {\SUFFICES} {\ASSUME} {\NEW} a2 \.{\in}
 {\mathcal{A}} \.{'}}%
 \@x{\@s{60.31} {\PROVE} ( \.{\land} aBal [ a2 ] \.{=} \.{-} 1
 \.{\implies} aVoted [ a2 ] \.{=} {\emptyset}}%
 \@x{\@s{93.38} \.{\land} \A\, r \.{\in} aVoted [ a2 ] \.{:} VotedForIn ( a2
 ,\, r . bal ,\, r . slot ,\, r . val ) \.{\land} r . bal \.{\leq} aBal [ a2
 ]}%
 \@x{\@s{93.38} \.{\land} \A\, b\@s{0.05} \.{\in} {\mathcal{B}} ,\, s \.{\in} {\mathcal{S}}
 ,\, v \.{\in} {\mathcal{V}} \.{:}}%
 \@x{\@s{106.27} \.{\land} VotedForIn ( a2 ,\, b ,\, s ,\, v ) \.{\implies}
 \E\, r \.{\in} aVoted [ a2 ] \.{:} r . bal \.{\geq} b \.{\land} r . slot
 \.{=} s}%
 \@x{\@s{106.27} \.{\land} b \.{>} MaxBalInSlot ( aVoted [ a2 ] ,\, s )
 \.{\implies} {\lnot} VotedForIn ( a2 ,\, b ,\, s ,\, v ) ) \.{'}}%
\@x{\@s{60.31} {\BY} {\DEF} AccInv}%
 \@x{\@s{12.0}\@pfstepnum{5}{1.}\  {\PICK} m \.{\in} msgs \.{:} Phase2b ( a )
 {\bang} ( m )\@s{2.0} {\BY}\@pfstepnum{4}{4}\  {\DEF} Phase2b}%
\begin{lcom}{14.5}%
\begin{cpar}{0}{F}{F}{0}{0}{}%
\@pfstepnum{5}{2} assumes that \ensuremath{a} received a \textsf{2a} message with a ballot lower than the highest it has seen, %annie: why write with negation in proof sc - i don't think its necessary but what I've written in <5>2 is the negation of <5>3. I remember that the solvers have trouble figuring out negations on quantified expressions but may be not in this one because there is no quantification.
thus triggering preemption. This case is simple as acceptor variables are unchanged.
\end{cpar}\end{lcom}
 \@x{\@s{12.0}\@pfstepnum{5}{2.} {\CASE} ( a2 \.{=} a \.{\land} {\lnot} ( m .
 bal \.{\geq} aBal [ a ] ) ) \.{\lor} a2 \.{\neq} a}%
 \@x{\@s{16.0}\@pfstepnum{6}{1.}\  \A\, b \.{\in} {\mathcal{B}} ,\, s \.{\in} {\mathcal{S}}
 ,\, v \.{\in} {\mathcal{V}} \.{:} VotedForIn ( a2 ,\, b ,\, s ,\, v ) \.{\equiv}
 VotedForIn ( a2 ,\, b ,\, s ,\, v ) \.{'}}%
 \@x{\@s{35.1} {\BY}\@pfstepnum{3}{1} ,\,\@pfstepnum{5}{1}
 ,\,\@pfstepnum{5}{2}\  {\DEF} Phase2b ,\, TypeOK ,\, {\mathcal{B}} ,\, Messages
 ,\, VotedForIn ,\, Send}%
 \@x{\@s{16.0}\@pfstepnum{6}{}\  {\QED} {\BY}\@pfstepnum{6}{1}
 ,\,\@pfstepnum{5}{1} ,\,\@pfstepnum{5}{2} ,\,\@pfstepnum{4}{4}
 ,\,\@pfstepnum{3}{1}\  {\DEF} Phase2b ,\, Send ,\, AccInv ,\, TypeOK ,\,
 Messages}%
\begin{lcom}{14.5}%
\begin{cpar}{0}{F}{F}{0}{0}{}%
\@pfstepnum{5}{3} assumes that \ensuremath{a} received a \textsf{2a} message with a ballot higher than or equal to the highest it has seen. Thus, it responds with a \textsf{2b} message and updates its variables as specified. Each of \@pfstepnum{6}{1-4} proves a conjunct of \ensuremath{AccInv}.
\end{cpar}\end{lcom}
 \@x{\@s{12.0}\@pfstepnum{5}{3.} {\CASE} a2 \.{=} a \.{\land} ( m . bal
 \.{\geq} aBal [ a ] )}%
 \@x{\@s{16.0}\@pfstepnum{6}{1.}\  ( aBal [ a2 ] \.{=} \.{-} 1 \.{\implies}
 aVoted [ a2 ] \.{=} {\emptyset} ) \.{'}\@s{2.0} {\BY}\@pfstepnum{5}{3}
 ,\,\@pfstepnum{4}{4} ,\,\@pfstepnum{3}{1}\  {\DEF} AccInv ,\, Phase2b ,\,
 Send ,\,}%
\@x{\@s{35.1} TypeOK ,\, Messages}%
 \@x{\@s{16.0}\@pfstepnum{6}{2.}\  ( \A\, r \.{\in} aVoted [ a2 ] \.{:}
 VotedForIn ( a2 ,\, r . bal ,\, r . slot ,\, r . val ) \.{\land} r . bal
 \.{\leq} aBal [ a2 ] ) \.{'}}%
 \@x{\@s{20.0}\@pfstepnum{7}{}\  {\SUFFICES} {\ASSUME} {\NEW} r \.{\in} (
 aVoted [ a2 ] ) \.{'}}%
 \@x{\@s{68.31} {\PROVE} ( VotedForIn ( a2 ,\, r . bal ,\, r . slot
 ,\, r . val ) \.{\land} r . bal \.{\leq} aBal [ a2 ] ) \.{'} \@s{4.0} {\OBVIOUS}}%
\begin{lcom}{22.5}%
\begin{cpar}{0}{F}{F}{0}{0}{}%
\@pfstepnum{7}{1} uses two cases. \@pfstepnum{8}{1} is for \ensuremath{r \in aVoted[a2]} and uses invariance lemma for  \ensuremath{Phase2b}. \@pfstepnum{8}{2} is for the increment \ensuremath{r \in aVoted'[a2] \setminus aVoted[a2]} and uses definition of \ensuremath{Phase2b}.
\end{cpar}\end{lcom}
 \@x{\@s{20.0}\@pfstepnum{7}{1.}\  VotedForIn ( a2 ,\, r . bal ,\, r . slot
 ,\, r . val ) \.{'}}%
\@x{\@s{24.0}\@pfstepnum{8}{1.} {\CASE} r \.{\in} aVoted [ a2 ]}%
 \@x{\@s{28.0}\@pfstepnum{9}{1.}\  VotedForIn ( a2 ,\, r . bal ,\, r . slot
 ,\, r . val )\@s{2.0} {\BY}\@pfstepnum{5}{3} ,\,\@pfstepnum{4}{4}
 ,\,\@pfstepnum{8}{1}\  {\DEF} AccInv}%
 \@x{\@s{28.0}\@pfstepnum{9}{}\  {\QED} {\BY}\@pfstepnum{9}{1} ,\,
 Phase2bVotedForInv ,\,\@pfstepnum{3}{1} ,\,\@pfstepnum{4}{4}\  {\DEF} TypeOK
 ,\, Messages}%
 \@x{\@s{24.0}\@pfstepnum{8}{2.} {\CASE} r \.{\in} aVoted \.{'} [ a2 ]
 \.{\,\backslash\,} aVoted [ a2 ]}%
 \@x{\@s{28.0}\@pfstepnum{9}{1.}\  \E\, m2 \.{\in} msgs \.{'} \.{:} m2 . type
 \.{=}\@w{2b} \.{\land} m2 . from \.{=} a2 \.{\land} m2 . bal \.{=} m . bal
 \.{\land} m2 . propSV \.{=} m . propSV}%
 \@x{\@s{47.1} {\BY}\@pfstepnum{3}{1} ,\,\@pfstepnum{8}{2}
 ,\,\@pfstepnum{5}{1} ,\,\@pfstepnum{5}{3}\  {\DEF} Send}%
 \@x{\@s{28.0}\@pfstepnum{9}{2.}\  r . bal \.{=} m . bal \.{\land} \E\, d
 \.{\in} m . propSV \.{:} r . slot \.{=} d . slot \.{\land} r . val \.{=} d
 . val\@s{2.0} {\BY}\@pfstepnum{5}{1} ,\,\@pfstepnum{5}{3}
 ,\,\@pfstepnum{3}{1} ,\,\@pfstepnum{8}{2}\ }%
 \@x{\@s{28.0}\@pfstepnum{9}{}\  {\QED} {\BY}\@pfstepnum{9}{1}
 ,\,\@pfstepnum{9}{2}\  {\DEF} Send ,\, TypeOK ,\, Messages ,\, VotedForIn}%
 \@x{\@s{24.0}\@pfstepnum{8}{}\  {\QED} {\BY}\@pfstepnum{8}{1}
 ,\,\@pfstepnum{8}{2}\ }%
 \@x{\@s{20.0}\@pfstepnum{7}{2.}\  ( r . bal \.{\leq} aBal [ a2 ] )
 \.{'}\@s{2.0} {\BY}\@pfstepnum{5}{1} ,\,\@pfstepnum{5}{3}
 ,\,\@pfstepnum{3}{1} ,\, aBal [ a ] \.{\leq} aBal \.{'} [ a ] ,\, r \.{\in}
 aVoted [ a ] \.{\implies} r . bal \.{\leq} aBal \.{'} [ a ] ,\,}%
 \@x{\@s{39.1} aVoted \.{'} [ a ] \.{=} \{ d \.{\in} aVoted [ a ] \.{:}
 {\nexists}\, d2 \.{\in} m . propSV \.{:} d . slot \.{=} d2 . slot \}
 \.{\cup}}%
 \@x{\@s{87.27} \{ [ bal \.{\mapsto} m . bal ,\, slot \.{\mapsto} d . slot
 ,\, val \.{\mapsto} d . val ] \.{:} d \.{\in} m . propSV \} ,\,}%
 \@x{\@s{39.1} r \.{\in} aVoted \.{'} [ a2 ] \.{\,\backslash\,} aVoted [ a2 ]
 \.{\implies} r . bal \.{=} m . bal {\DEF} AccInv ,\, Send ,\, TypeOK ,\, Messages}%
 \@x{\@s{20.0}\@pfstepnum{7}{}\  {\QED} {\BY}\@pfstepnum{7}{1}
 ,\,\@pfstepnum{7}{2}\ }%
 \@x{\@s{16.0}\@pfstepnum{6}{3.}\  ( \A\, b \.{\in} {\mathcal{B}} ,\, s \.{\in}
 {\mathcal{S}} ,\, v \.{\in} {\mathcal{V}} \.{:} VotedForIn ( a2 ,\, b ,\, s ,\, v )
 \.{\implies} \E\, r \.{\in} aVoted [ a2 ] \.{:} r . bal \.{\geq} b \.{\land}
 r . slot \.{=} s ) \.{'}}%
 \@x{\@s{20.0}\@pfstepnum{7}{}\  {\SUFFICES} {\ASSUME} {\NEW} b \.{\in}
 {\mathcal{B}} \.{'} ,\, {\NEW} s \.{\in} {\mathcal{S}} \.{'} ,\, {\NEW} v \.{\in} {\mathcal{V}}
 \.{'} ,\, VotedForIn ( a2 ,\, b ,\, s ,\, v ) \.{'}}%
 \@x{\@s{68.31} {\PROVE} ( \E\, r \.{\in} aVoted [ a2 ] \.{:} r . bal
 \.{\geq} b \.{\land} r . slot \.{=} s ) \.{'} \@s{4.0} {\OBVIOUS}}%
\@x{\@s{20.0}\@pfstepnum{7}{1.} {\CASE} VotedForIn ( a2 ,\, b ,\, s ,\, v )}%
 \@x{\@s{24.0}\@pfstepnum{8}{1.}\  {\PICK} r \.{\in} aVoted [ a2 ] \.{:} r .
 bal \.{\geq} b \.{\land} r . slot \.{=} s\@s{2.0} {\BY}\@pfstepnum{7}{1}\ 
 {\DEF} AccInv ,\, TypeOK ,\, Messages}%
 \@x{\@s{24.0}\@pfstepnum{8}{2.}\  m . bal \.{\geq} b\@s{2.0}
 {\BY}\@pfstepnum{8}{1} ,\,\@pfstepnum{5}{3}\  {\DEF} TypeOK ,\, Messages ,\,
 AccInv}%
 \@x{\@s{24.0}\@pfstepnum{8}{3.} {\CASE} \E\, d \.{\in} m . propSV \.{:} d .
 slot \.{=} s}%
 \@x{\@s{28.0}\@pfstepnum{9}{1.}\  \E\, r2 \.{\in} aVoted \.{'} [ a ] \.{:} r2
 . bal \.{=} m . bal \.{\land} r2 . slot \.{=} s\@s{2.0} {\BY}\@pfstepnum{5}{1}
 ,\,\@pfstepnum{5}{3} ,\,}%
 \@x{\@s{47.1} aVoted \.{'} [ a ] \.{=} \{ d \.{\in} aVoted [ a ] \.{:}
 {\nexists}\, d2 \.{\in} m . propSV \.{:} d . slot \.{=} d2 . slot \}
 \.{\cup}}%
 \@x{\@s{47.1} \{ [ bal \.{\mapsto} m . bal ,\, slot \.{\mapsto} d . slot ,\,
 val \.{\mapsto} d . val ] \.{:} d \.{\in} m . propSV \}
 ,\,\@pfstepnum{8}{3}\  {\DEF} TypeOK}%
 \@x{\@s{28.0}\@pfstepnum{9}{}\  {\QED} {\BY}\@pfstepnum{9}{1}
 ,\,\@pfstepnum{5}{3} ,\,\@pfstepnum{8}{2}\ }%
 \@x{\@s{24.0}\@pfstepnum{8}{4.} {\CASE} {\nexists}\, d \.{\in} m . propSV
 \.{:} d . slot \.{=} s\@s{2.0} {\BY}\@pfstepnum{8}{4} ,\,\@pfstepnum{8}{1}
 ,\,\@pfstepnum{5}{1} ,\,\@pfstepnum{5}{3} ,\, r \.{\in} aVoted \.{'} [ a2 ]}%
 \@x{\@s{24.0}\@pfstepnum{8}{}\  {\QED} {\BY}\@pfstepnum{8}{3}
 ,\,\@pfstepnum{8}{4}\ }%
 \@x{\@s{20.0}\@pfstepnum{7}{2.} {\CASE} {\lnot} VotedForIn ( a2 ,\, b ,\, s
 ,\, v )}%
 \@x{\@s{24.0}\@pfstepnum{8}{1.}\  \E\, d \.{\in} m . propSV \.{:} d . slot
 \.{=} s\@s{2.0} {\BY}\@pfstepnum{5}{1} ,\,\@pfstepnum{7}{2}\  {\DEF} Send
 ,\, TypeOK ,\, Messages ,\, VotedForIn}%
 \@x{\@s{24.0}\@pfstepnum{8}{2.}\  \E\, r\@s{0.68} \.{\in} aVoted \.{'} [ a2 ]
 \.{:} r . bal \.{=} m . bal \.{\land} r . slot \.{=} s\@s{2.0}
 {\BY}\@pfstepnum{5}{1} ,\,\@pfstepnum{8}{1} ,\,\@pfstepnum{2}{1}
 ,\,\@pfstepnum{5}{3}\  {\DEF} Send ,\, TypeOK ,\,}%
\@x{\@s{43.1} Messages}%
 \@x{\@s{24.0}\@pfstepnum{8}{3.}\  b \.{=} m . bal\@s{2.0}
 {\BY}\@pfstepnum{5}{1} ,\,\@pfstepnum{7}{2} ,\,\@pfstepnum{5}{3}\  {\DEF}
 VotedForIn ,\, Send}%
 \@x{\@s{24.0}\@pfstepnum{8}{}\  {\QED} {\BY}\@pfstepnum{8}{2}
 ,\,\@pfstepnum{8}{3}\ }%
 \@x{\@s{20.0}\@pfstepnum{7}{}\  {\QED} {\BY}\@pfstepnum{7}{1}
 ,\,\@pfstepnum{7}{2}\ }%
 \@x{\@s{16.0}\@pfstepnum{6}{4.}\  ( \A\, b \.{\in} {\mathcal{B}} ,\, s \.{\in}
 {\mathcal{S}} ,\, v \.{\in} {\mathcal{V}} \.{:} b \.{>} MaxBalInSlot ( aVoted [ a2 ] ,\,
 s ) \.{\implies} {\lnot} VotedForIn ( a2 ,\, b ,\, s ,\, v ) ) \.{'}}%
 \@x{\@s{20.0}\@pfstepnum{7}{}\  {\SUFFICES} {\ASSUME} {\NEW} b \.{\in}
 {\mathcal{B}} \.{'} ,\, {\NEW} s \.{\in} {\mathcal{S}} \.{'} ,\, {\NEW} v \.{\in} {\mathcal{V}}
 \.{'} ,\, ( VotedForIn ( a2 ,\, b ,\, s ,\, v ) ) \.{'} ,\,}%
\@x{\@s{98.91} b \.{>} MaxBalInSlot ( aVoted' [ a2 ] ,\, s ) \@s{2.0} {\PROVE} {\FALSE} \@s{4.0} {\OBVIOUS}}%
 \@x{\@s{20.0}\@pfstepnum{7}{1.}\  {\nexists}\, d \.{\in} aVoted \.{'} [ a2 ]
 \.{:} d . slot \.{=} s \.{\land} d . bal \.{>} MaxBalInSlot ( aVoted [ a2
 ] ,\, s ) \.{'}}%
\@x{\@s{39.1} {\BY} MaxBinSNoMore ,\,\@pfstepnum{3}{1}\  {\DEF} TypeOK}%
 \@x{\@s{20.0}\@pfstepnum{7}{}\  {\QED} {\BY}\@pfstepnum{7}{1}
 ,\,\@pfstepnum{6}{3} ,\, \E\, r \.{\in} aVoted \.{'} [ a2 ] \.{:} r . bal
 \.{\geq} b \.{\land} r . slot \.{=} s ,\, MaxBinSType ,\,\@pfstepnum{3}{1}\ 
 {\DEF} Send ,\,}%
\@x{\@s{32.89} TypeOK ,\, Messages}%
 \@x{\@s{16.0}\@pfstepnum{6}{}\  {\QED} {\BY}\@pfstepnum{6}{1}
 ,\,\@pfstepnum{6}{2} ,\,\@pfstepnum{6}{3} ,\,\@pfstepnum{6}{4}\ }%
 \@x{\@s{12.0}\@pfstepnum{5}{}\  {\QED} {\BY}\@pfstepnum{5}{2}
 ,\,\@pfstepnum{5}{3}\ }%
 \@x{\@s{8.0}\@pfstepnum{4}{5.} {\CASE} \E\, p \.{\in} {\mathcal{P}} \.{:}
 Preempt ( p )}%
 \@x{\@s{12.0}\@pfstepnum{5}{1.}\  \A\, a \.{\in} {\mathcal{A}} ,\, s \.{\in}
 {\mathcal{S}} \.{:} MaxBalInSlot ( aVoted [ a ] ,\, s ) \.{=} MaxBalInSlot (
 aVoted [ a ] ,\, s ) \.{'}}%
 \@x{\@s{31.1} {\BY}\@pfstepnum{3}{1} ,\,\@pfstepnum{4}{5}\  {\DEF} Preempt
 ,\, MaxBalInSlot}%
 \@x{\@s{12.0}\@pfstepnum{5}{}\  {\QED} {\BY}\@pfstepnum{4}{5}
 ,\,\@pfstepnum{3}{1} ,\, PreemptVotedForInv ,\,\@pfstepnum{5}{1}\  {\DEF}
 AccInv ,\, TypeOK ,\, Preempt ,\, Send}%
 \@x{\@s{8.0}\@pfstepnum{4}{} . {\QED} {\BY}\@pfstepnum{4}{1}
 ,\,\@pfstepnum{4}{2} ,\,\@pfstepnum{4}{3} ,\,\@pfstepnum{4}{4}
 ,\,\@pfstepnum{4}{5} ,\,\@pfstepnum{2}{1}\  {\DEF} Next}%
%\@pvspace{8.0pt}%
\begin{lcom}{6.5}%
\begin{cpar}{0}{F}{F}{0}{0}{}%
 \@pfstepnum{3}{3} proves \ensuremath{MsgInv'} for \ensuremath{Next}. 
 %It uses the same strategy as for proving \ensuremath{TypeOK'} in \@pfstepnum{3}{1}.
 Each of \@pfstepnum{4}{1-4} assumes the action of a phase and proves \ensuremath{MsgInv'} for that case.
\end{cpar}\end{lcom}
\@x{\@s{4.0}\@pfstepnum{3}{3.}\  MsgInv \.{'}}%
 \@x{\@s{8.0}\@pfstepnum{4}{1.} {\CASE} \E\, p \.{\in} {\mathcal{P}} \.{:}
 Phase1a ( p )\@s{2.0} {\BY}\@pfstepnum{4}{1} ,\, Phase1aVotedForInv
 ,\,\@pfstepnum{3}{1} ,\, SafeAtStable ,\,\@pfstepnum{2}{1}\ }%
 \@x{\@s{18.0} {\DEF} Phase1a ,\, MsgInv ,\, Send ,\, TypeOK ,\, Messages ,\,
 MsgInv1b ,\, MsgInv2a ,\, MsgInv2b}%
\begin{lcom}{10.5}%
\begin{cpar}{0}{F}{F}{0}{0}{}%
 \@pfstepnum{4}{2} proves \ensuremath{MsgInv'} for \ensuremath{Phase1b}. \@pfstepnum{5}{13,14} conclude the proof while \@pfstepnum{5}{1-12} prove intermediate facts. Each of \@pfstepnum{5}{2,4,12} proves a conjunct of \ensuremath{MsgInv1b} for the increment \ensuremath{m1}---the new \ensuremath{1b} message sent in \ensuremath{Phase1b(a)}.
\end{cpar}\end{lcom}
 \@x{\@s{8.0}\@pfstepnum{4}{2.}\  {\ASSUME} {\NEW} a \.{\in} {\mathcal{A}} ,\,
 Phase1b ( a )\@s{2.0} {\PROVE} MsgInv \.{'}}%
 \@x{\@s{12.0}\@pfstepnum{5}{1.}\  {\PICK} m \.{\in} msgs \.{:} Phase1b ( a )
 {\bang} ( m )\@s{2.0} {\BY}\@pfstepnum{4}{2}\  {\DEF} Phase1b}%
 \@x{\@s{12.0}\@pfstepnum{5}{}\  {\DEFINE} m1 \.{\defeq} [ type
 \.{\mapsto}\@w{1b} ,\, from \.{\mapsto} a ,\, bal \.{\mapsto} m . bal ,\,
 voted \.{\mapsto} aVoted [ a ] ]}%
 \@x{\@s{12.0}\@pfstepnum{5}{2.}\  ( m1 . bal \.{\leq} aBal [ m1 . from ] )
 \.{'}\@s{2.0} {\BY}\@pfstepnum{5}{1} ,\,\@pfstepnum{3}{1}\  {\DEF} Phase1b
 ,\, Send ,\, TypeOK ,\, MsgInv ,\, Messages}%
 \@x{\@s{12.0}\@pfstepnum{5}{3.}\  m1 . voted \.{=} aVoted [ m1 . from
 ]\@s{2.0} {\BY}\@pfstepnum{5}{1} ,\,\@pfstepnum{3}{1}\  {\DEF} Phase1b ,\,
 Send ,\, TypeOK ,\, Messages}%
 \@x{\@s{12.0}\@pfstepnum{5}{4.}\  ( \A\, r \.{\in} m1 . voted \.{:}
 VotedForIn ( m1 . from ,\, r . bal ,\, r . slot ,\, r . val ) ) \.{'}}%
 \@x{\@s{31.1} {\BY}\@pfstepnum{5}{1} ,\,\@pfstepnum{4}{2} ,\,
 Phase1bVotedForInv ,\,\@pfstepnum{3}{1} ,\,\@pfstepnum{5}{3}\  {\DEF} TypeOK
 ,\, Messages ,\, AccInv}%
 \@x{\@s{12.0}\@pfstepnum{5}{5.}\  ( \A\, b \.{\in} {\mathcal{B}} ,\, s \.{\in}
 {\mathcal{S}} ,\, v \.{\in} {\mathcal{V}} \.{:} b \.{>} MaxBalInSlot ( m1 . voted ,\, s
 ) \.{\implies} {\lnot} VotedForIn ( m1 . from ,\, b ,\, s ,\, v ) ) \.{'}}%
 \@x{\@s{31.1} {\BY} Phase1bVotedForInv ,\,\@pfstepnum{4}{2}
 ,\,\@pfstepnum{5}{3} ,\,\@pfstepnum{5}{1} ,\,\@pfstepnum{3}{2}
 ,\,\@pfstepnum{3}{1}\  {\DEF} AccInv ,\, MsgInv ,\, Send ,\, TypeOK ,\,
 Messages}%
 \@x{\@s{12.0}\@pfstepnum{5}{6.}\  \A\, s \.{\in} {\mathcal{S}} \.{:} MaxBalInSlot
 ( m1 . voted ,\, s ) \.{\in} {\mathcal{B}} \.{\cup} \{ \.{-} 1 \}\@s{2.0}
 {\BY}\@pfstepnum{3}{1} ,\, MaxBinSType {\DEF} TypeOK ,\, Messages}%
 \@x{\@s{12.0}\@pfstepnum{5}{7.}\  \A\, s \.{\in} {\mathcal{S}} \.{:} \.{\land} MaxBalInSlot
 ( m1 . voted ,\, s ) \.{+} 1 \.{\in} {\mathcal{B}}}%
 \@x{\@s{63.0} \.{\land} MaxBalInSlot
 ( m1 . voted ,\, s ) \.{+} 1 \.{>} MaxBalInSlot ( m1 . voted ,\, s )}%
\@x{\@s{16.0}\@pfstepnum{6}{}\  {\SUFFICES} {\ASSUME} {\NEW} s \.{\in} {\mathcal{S}}}%
 \@x{\@s{64.31} {\PROVE} \.{\land} MaxBalInSlot
 ( m1 . voted ,\, s ) \.{+} 1 \.{>} MaxBalInSlot ( m1 . voted ,\, s )}%
 \@x{\@s{95.0} \.{\land} MaxBalInSlot ( m1 . voted ,\, s ) \.{+} 1
 \.{\in} {\mathcal{B}}\@s{4.0} {\OBVIOUS}}%
 \@x{\@s{16.0}\@pfstepnum{6}{1.} {\CASE} MaxBalInSlot ( m1 . voted ,\, s )
 \.{=} \.{-} 1\@s{2.0} {\BY}\@pfstepnum{6}{1}\ }%
 \@x{\@s{16.0}\@pfstepnum{6}{2.} {\CASE} MaxBalInSlot ( m1 . voted ,\, s )
 \.{\in} {\mathcal{B}}\@s{2.0} {\BY} \A\, x \.{\in} {\mathcal{B}} \.{:} x \.{+} 1 \.{>} x
 ,\,\@pfstepnum{6}{2}\ }%
 \@x{\@s{16.0}\@pfstepnum{6}{}\  {\QED} {\BY}\@pfstepnum{6}{1}
 ,\,\@pfstepnum{6}{2} ,\,\@pfstepnum{5}{6}\ }%
 \@x{\@s{12.0}\@pfstepnum{5}{9.}\  m1 . bal \.{\in} {\mathcal{B}}\@s{2.0}
 {\BY}\@pfstepnum{3}{1}\  {\DEF} TypeOK ,\, Messages}%
 \@x{\@s{12.0}\@pfstepnum{5}{10.}\  \A\, b \.{\in} {\mathcal{B}} ,\, s \.{\in}
 {\mathcal{S}} \.{:} b \.{\in} MaxBalInSlot ( m1 . voted ,\, s ) \.{+} 1
 \.{\dotdot} m1 . bal \.{-} 1 \.{\implies} b \.{>} MaxBalInSlot ( m1 . voted ,\, s )}%
 \@x{\@s{16.0}\@pfstepnum{6}{}\  {\SUFFICES} {\ASSUME} {\NEW} b \.{\in}
 {\mathcal{B}} ,\, {\NEW} s \.{\in} {\mathcal{S}} ,\, b \.{\in} MaxBalInSlot ( m1 .
 voted ,\, s ) \.{+} 1 \.{\dotdot} m1 . bal \.{-} 1}%
\@x{\@s{64.31} {\PROVE} b \.{>} MaxBalInSlot ( m1 . voted ,\, s ) \@s{4.0} {\OBVIOUS}}%
\@x{\@s{16.0}\@pfstepnum{6}{}\  {\HIDE} {\DEF} m1}%
 \@x{\@s{16.0}\@pfstepnum{6}{}\  {\DEFINE} x \.{\defeq} MaxBalInSlot ( m1 .
 voted ,\, s )\@s{2.0} y \.{\defeq} m1 . bal \.{-} 1}%
 \@x{\@s{16.0}\@pfstepnum{6}{1.}\  x \.{\in} {\mathcal{B}} \.{\cup} \{ \.{-} 1
 \}\@s{2.0} {\BY}\@pfstepnum{5}{6}\ }%
 \@x{\@s{16.0}\@pfstepnum{6}{2.}\  y\@s{0.08} \.{\in} {\mathcal{B}} \.{\cup} \{
 \.{-} 1 \}\@s{2.0} {\BY}\@pfstepnum{5}{9}\ }%
\@x{\@s{16.0}\@pfstepnum{6}{}\  {\HIDE} {\DEF} x ,\, y}%
\@x{\@s{16.0}\@pfstepnum{6}{3.} {\CASE} x \.{+} 1 \.{>} y}%
 \@x{\@s{20.0}\@pfstepnum{7}{1.}\  \A\, e \.{\in} {\mathcal{B}} \.{:} e \.{\notin} x
 \.{+} 1 \.{\dotdot} y\@s{2.0} {\BY}\@pfstepnum{6}{3} ,\,\@pfstepnum{6}{1}
 ,\,\@pfstepnum{6}{2}\ }%
 \@x{\@s{20.0}\@pfstepnum{7}{}\  {\QED} {\BY}\@pfstepnum{6}{3}
 ,\,\@pfstepnum{7}{1}\  {\DEF} x ,\, y}%
 \@x{\@s{16.0}\@pfstepnum{6}{4.} {\CASE} x \.{+} 1 \.{=} y\@s{2.0}
 {\BY}\@pfstepnum{6}{4} ,\,\@pfstepnum{5}{7}
 ,\,\@pfstepnum{3}{1} ,\,\@pfstepnum{5}{9}\  {\DEF} x ,\, y}%
\@x{\@s{16.0}\@pfstepnum{6}{5.} {\CASE} x \.{+} 1 \.{<} y}%
 \@x{\@s{20.0}\@pfstepnum{7}{1.}\  \A\, e \.{\in} {\mathcal{B}} \.{:} e \.{\in} x
 \.{+} 1 \.{\dotdot} y \.{\implies} e \.{>} x\@s{2.0} {\BY}\@pfstepnum{6}{5}
 ,\,\@pfstepnum{6}{1} ,\,\@pfstepnum{6}{2}\ }%
 \@x{\@s{20.0}\@pfstepnum{7}{2.}\  b \.{\in} x \.{+} 1 \.{\dotdot} y\@s{2.0}
 {\BY} {\DEF} x ,\, y}%
 \@x{\@s{20.0}\@pfstepnum{7}{3.}\  b \.{>} x\@s{2.0} {\BY}\@pfstepnum{7}{2}
 ,\,\@pfstepnum{7}{1}\ }%
\@x{\@s{20.0}\@pfstepnum{7}{}\  {\QED} {\BY}\@pfstepnum{7}{3}\  {\DEF} x}%
 \@x{\@s{16.0}\@pfstepnum{6}{}\  {\QED} {\BY}\@pfstepnum{6}{3}
 ,\,\@pfstepnum{6}{4} ,\,\@pfstepnum{6}{5} ,\,\@pfstepnum{6}{1}
 ,\,\@pfstepnum{6}{2}\ }%
 \@x{\@s{12.0}\@pfstepnum{5}{11.}\  ( \A\, b \.{\in} {\mathcal{B}} ,\, s \.{\in}
 {\mathcal{S}} \.{:} b \.{\in} MaxBalInSlot ( m1 . voted ,\, s ) \.{+} 1
 \.{\dotdot} m1 . bal \.{-} 1 \.{\implies}}%
 \@x{\@s{40.22} b \.{>} MaxBalInSlot ( m1 . voted ,\, s ) ) \.{'}\@s{2.0}
 {\BY}\@pfstepnum{5}{10} ,\,\@pfstepnum{5}{1}\ }%
 \@x{\@s{12.0}\@pfstepnum{5}{12.}\  ( \A\, b \.{\in} {\mathcal{B}} ,\, s \.{\in}
 {\mathcal{S}} ,\, v \.{\in} {\mathcal{V}} \.{:} b \.{\in} MaxBalInSlot ( m1 . voted ,\,
 s ) \.{+} 1 \.{\dotdot} m1 . bal \.{-} 1 \.{\implies}}%
\@x{\@s{40.22} {\lnot} VotedForIn ( m1 . from ,\, b ,\, s ,\, v ) ) \.{'}}%
 \@x{\@s{16.0}\@pfstepnum{6}{}\  {\SUFFICES} {\ASSUME} {\NEW} b \.{\in}
 {\mathcal{B}} \.{'} ,\, {\NEW} s \.{\in} {\mathcal{S}} \.{'} ,\, {\NEW} v \.{\in} {\mathcal{V}}
 \.{'}}%
 \@x{\@s{64.31} {\PROVE} ( b \.{\in} MaxBalInSlot ( m1 . voted ,\,
 s ) \.{+} 1 \.{\dotdot} m1 . bal \.{-} 1 \.{\implies}}%
\@x{\@s{102.93} {\lnot} VotedForIn ( m1 . from ,\, b ,\, s ,\, v ) ) \.{'} \@s{4.0} {\OBVIOUS}}%
 \@x{\@s{16.0}\@pfstepnum{6}{1.} {\CASE} {\nexists}\, x \.{\in} {\mathcal{B}} \.{:}
 x \.{\in} ( MaxBalInSlot ( m1 . voted ,\, s ) \.{+} 1 \.{\dotdot} m1 .
 bal \.{-} 1 ) \.{'}\@s{2.0} {\BY}\@pfstepnum{6}{1}\ }%
 \@x{\@s{16.0}\@pfstepnum{6}{2.} {\CASE} \E\, x \.{\in} {\mathcal{B}} \.{:} x
 \.{\in} ( MaxBalInSlot ( m1 . voted ,\, s ) \.{+} 1 \.{\dotdot} m1 . bal
 \.{-} 1 ) \.{'}\@s{2.0} {\BY}\@pfstepnum{6}{2} ,\,\@pfstepnum{5}{5}
 ,\,\@pfstepnum{5}{11}\ }%
 \@x{\@s{16.0}\@pfstepnum{6}{}\  {\QED} {\BY}\@pfstepnum{6}{1}
 ,\,\@pfstepnum{6}{2}\ }%
\begin{lcom}{14.5}%
\begin{cpar}{0}{F}{F}{0}{0}{}%
\@pfstepnum{5}{13} is for \textsf{1a} message $m$ having a higher ballot than the highest seen, thus generating a \textsf{1b} message. %annie: what about 2b? sc - this is Phase1b, so no new 2b or 2a messages, right?
\end{cpar}\end{lcom}
\@x{\@s{12.0}\@pfstepnum{5}{13.} {\CASE} m . bal \.{>} aBal [ a ]}%
 \@x{\@s{16.0}\@pfstepnum{6}{}\  {\SUFFICES} {\ASSUME} {\NEW} m2 \.{\in} msgs
 \.{'}}%
 \@x{\@s{64.31} {\PROVE} ( \.{\land} ( m2 . type \.{=}\@w{1b} )
 \.{\implies} MsgInv1b ( m2 ) \.{\land} ( m2 . type\@s{1.1} \.{=}\@w{2a} )\@s{0.29}
 \.{\implies} MsgInv2a ( m2 )}%
 \@x{\@s{97.38} \.{\land} ( m2 . type\@s{1.1} \.{=}\@w{2b} ) \.{\implies}
 MsgInv2b ( m2 ) ) \.{'} \@s{4.0} {\BY} {\DEF} MsgInv}%
\begin{lcom}{18.5}%
\begin{cpar}{0}{F}{F}{0}{0}{}%
 Proves \ensuremath{MsgInv1b} using two cases. \@pfstepnum{7}{1} is for \ensuremath{m2 \in msgs}. \@pfstepnum{7}{2} is for the increment \ensuremath{m2 \in msgs' \setminus msgs}. 
\end{cpar}\end{lcom}
 \@x{\@s{16.0}\@pfstepnum{6}{1.}\  ( m2 . type \.{=}\@w{1b} \.{\implies}
 MsgInv1b ( m2 ) ) \.{'}}%
 \@x{\@s{20.0}\@pfstepnum{7}{1.} {\CASE} m2\@s{0.83} \.{\in} msgs\@s{2.0}
 {\BY}\@pfstepnum{7}{1} ,\,\@pfstepnum{5}{13} ,\,\@pfstepnum{5}{1} ,\,
 Phase1bVotedForInv ,\,\@pfstepnum{4}{2}\  {\DEF} MsgInv ,\, MsgInv1b ,\,}%
\@x{\@s{36.44} TypeOK ,\, Messages}%
 \@x{\@s{20.0}\@pfstepnum{7}{2.} {\CASE} m2 \.{\in} msgs \.{'}
 \.{\,\backslash\,} msgs}%
 \@x{\@s{24.0}\@pfstepnum{8}{1.}\  m2 \.{=} m1\@s{2.0} {\BY}\@pfstepnum{5}{1}
 ,\,\@pfstepnum{7}{2} ,\,\@pfstepnum{5}{13}\  {\DEF} Send}%
 \@x{\@s{24.0}\@pfstepnum{8}{}\  {\QED} {\BY}\@pfstepnum{7}{2}
 ,\,\@pfstepnum{5}{13} ,\, Phase1bVotedForInv ,\,\@pfstepnum{4}{2}
 ,\,\@pfstepnum{5}{2} ,\,\@pfstepnum{5}{4} ,\,\@pfstepnum{5}{12}
 ,\,\@pfstepnum{5}{1} ,\,\@pfstepnum{3}{1} ,\,\@pfstepnum{2}{1} ,\,}%
 \@x{\@s{36.89}\@pfstepnum{8}{1}\  {\DEF} Send ,\, TypeOK ,\, MsgInv ,\,
 Messages ,\, MsgInv1b}%
 \@x{\@s{20.0}\@pfstepnum{7}{}\  {\QED} {\BY}\@pfstepnum{7}{1}
 ,\,\@pfstepnum{7}{2}\ }%
\begin{lcom}{18.5}%
\begin{cpar}{0}{F}{F}{0}{0}{}%
Proves \ensuremath{MsgInv2a} and \ensuremath{MsgInv2b} using invariance lemma for \ensuremath{Phase1b} because it does not send \textsf{2a} or \textsf{2b} messages.
\end{cpar}\end{lcom}
 \@x{\@s{16.0}\@pfstepnum{6}{2.}\  ( ( m2 . type \.{=}\@w{2a} \.{\implies}
 MsgInv2a ( m2 ) ) \.{\land} ( m2 . type \.{=}\@w{2b} \.{\implies} MsgInv2b (
 m2 ) ) ) \.{'}}%
 \@x{\@s{35.1} {\BY}\@pfstepnum{5}{13} ,\, Phase1bVotedForInv
 ,\,\@pfstepnum{5}{1} ,\,\@pfstepnum{4}{2} ,\,\@pfstepnum{3}{1} ,\,
 SafeAtStable ,\,\@pfstepnum{2}{1}\  {\DEF} Send ,\, TypeOK ,\,}%
\@x{\@s{35.1} MsgInv ,\, Messages ,\, MsgInv2a ,\, MsgInv2b}%
 \@x{\@s{16.0}\@pfstepnum{6}{}\  {\QED} {\BY}\@pfstepnum{6}{1}
 ,\,\@pfstepnum{6}{2}\ }%
\begin{lcom}{14.5}%
\begin{cpar}{0}{F}{F}{0}{0}{}%
 \@pfstepnum{5}{14} is the simple case of preemption and uses the invariance lemmas for \ensuremath{Phase1b} and \ensuremath{SafeAt}.
\end{cpar}\end{lcom}
 \@x{\@s{12.0}\@pfstepnum{5}{14.} {\CASE} {\lnot} ( m . bal \.{>} aBal [ a ] )\,
 {\BY}\@pfstepnum{5}{14} ,\, Phase1bVotedForInv ,\,\@pfstepnum{4}{2}
 ,\,\@pfstepnum{5}{2} ,\,\@pfstepnum{5}{4} ,\,\@pfstepnum{5}{12}
 ,\,\@pfstepnum{5}{1} ,\,}%
 \@x{\@s{32.44} SafeAtStable ,\,\@pfstepnum{3}{1} ,\,\@pfstepnum{2}{1}\ 
 {\DEF} Send ,\, TypeOK ,\, MsgInv ,\, Messages ,\, MsgInv1b ,\, MsgInv2a ,\,
 MsgInv2b}%
 \@x{\@s{12.0}\@pfstepnum{5}{}\  {\QED} {\BY}\@pfstepnum{5}{13}
 ,\,\@pfstepnum{5}{14}\ }%
\begin{lcom}{10.5}%
\begin{cpar}{0}{F}{F}{0}{0}{}%
 \@pfstepnum{4}{3} proves \ensuremath{MsgInv'} for \ensuremath{Phase2a}. Each of \@pfstepnum{5}{4-6} proves a conjunct of \ensuremath{MsgInv}.
\end{cpar}\end{lcom}
 \@x{\@s{8.0}\@pfstepnum{4}{3.}\  {\ASSUME} {\NEW} p \.{\in} {\mathcal{P}} ,\,
 Phase2a ( p )\@s{2.0} {\PROVE} MsgInv \.{'}}%
 \@x{\@s{12.0}\@pfstepnum{5}{}\  {\SUFFICES} {\ASSUME} {\NEW} m \.{\in} msgs
 \.{'}}%
 \@x{\@s{60.31} {\PROVE} ( \.{\land} ( m . type \.{=}\@w{1b}
 \.{\implies} MsgInv1b ( m )) \.{\land} ( m . type\@s{1.1} \.{=}\@w{2a} \@s{0.29}
 \.{\implies} MsgInv2a ( m ))}%
 \@x{\@s{93.38} \.{\land} ( m . type\@s{1.1} \.{=}\@w{2b} \.{\implies}
 MsgInv2b ( m )) ) \.{'} \@s{4.0} {\BY} {\DEF} MsgInv}%
\@x{\@s{12.0}\@pfstepnum{5}{}\  {\DEFINE} b \.{\defeq} pBal [ p ]}%
 \@x{\@s{12.0}\@pfstepnum{5}{1.}\  {\PICK} Q \.{\in} {\mathcal{Q}} ,\, S \.{\in}
 {\SUBSET} \{ m2 \.{\in} msgs \.{:} ( m2 . type \.{=}\@w{1b} ) \.{\land} ( m2
 . bal \.{=} b ) \} \.{:}}%
 \@x{\@s{35.1} \.{\land} \A\, a \.{\in} Q \.{:} \E\, m2 \.{\in} S \.{:} m2 .
 from \.{=} a}%
 \@x{\@s{35.1} \.{\land} Send ( [ type \.{\mapsto}\@w{2a} ,\, bal \.{\mapsto}
 b ,\, from \.{\mapsto} p ,\, propSV \.{\mapsto} PropSV ( {\UNION}
 \{ m2 . voted \.{:} m2 \.{\in} S \} ) ] )}%
\@x{\@s{31.1} {\BY}\@pfstepnum{4}{3}\  {\DEF} Phase2a}%
 \@x{\@s{12.0}\@pfstepnum{5}{2.}\  b \.{=} pBal \.{'} [ p ] \.{\land} b
 \.{\in} {\mathcal{B}}\@s{2.0} {\BY}\@pfstepnum{4}{3}\  {\DEF} Phase2a ,\, TypeOK}%
 \@x{\@s{12.0}\@pfstepnum{5}{3.}\  \A\, m2 \.{\in} msgs \.{'}
 \.{\,\backslash\,} msgs \.{:} m2 . type \.{=}\@w{2a} \.{\land} m2 . bal
 \.{=} b\@s{2.0} {\BY}\@pfstepnum{5}{1}\  {\DEF} Send}%
\begin{lcom}{14.5}%
\begin{cpar}{0}{F}{F}{0}{0}{}%
 \@pfstepnum{5}{4} proves \ensuremath{MsgInv1b'}. % for \ensuremath{Phase2a}. 
 It uses invariance lemma for \ensuremath{Phase2a}, because \ensuremath{Phase2a} does not send 1b messages.
\end{cpar}\end{lcom}
 \@x{\@s{12.0}\@pfstepnum{5}{4.}\  ( m . type \.{=}\@w{1b} \.{\implies}
 MsgInv1b ( m ) ) \.{'}}%
 \@x{\@s{16.0}\@pfstepnum{6}{}\  {\SUFFICES} {\ASSUME} ( m . type \.{=}\@w{1b}
 ) \.{'}\@s{2.0} {\PROVE} MsgInv1b ( m ) \.{'} \@s{4.0} {\OBVIOUS}}%
 \@x{\@s{16.0}\@pfstepnum{6}{1.}\  ( m . bal \.{\leq} aBal [ m . from ] )
 \.{'}\@s{2.0} {\BY}\@pfstepnum{4}{3} ,\,\@pfstepnum{5}{3}
 ,\,\@pfstepnum{3}{1} ,\, Phase2aVotedForInv {\DEF} TypeOK ,\, Messages ,\,}%
\@x{\@s{35.1} MsgInv ,\, Phase2a ,\, MsgInv1b}%
 \@x{\@s{16.0}\@pfstepnum{6}{2.}\  ( \A\, r \.{\in} m . voted \.{:} VotedForIn
 ( m . from ,\, r . bal ,\, r . slot ,\, r . val ) ) \.{'}\@s{2.0}
 {\BY}\@pfstepnum{4}{3} ,\,\@pfstepnum{5}{3} ,\,\@pfstepnum{3}{1} ,\,}%
 \@x{\@s{35.1} Phase2aVotedForInv {\DEF} TypeOK ,\, Messages ,\, MsgInv ,\,
 MsgInv1b}%
 \@x{\@s{16.0}\@pfstepnum{6}{3.}\  \A\, s \.{\in} {\mathcal{S}} \.{:} MaxBalInSlot
 ( m . voted ,\, s ) \.{=} MaxBalInSlot ( m . voted ,\, s ) \.{'}\@s{2.0}
 {\BY} {\DEF} MaxBalInSlot}%
 \@x{\@s{16.0}\@pfstepnum{6}{4.}\  ( \A\, b2 \.{\in} {\mathcal{B}} ,\, s \.{\in}
 {\mathcal{S}} ,\, v \.{\in} {\mathcal{V}} \.{:} b2 \.{\in} MaxBalInSlot ( m . voted ,\,
 s ) \.{+} 1 \.{\dotdot} m . bal \.{-} 1 \.{\implies}}%
\@x{\@s{40.22} {\lnot} VotedForIn ( m . from ,\, b2 ,\, s ,\, v ) ) \.{'}}%
 \@x{\@s{35.1} {\BY}\@pfstepnum{4}{3} ,\,\@pfstepnum{5}{3}
 ,\,\@pfstepnum{3}{1} ,\, Phase2aVotedForInv ,\,\@pfstepnum{6}{3}\  {\DEF}
 TypeOK ,\, Messages ,\, MsgInv ,\, MsgInv1b}%
 \@x{\@s{16.0}\@pfstepnum{6}{}\  {\QED} {\BY}\@pfstepnum{6}{1}
 ,\,\@pfstepnum{6}{2} ,\,\@pfstepnum{6}{4}\  {\DEF} MsgInv1b}%
\begin{lcom}{14.5}%
\begin{cpar}{0}{F}{F}{0}{0}{}%
 \@pfstepnum{5}{5} proves \ensuremath{MsgInv2a'}. Each of \@pfstepnum{6}{2-4} proves a conjunct of \ensuremath{MsgInv2a} using the increment approach. The increment is \ensuremath{m2} in \@pfstepnum{8}{2} of \@pfstepnum{7}{9}.
%annie: not clear which is an increment case. sc-all of them are. Should I mention the exact step number in the comments for each case?
\end{cpar}\end{lcom}
 \@x{\@s{12.0}\@pfstepnum{5}{5.}\  ( m . type \.{=}\@w{2a} \.{\implies}
 MsgInv2a ( m ) ) \.{'}}%
 \@x{\@s{16.0}\@pfstepnum{6}{}\  {\SUFFICES} {\ASSUME} ( m . type \.{=}\@w{2a}
 ) \.{'}\@s{2.0} {\PROVE} MsgInv2a ( m ) \.{'} \@s{4.0} {\OBVIOUS}}%
 \@x{\@s{16.0}\@pfstepnum{6}{}\  {\DEFINE} VS \.{\defeq} {\UNION} \{ m2 .
 voted \.{:} m2 \.{\in} S \}}%
 \@x{\@s{16.0}\@pfstepnum{6}{1.}\  \A\, a \.{\in} Q \.{:} aBal [ a ] \.{\geq}
 b\@s{2.0} {\BY}\@pfstepnum{5}{1} ,\,\@pfstepnum{3}{2} ,\,\@pfstepnum{3}{1}\ 
 {\DEF} MsgInv ,\, TypeOK ,\, Messages ,\, MsgInv1b}%
 \@x{\@s{16.0}\@pfstepnum{6}{2.}\  ( \A\, d \.{\in} m . propSV \.{:} SafeAt (
 m . bal ,\, d . slot ,\, d . val ) ) \.{'}}%
 \@x{\@s{20.0}\@pfstepnum{7}{1.}\  \A\, d \.{\in} [ slot \.{:} UnusedS ( VS
 ) ,\, val \.{:} {\mathcal{V}} ]  \.{:} SafeAt ( b ,\, d .
 slot ,\, d . val )}%
 \@x{\@s{24.0}\@pfstepnum{8}{}\  {\SUFFICES} {\ASSUME} {\NEW} d \.{\in} [ slot
 \.{:} UnusedS ( VS ) ,\, val \.{:} {\mathcal{V}} ] 
 \@s{2.0} {\PROVE} SafeAt ( b ,\, d . slot ,\, d . val )}%
\@x{\@s{36.89} {\OBVIOUS}}%
 \@x{\@s{24.0}\@pfstepnum{8}{1.}\  \A\, m2 \.{\in} S \.{:} {\nexists}\, d2
 \.{\in} m2 . voted \.{:} d . slot \.{=} d2 . slot\@s{2.0} {\BY} {\DEF}
 UnusedS}%
 \@x{\@s{24.0}\@pfstepnum{8}{2.}\  \A\, m2 \.{\in} S \.{:} MaxBalInSlot (
 m2 . voted ,\, d . slot ) \.{+} 1 \.{=} 0\@s{2.0} {\BY}\@pfstepnum{8}{1}\ 
 {\DEF} MaxBalInSlot}%
 \@x{\@s{24.0}\@pfstepnum{8}{3.}\  \A\, m2 \.{\in} S ,\, b2 \.{\in} {\mathcal{B}}
 ,\, s \.{\in} {\mathcal{S}} ,\, v \.{\in} {\mathcal{V}} \.{:} b2 \.{\in} MaxBalInSlot (
 m2 . voted ,\, s ) \.{+} 1 \.{\dotdot} m2 . bal \.{-} 1 \.{\implies}}%
 \@x{\@s{47.1} {\lnot} VotedForIn ( m2 . from ,\, b2 ,\, s ,\, v )\@s{2.0}
 {\BY} {\DEF} MsgInv ,\, MsgInv1b}%
 \@x{\@s{24.0}\@pfstepnum{8}{4.}\  \A\, v \.{\in} {\mathcal{V}} ,\, b2 \.{\in}
 {\mathcal{B}} ,\, a \.{\in} Q \.{:} b2 \.{\in} 0 \.{\dotdot} b \.{-} 1
 \.{\implies} {\lnot} VotedForIn ( a ,\, b2 ,\, d . slot ,\, v )}%
 \@x{\@s{43.1} {\BY}\@pfstepnum{5}{1} ,\,\@pfstepnum{8}{2}
 ,\,\@pfstepnum{8}{3}\  {\DEF} UnusedS ,\, TypeOK ,\, Messages}%
 \@x{\@s{24.0}\@pfstepnum{8}{}\  {\QED} {\BY}\@pfstepnum{8}{4}
 ,\,\@pfstepnum{3}{1} ,\,\@pfstepnum{6}{1}\  {\DEF} SafeAt ,\, NewSV
 ,\, UnusedS ,\, WontVoteIn ,\, TypeOK ,\, Messages}%
 \@x{\@s{20.0}\@pfstepnum{7}{2.}\  \A\, d \.{\in} [ slot \.{:} UnusedS ( VS
 ) ,\, val \.{:} {\mathcal{V}} ]  \.{:} SafeAt ( b ,\, d .
 slot ,\, d . val ) \.{'}\@s{2.0} {\BY}\@pfstepnum{7}{1} ,\, SafeAtStable
 ,\,}%
 \@x{\@s{39.1}\@pfstepnum{3}{1} ,\,\@pfstepnum{2}{1} ,\,\@pfstepnum{5}{2}\ 
 {\DEF} NewSV ,\, UnusedS ,\, TypeOK ,\, Messages}%
 \@x{\@s{20.0}\@pfstepnum{7}{3.}\  \A\, d \.{\in} NewSV ( VS ) \.{:}
 SafeAt ( b ,\, d . slot ,\, d . val ) \.{'}\@s{2.0} {\BY}\@pfstepnum{7}{2}
 ,\, Misc {\DEF} NewSV}%
 \@x{\@s{20.0}\@pfstepnum{7}{4.}\  \A\, d \.{\in} MaxBSV ( VS ) \.{:}
 SafeAt ( b ,\, d . slot ,\, d . val )}%
 \@x{\@s{24.0}\@pfstepnum{8}{}\  {\SUFFICES} {\ASSUME} {\NEW} d \.{\in}
 MaxBSV ( VS ) ,\, {\NEW} b2 \.{\in} {\mathcal{B}} ,\, b2 \.{\in} 0
 \.{\dotdot} ( b \.{-} 1 )}%
 \@x{\@s{72.31} {\PROVE} \E\, Q2 \.{\in} {\mathcal{Q}} \.{:} \A\, a \.{\in}
 Q2 \.{:} \.{\lor} VotedForIn ( a ,\, b2 ,\, d . slot ,\, d . val )}%
 \@x{\@s{180.63}  \.{\lor} WontVoteIn ( a ,\, b2 ,\, d . slot ) \@s{4.0} {\BY} {\DEF} SafeAt}%
 \@x{\@s{24.0}\@pfstepnum{8}{}\  {\DEFINE} max \.{\defeq} MaxBalInSlot ( VS
 ,\, d . slot )}%
\@x{\@s{24.0}\@pfstepnum{8}{}\  {\USE} {\DEF} MaxBSV}%
 \@x{\@s{24.0}\@pfstepnum{8}{1.}\  max \.{\in} {\mathcal{B}}\@s{2.0} {\BY} MaxBinSType
 ,\, MaxBinSNoSlot {\DEF} TypeOK ,\, Messages}%
 \@x{\@s{24.0}\@pfstepnum{8}{2.}\  \A\, m2 \.{\in} S \.{:} MaxBalInSlot (
 m2 . voted ,\, d . slot ) \.{\leq} max}%
 \@x{\@s{43.1} {\BY} \A\, m2 \.{\in} S \.{:} m2 . voted \.{\subseteq} VS ,\,
 MaxBinSSubsets {\DEF} MaxBSV ,\, TypeOK ,\, Messages}%
 \@x{\@s{24.0}\@pfstepnum{8}{3.}\  {\nexists}\, d2 \.{\in} VS \.{:} ( d2 . bal
 \.{>} d . bal \.{\land} d2 . slot \.{=} d . slot )}%
 \@x{\@s{43.1} {\BY} \A\, d2 \.{\in} VS \.{:} {\lnot} ( {\lnot} ( d2 . bal
 \.{\leq} d . bal ) \.{\land} d2 . slot \.{=} d . slot ) {\DEF} MaxBSV
 ,\, TypeOK ,\, Messages}%
 \@x{\@s{24.0}\@pfstepnum{8}{4.}\  VS \.{\subseteq} [ bal \.{:} {\mathcal{B}}
 ,\, slot \.{:} {\mathcal{S}} ,\, val \.{:} {\mathcal{V}} ]\@s{2.0} {\BY}\@pfstepnum{3}{1}\ 
 {\DEF} TypeOK ,\, Messages}%
\@x{\@s{24.0}\@pfstepnum{8}{5.}\  max \.{=} d . bal}%
 \@x{\@s{28.0}\@pfstepnum{9}{}\  {\SUFFICES} {\ASSUME} max \.{\neq} d .
 bal\@s{2.0} {\PROVE} {\FALSE} \@s{4.0} {\OBVIOUS}}%
\@x{\@s{28.0}\@pfstepnum{9}{1.} {\CASE} max \.{>} d . bal}%
\@x{\@s{32.0}\@pfstepnum{10}{}\  {\HIDE} {\DEF} VS}%
 \@x{\@s{32.0}\@pfstepnum{10}{1.}\  \E\, d2 \.{\in} VS \.{:} d2 . bal \.{=}
 max \.{\land} d2 . slot \.{=} d . slot\@s{2.0} {\BY}\@pfstepnum{8}{4}
 ,\,\@pfstepnum{8}{1} ,\, MaxBinSExists}%
 \@x{\@s{32.0}\@pfstepnum{10}{}\  {\QED} {\BY}\@pfstepnum{10}{1}
 ,\,\@pfstepnum{8}{3} ,\,\@pfstepnum{9}{1}\ }%
 \@x{\@s{28.0}\@pfstepnum{9}{2.} {\CASE} max \.{<} d . bal\@s{2.0} {\BY}
 MaxBinSNoMore ,\,\@pfstepnum{9}{2}\  {\DEF} MaxBSV ,\, TypeOK ,\,
 Messages}%
 \@x{\@s{28.0}\@pfstepnum{9}{}\  {\QED} {\BY}\@pfstepnum{9}{1}
 ,\,\@pfstepnum{9}{2} ,\,\@pfstepnum{8}{1}\  {\DEF} {\mathcal{B}} ,\, TypeOK ,\,
 Messages}%
 \@x{\@s{24.0}\@pfstepnum{8}{6.} {\CASE} b2 \.{\in} max \.{+} 1 \.{\dotdot} b
 \.{-} 1}%
\@x{\@s{28.0}\@pfstepnum{9}{}\  {\HIDE} {\DEF} max}%
 \@x{\@s{28.0}\@pfstepnum{9}{1.}\  \A\, m2 \.{\in} S ,\, b3 \.{\in} {\mathcal{B}}
 ,\, v \.{\in} {\mathcal{V}} \.{:} b3 \.{\in} MaxBalInSlot ( m2 . voted ,\, d .
 slot ) \.{+} 1 \.{\dotdot} b \.{-} 1 \.{\implies}}%
\@x{\@s{51.1} {\lnot} VotedForIn ( m2 . from ,\, b3 ,\, d . slot ,\, v )   \@s{4.0} {\BY} {\DEF} MsgInv ,\, TypeOK ,\, Messages ,\, MsgInv1b}%
 \@x{\@s{28.0}\@pfstepnum{9}{2.}\  \A\, m2 \.{\in} S ,\, v \.{\in} {\mathcal{V}}
 \.{:} {\lnot} VotedForIn ( m2 . from ,\, b2 ,\, d . slot ,\, v )}%
 \@x{\@s{47.1} {\BY}\@pfstepnum{8}{6} ,\,\@pfstepnum{9}{1}
 ,\,\@pfstepnum{8}{2} ,\,\@pfstepnum{8}{1} ,\, MaxBinSType {\DEF} TypeOK ,\,
 Messages ,\, Send}%
 \@x{\@s{28.0}\@pfstepnum{9}{}\  {\QED} {\BY}\@pfstepnum{5}{1}
 ,\,\@pfstepnum{8}{6} ,\,\@pfstepnum{6}{1} ,\,\@pfstepnum{9}{2}\  {\DEF}
 MsgInv ,\, MsgInv1b ,\, TypeOK ,\, Messages ,\, WontVoteIn ,\,}%
\@x{\@s{40.89} MaxBSV}%
\@x{\@s{24.0}\@pfstepnum{8}{7.} {\CASE} b2 \.{=} max}%
 \@x{\@s{32.0}\@pfstepnum{9}{1.}\  \E\, a \.{\in} {\mathcal{A}} ,\, m2 \.{\in} S
 \.{:} m2 . from \.{=} a \.{\land} \E\, d2 \.{\in} m2 . voted \.{:}}%
 \@x{\@s{55.1} d2 . bal \.{=} d . bal \.{\land} d2 . slot \.{=} d . slot
 \.{\land} d2 . val \.{=} d . val}%
\@x{\@s{51.1} {\BY} {\DEF} MaxBalInSlot ,\, TypeOK ,\, Messages}%
 \@x{\@s{32.0}\@pfstepnum{9}{2.}\  \E\, a \.{\in} {\mathcal{A}} \.{:} VotedForIn (
 a ,\, b2 ,\, d . slot ,\, d . val )}%
 \@x{\@s{51.1} {\BY}\@pfstepnum{8}{7} ,\,\@pfstepnum{9}{1}
 ,\,\@pfstepnum{8}{5}\  {\DEF} MsgInv ,\, TypeOK ,\, Messages ,\, MsgInv1b}%
 \@x{\@s{32.0}\@pfstepnum{9}{3.}\  \A\, q \.{\in} Q ,\, v2 \.{\in} {\mathcal{V}}
 \.{:} VotedForIn ( q ,\, b2 ,\, d . slot ,\, v2 ) \.{\implies} v2 \.{=} d .
 val}%
 \@x{\@s{51.1} {\BY}\@pfstepnum{9}{2} ,\, VotedOnce ,\, QuorumAssumption
 {\DEF} TypeOK ,\, Messages}%
 \@x{\@s{32.0}\@pfstepnum{9}{4.}\  \A\, q \.{\in} Q \.{:} aBal [ q ] \.{>}
 b2\@s{2.0} {\BY}\@pfstepnum{5}{1}\  {\DEF} MsgInv ,\, TypeOK ,\, Messages
 ,\, MsgInv1b}%
 \@x{\@s{32.0}\@pfstepnum{9}{}\  {\QED} {\BY}\@pfstepnum{8}{7}
 ,\,\@pfstepnum{9}{3} ,\,\@pfstepnum{9}{4}\  {\DEF} WontVoteIn}%
\@x{\@s{24.0}\@pfstepnum{8}{8.} {\CASE} b2 \.{\in} 0 \.{\dotdot} max \.{-} 1}%
 \@x{\@s{32.0}\@pfstepnum{9}{1.}\  \E\, a \.{\in} {\mathcal{A}} \.{:} VotedForIn (
 a ,\, d . bal ,\, d . slot ,\, d . val )}%
 \@x{\@s{51.1} {\BY}\@pfstepnum{8}{8} ,\,\@pfstepnum{8}{2}\  {\DEF} MsgInv
 ,\, TypeOK ,\, Messages ,\, MsgInv1b}%
 \@x{\@s{32.0}\@pfstepnum{9}{2.}\  SafeAt ( d . bal ,\, d . slot ,\, d . val
 )\@s{2.0} {\BY}\@pfstepnum{9}{1} ,\, VotedInv {\DEF} TypeOK ,\, Messages}%
 \@x{\@s{32.0}\@pfstepnum{9}{}\  {\QED} {\BY}\@pfstepnum{8}{8}
 ,\,\@pfstepnum{9}{2} ,\,\@pfstepnum{8}{5}\  {\DEF} SafeAt ,\, MsgInv ,\,
 TypeOK ,\, Messages ,\, MaxBalInSlot}%
 \@x{\@s{24.0}\@pfstepnum{8}{}\  {\QED} {\BY}\@pfstepnum{8}{6}
 ,\,\@pfstepnum{8}{7} ,\,\@pfstepnum{8}{8} ,\,\@pfstepnum{8}{1}\ }%
 \@x{\@s{20.0}\@pfstepnum{7}{5.}\  MaxBSV ( VS ) \.{\subseteq} [ bal
 \.{:} {\mathcal{B}} ,\, slot \.{:} {\mathcal{S}} ,\, val \.{:} {\mathcal{V}} ]\@s{2.0} {\BY}
 {\DEF} MaxBSV ,\, TypeOK ,\, Messages}%
 \@x{\@s{20.0}\@pfstepnum{7}{6.}\  \A\, d \.{\in} MaxBSV ( VS ) \.{:}
 SafeAt ( b ,\, d . slot ,\, d . val ) \.{'}\@s{2.0} {\BY}\@pfstepnum{7}{4}
 ,\, SafeAtStable ,\,\@pfstepnum{3}{1} ,\,\@pfstepnum{7}{5}
 ,\,\@pfstepnum{2}{1} ,\,\@pfstepnum{5}{2}\ }%
 \@x{\@s{20.0}\@pfstepnum{7}{7.}\  \A\, d \.{\in} MaxSV ( VS ) \.{:} SafeAt ( b
 ,\, d . slot ,\, d . val ) \.{'}\@s{2.0} {\BY}\@pfstepnum{7}{6}
 ,\,\@pfstepnum{7}{5}\  {\DEF} MaxSV}%
 \@x{\@s{20.0}\@pfstepnum{7}{8.}\  \A\, d \.{\in} MaxSV ( VS ) \.{\cup}
 NewSV ( VS ) \.{:} SafeAt ( b ,\, d . slot ,\, d . val )
 \.{'}\@s{2.0} {\BY}\@pfstepnum{7}{7} ,\,\@pfstepnum{7}{3}\ }%
 \@x{\@s{20.0}\@pfstepnum{7}{9.}\  ( \A\, m2 \.{\in} msgs \.{:} m2 . type
 \.{=}\@w{2a} \.{\implies} \A\, d \.{\in} m2 . propSV \.{:} SafeAt ( m2 .
 bal ,\, d . slot ,\, d . val ) ) \.{'}}%
 \@x{\@s{24.0}\@pfstepnum{8}{}\  {\SUFFICES} {\ASSUME} {\NEW} m2 \.{\in} msgs
 \.{'} ,\, ( m2 . type \.{=}\@w{2a} ) \.{'} ,\, {\NEW} d \.{\in} m2 .
 propSV}%
 \@x{\@s{72.31} {\PROVE} ( SafeAt ( m2 . bal ,\, d . slot ,\, d . val
 ) ) \.{'} \@s{4.0} {\OBVIOUS}}%
 \@x{\@s{24.0}\@pfstepnum{8}{1.} {\CASE} m2 \.{\in} msgs\@s{2.0}
 {\BY}\@pfstepnum{3}{1} ,\, SafeAtStable ,\,\@pfstepnum{8}{1}
 ,\,\@pfstepnum{2}{1}\  {\DEF} MsgInv ,\, MsgInv2a ,\, Messages ,\, TypeOK}%
 \@x{\@s{24.0}\@pfstepnum{8}{2.} {\CASE} m2 \.{\in} msgs \.{'}
 \.{\,\backslash\,} msgs}%
 \@x{\@s{28.0}\@pfstepnum{9}{1.}\  SafeAt ( m2 . bal ,\, d . slot ,\, d . val
 ) \.{'}\@s{2.0} {\BY}\@pfstepnum{7}{8} ,\,\@pfstepnum{8}{2}
 ,\,\@pfstepnum{5}{1} ,\,\@pfstepnum{5}{2}\  {\DEF} Send ,\, PropSV}%
 \@x{\@s{28.0}\@pfstepnum{9}{}\  {\QED} {\BY}\@pfstepnum{3}{1}
 ,\,\@pfstepnum{9}{1}\  {\DEF} Send ,\, TypeOK ,\, Messages}%
 \@x{\@s{24.0}\@pfstepnum{8}{}\  {\QED} {\BY}\@pfstepnum{8}{1}
 ,\,\@pfstepnum{8}{2}\ }%
\@x{\@s{20.0}\@pfstepnum{7}{}\  {\QED} {\BY}\@pfstepnum{7}{9}\ }%
\begin{lcom}{18.5}%
\begin{cpar}{0}{F}{F}{0}{0}{}%
The increment is \ensuremath{m2} in \@pfstepnum{7}{1}.
\end{cpar}\end{lcom}
 \@x{\@s{16.0}\@pfstepnum{6}{3.}\  ( \A\, d1 ,\, d2 \.{\in} m . propSV \.{:}
 d1 . slot \.{=} d2 . slot \.{\implies} d1 \.{=} d2 ) \.{'}}%
 \@x{\@s{20.0}\@pfstepnum{7}{1.}\  \A\, m2 \.{\in} msgs \.{'}
 \.{\,\backslash\,} msgs \.{:} \A\, d1 ,\, d2 \.{\in} m2 . propSV\@s{0.84}
 \.{:} d1 . slot \.{=} d2 . slot \.{\implies} d1 \.{=} d2}%
 \@x{\@s{24.0}\@pfstepnum{8}{1.}\  VS \.{\in} {\SUBSET} [ bal \.{:} {\mathcal{B}}
 ,\, slot \.{:} {\mathcal{S}} ,\, val \.{:} {\mathcal{V}} ]\@s{2.0} {\BY} {\DEF} Messages
 ,\, TypeOK}%
 \@x{\@s{24.0}\@pfstepnum{8}{2.}\  \A\, r1 ,\, r2 \.{\in} MaxBSV ( VS )
 \.{:} r1 . slot \.{=} r2 . slot \.{\implies} r1 . bal \.{=} r2 . bal\@s{2.0}
 {\BY}\@pfstepnum{8}{1}\  {\DEF} MaxBSV}%
 \@x{\@s{24.0}\@pfstepnum{8}{3.}\  MaxBSV ( VS ) \.{\subseteq} VS\@s{2.0}
 {\BY}\@pfstepnum{8}{1}\  {\DEF} MaxBSV}%
 \@x{\@s{24.0}\@pfstepnum{8}{4.}\  \A\, r1 ,\, r2 \.{\in} MaxBSV ( VS )
 \.{:} r1 . bal \.{=} r2 . bal \.{\land} r1 . slot \.{=} r2 . slot
 \.{\implies} r1 . val \.{=} r2 . val}%
\@x{\@s{43.1} {\BY}\@pfstepnum{8}{3} ,\, VotedUnion}%
 \@x{\@s{24.0}\@pfstepnum{8}{5.}\  \A\, r1 ,\, r2 \.{\in} MaxBSV ( VS )
 \.{:} r1 . slot \.{=} r2 . slot \.{\implies} r1 . bal \.{=} r2 . bal
 \.{\land} r1 . val \.{=} r2 . val}%
 \@x{\@s{43.1} {\BY}\@pfstepnum{8}{4} ,\,\@pfstepnum{8}{2}
 ,\,\@pfstepnum{8}{3} ,\,\@pfstepnum{8}{1}\ }%
 \@x{\@s{24.0}\@pfstepnum{8}{6.}\  \A\, r1 ,\, r2 \.{\in} MaxSV ( VS ) \.{:} r1
 . slot \.{=} r2 . slot \.{\implies} r1 \.{=} r2\@s{2.0}
 {\BY}\@pfstepnum{8}{5}\  {\DEF} MaxSV}%
 \@x{\@s{24.0}\@pfstepnum{8}{}\  {\QED} {\BY}\@pfstepnum{8}{6} ,\, Misc
 ,\,\@pfstepnum{5}{1}\  {\DEF} PropSV ,\, Send}%
 \@x{\@s{20.0}\@pfstepnum{7}{}\  {\QED} {\BY}\@pfstepnum{7}{1}\  {\DEF} MsgInv
 ,\, MsgInv2a}%
\begin{lcom}{18.5}%
\begin{cpar}{0}{F}{F}{0}{0}{}%
The increment is \ensuremath{m1,m2} in \@pfstepnum{7}{2}.
\end{cpar}\end{lcom}
 \@x{\@s{16.0}\@pfstepnum{6}{4.}\  ( \A\, m2 \.{\in} msgs \.{:} ( m2 . type
 \.{=}\@w{2a} \.{\land} m2 . bal \.{=} m . bal ) \.{\implies} ( m2 \.{=} m )
 ) \.{'}}%
 \@x{\@s{20.0}\@pfstepnum{7}{1.}\  \A\, m2 \.{\in} msgs \.{:} ( m2 . type
 \.{=}\@w{2a} ) \.{\implies} ( m2 . bal \.{\neq} b )\@s{2.0}
 {\BY}\@pfstepnum{4}{3}\  {\DEF} Phase2a}%
 \@x{\@s{20.0}\@pfstepnum{7}{2.}\  \A\, m1 ,\, m2 \.{\in} msgs \.{'}
 \.{\,\backslash\,} msgs \.{:} m1 \.{=} m2\@s{2.0} {\BY}\@pfstepnum{4}{3}\ 
 {\DEF} Phase2a ,\, Send}%
 \@x{\@s{20.0}\@pfstepnum{7}{}\  {\QED} {\BY}\@pfstepnum{7}{1}
 ,\,\@pfstepnum{7}{2} ,\,\@pfstepnum{5}{3} ,\,\@pfstepnum{2}{1}
 ,\,\@pfstepnum{3}{1}\  {\DEF} MsgInv ,\, MsgInv2a}%
 \@x{\@s{16.0}\@pfstepnum{6}{}\  {\QED} {\BY}\@pfstepnum{6}{2}
 ,\,\@pfstepnum{6}{3} ,\,\@pfstepnum{6}{4}\  {\DEF} MsgInv2a}%
\iffalse
\begin{lcom}{14.5}%
\begin{cpar}{0}{F}{F}{0}{0}{}%
 \@pfstepnum{5}{6} proves \ensuremath{MsgInv2b'}. Since \ensuremath{Phase2a} sends \ensuremath{2a} messages, the proof is simple.  %annie: the reason does say anything really. 
\end{cpar}\end{lcom}
\fi
 \@x{\@s{12.0}\@pfstepnum{5}{6.}\  ( ( m . type \.{=}\@w{2b} ) \.{\implies}
 MsgInv2b ( m ) ) \.{'}\@s{2.0} {\BY}\@pfstepnum{5}{3} ,\,\@pfstepnum{5}{1}
 ,\, m . type \.{=}\@w{2b} \.{\implies} m \.{\in} msgs ,\,}%
 \@x{\@s{31.1}\@pfstepnum{3}{1} ,\,\@pfstepnum{4}{3}\  {\DEF} TypeOK ,\,
 Messages ,\, MsgInv ,\, Phase2a ,\, Send ,\, MsgInv2b}%
 \@x{\@s{12.0}\@pfstepnum{5}{}\  {\QED} {\BY}\@pfstepnum{5}{4}
 ,\,\@pfstepnum{5}{5} ,\,\@pfstepnum{5}{6}\ }%
\begin{lcom}{10.5}%
\begin{cpar}{0}{F}{F}{0}{0}{}%
 \@pfstepnum{4}{4} proves \ensuremath{MsgInv'} for \ensuremath{Phase2b}. Each of \@pfstepnum{5}{2-4} proves a conjunct of \ensuremath{MsgInv}.
\end{cpar}\end{lcom}
 \@x{\@s{8.0}\@pfstepnum{4}{4.}\  {\ASSUME} {\NEW} a \.{\in} {\mathcal{A}} ,\,
 Phase2b ( a )\@s{2.0} {\PROVE} MsgInv \.{'}}%
 \@x{\@s{12.0}\@pfstepnum{5}{}\  {\SUFFICES} {\ASSUME} {\NEW} m \.{\in} msgs
 \.{'}}%
 \@x{\@s{60.31} {\PROVE} ( \.{\land} ( m . type \.{=}\@w{1b} )
 \.{\implies} MsgInv1b ( m ) \.{\land} ( m . type\@s{1.1} \.{=}\@w{2a} )\@s{0.29}
 \.{\implies} MsgInv2a ( m )}%
 \@x{\@s{93.38} \.{\land} ( m . type\@s{1.1} \.{=}\@w{2b} ) \.{\implies}
 MsgInv2b ( m ) ) \.{'} \@s{4.0} {\BY} {\DEF} MsgInv}%
 \@x{\@s{12.0}\@pfstepnum{5}{1.}\  {\PICK} m1 \.{\in} msgs \.{:} Phase2b ( a )
 {\bang} ( m1 )\@s{2.0} {\BY}\@pfstepnum{4}{4}\  {\DEF} Phase2b}%
\begin{lcom}{14.5}%
\begin{cpar}{0}{F}{F}{0}{0}{}%
 \@pfstepnum{5}{2} proves \ensuremath{MsgInv1b'} for\ensuremath{Phase2b}. Invariance lemmas do not apply because the 3rd conjunct in \ensuremath{MsgInv1b} quantifies over \ensuremath{2b} messages negatively 
 %annie: does it matter to be negative
 %sc: if it was not negative, the invariance lemma would have worked. 
 %annie: i remember this one now.  thanks!
 ---\ensuremath{VotedForIn(a, b, s, v)} means acceptor \ensuremath{a} has sent a \textsf{2b} message voting \ensuremath{\langle b, s, v\rangle}.
 %\ensuremath{Phase2bVotedForInv} fails to work here for \@pfstepnum{6}{2}.
\end{cpar}\end{lcom}
 \@x{\@s{12.0}\@pfstepnum{5}{2.}\  ( ( m . type \.{=}\@w{1b} ) \.{\implies}
 MsgInv1b ( m ) ) \.{'}}%
 \@x{\@s{16.0}\@pfstepnum{6}{}\  {\SUFFICES} {\ASSUME} ( m . type \.{=}\@w{1b}
 ) \.{'}\@s{2.0} {\PROVE} MsgInv1b ( m ) \.{'} \@s{4.0} {\OBVIOUS}}%
 \@x{\@s{16.0}\@pfstepnum{6}{1.}\  ( m . bal \.{\leq} aBal [ m . from ]
 \.{\land} \A\, r \.{\in} m . voted \.{:} VotedForIn ( m . from ,\, r . bal
 ,\, r . slot ,\, r . val ) ) \.{'}}%
 \@x{\@s{35.1} {\BY}\@pfstepnum{5}{1} ,\,\@pfstepnum{3}{1}
 ,\,\@pfstepnum{4}{4} ,\, Phase2bVotedForInv {\DEF} MsgInv ,\, MsgInv1b ,\,
 TypeOK ,\, Messages ,\, Send}%
%\@x{\@s{35.1} Messages ,\, Send}%
 \@x{\@s{16.0}\@pfstepnum{6}{2.}\  ( \A\, b \.{\in} {\mathcal{B}} ,\, s \.{\in}
 {\mathcal{S}} ,\, v \.{\in} {\mathcal{V}} \.{:} b \.{\in} MaxBalInSlot ( m . voted ,\, s
 ) \.{+} 1 \.{\dotdot} m . bal \.{-} 1 \.{\implies}}%
\@x{\@s{40.22} {\lnot} VotedForIn ( m . from ,\, b ,\, s ,\, v ) ) \.{'}}%
 \@x{\@s{20.0}\@pfstepnum{7}{}\  {\SUFFICES} {\ASSUME} {\NEW} b \.{\in}
 {\mathcal{B}} \.{'} ,\, {\NEW} s \.{\in} {\mathcal{S}} \.{'} ,\, {\NEW} v \.{\in} {\mathcal{V}}
 \.{'} ,\,}%
 \@x{\@s{98.91} ( b \.{\in} MaxBalInSlot ( m . voted ,\, s ) \.{+} 1
 \.{\dotdot} m . bal \.{-} 1 ) \.{'}}%
 \@x{\@s{68.31} {\PROVE} ( {\lnot} VotedForIn ( m . from ,\, b ,\, s
 ,\, v ) ) \.{'} \@s{4.0} {\OBVIOUS}}%
 \@x{\@s{20.0}\@pfstepnum{7}{1.}\  {\lnot} VotedForIn ( m . from ,\, b ,\, s
 ,\, v )\@s{2.0} {\BY}\@pfstepnum{5}{1}\  {\DEF} Send ,\, MsgInv ,\, TypeOK
 ,\, Messages ,\, MsgInv1b}%
 \@x{\@s{20.0}\@pfstepnum{7}{2.} {\CASE} m . from \.{\neq} a \.{\lor} {\lnot}
 ( m1 . bal \.{\geq} aBal [ a ] )\@s{2.0} {\BY}\@pfstepnum{5}{1}
 ,\,\@pfstepnum{3}{1} ,\,\@pfstepnum{7}{2} ,\,\@pfstepnum{7}{1}\  {\DEF}
 VotedForIn ,\,}%
\@x{\@s{38.44} TypeOK ,\, Messages ,\, Send}%
 \@x{\@s{20.0}\@pfstepnum{7}{3.} {\CASE} m . from \.{=} a \.{\land} ( m1 . bal
 \.{\geq} aBal [ a ] )}%
 \@x{\@s{24.0}\@pfstepnum{8}{1.}\  \A\, m2 \.{\in} msgs \.{'}
 \.{\,\backslash\,} msgs \.{:} m2 . bal \.{=} m1 . bal\@s{2.0}
 {\BY}\@pfstepnum{5}{1} ,\,\@pfstepnum{7}{3} ,\,\@pfstepnum{6}{1}
 ,\,\@pfstepnum{7}{3}\  {\DEF} Send ,\, TypeOK}%
 \@x{\@s{24.0}\@pfstepnum{8}{2.}\  \A\, m2 \.{\in} msgs \.{'}
 \.{\,\backslash\,} msgs \.{:} m2 . bal \.{\neq} b\@s{2.0}
 {\BY}\@pfstepnum{5}{1} ,\,\@pfstepnum{7}{3} ,\,\@pfstepnum{6}{1}
 ,\,\@pfstepnum{7}{3} ,\,\@pfstepnum{8}{1}\  {\DEF} TypeOK ,\, Messages}%
 \@x{\@s{24.0}\@pfstepnum{8}{}\  {\QED} {\BY}\@pfstepnum{7}{3}
 ,\,\@pfstepnum{7}{1} ,\,\@pfstepnum{8}{2}\  {\DEF} VotedForIn ,\, TypeOK ,\,
 Messages}%
 \@x{\@s{20.0}\@pfstepnum{7}{}\  {\QED} {\BY}\@pfstepnum{7}{2}
 ,\,\@pfstepnum{7}{3}\  {\DEF} TypeOK ,\, Messages}%
 \@x{\@s{16.0}\@pfstepnum{6}{}\  {\QED} {\BY}\@pfstepnum{6}{1}
 ,\,\@pfstepnum{6}{2}\  {\DEF} MsgInv1b}%
\iffalse
\begin{lcom}{14.5}%
\begin{cpar}{0}{F}{F}{0}{0}{}%
 \@pfstepnum{5}{3} proves \ensuremath{MsgInv2a'}. Because \ensuremath{Phase2b} sends \ensuremath{2b} messages, it uses invariance lemma for \ensuremath{SafeAt}.  %annie: the reason seems to be say anything.
\end{cpar}\end{lcom}
\fi
 \@x{\@s{12.0}\@pfstepnum{5}{3.}\  ( ( m . type \.{=}\@w{2a} ) \.{\implies}
 MsgInv2a ( m ) ) \.{'}\@s{2.0} {\BY} SafeAtStable ,\,\@pfstepnum{3}{1}
 ,\,\@pfstepnum{4}{4} ,\,\@pfstepnum{2}{1}\  {\DEF} MsgInv ,\, MsgInv2a ,\,}%
\@x{\@s{31.1} TypeOK ,\, Messages ,\, Phase2b ,\, Send}%
\begin{lcom}{14.5}%
\begin{cpar}{0}{F}{F}{0}{0}{}%
 \@pfstepnum{5}{4} proves \ensuremath{MsgInv2b'}. It uses two cases: the second case, \@pfstepnum{6}{2}, is for the increment \ensuremath{m}.
\end{cpar}\end{lcom}
 \@x{\@s{12.0}\@pfstepnum{5}{4.}\  ( ( m . type \.{=}\@w{2b} ) \.{\implies}
 MsgInv2b ( m ) ) \.{'}}%
 \@x{\@s{16.0}\@pfstepnum{6}{1.} {\CASE} {\lnot} ( m1 . bal \.{\geq} aBal [ a
 ] ) \.{\lor} m \.{\in} msgs\@s{2.0} {\BY}\@pfstepnum{5}{1}
 ,\,\@pfstepnum{3}{1} ,\,\@pfstepnum{6}{1}\  {\DEF} TypeOK ,\, Messages ,\,}%
\@x{\@s{34.44} Send ,\, MsgInv ,\, MsgInv2b}%
 \@x{\@s{16.0}\@pfstepnum{6}{2.} {\CASE} m1 . bal \.{\geq} aBal [ a ]
 \.{\land} m \.{\in} msgs \.{'} \.{\,\backslash\,} msgs\@s{2.0}
 {\BY}\@pfstepnum{5}{1} ,\,\@pfstepnum{3}{1} ,\,\@pfstepnum{6}{2}\  {\DEF}
 TypeOK ,\, Send ,\, MsgInv2b}%
 \@x{\@s{16.0}\@pfstepnum{6}{}\  {\QED} {\BY}\@pfstepnum{6}{1}
 ,\,\@pfstepnum{6}{2}\ }%
 \@x{\@s{12.0}\@pfstepnum{5}{}\  {\QED} {\BY}\@pfstepnum{5}{2}
 ,\,\@pfstepnum{5}{3} ,\,\@pfstepnum{5}{4}\  {\DEF} MsgInv2b}%
\begin{lcom}{10.5}%
\begin{cpar}{0}{F}{F}{0}{0}{}%
 \@pfstepnum{4}{5} proves \ensuremath{MsgInv'} for \ensuremath{Preempt}.  It uses the invariance lemma for \ensuremath{Preempt}, since \ensuremath{Preempt} does not send messages.
\end{cpar}\end{lcom}
 \@x{\@s{8.0}\@pfstepnum{4}{5.} {\CASE} \E\, p \.{\in} {\mathcal{P}} \.{:}
 Preempt ( p )\@s{2.0} {\BY}\@pfstepnum{4}{5} ,\, PreemptVotedForInv
 ,\,\@pfstepnum{3}{1} ,\, SafeAtStable ,\,}%
 \@x{\@s{26.44}\@pfstepnum{2}{1}\  {\DEF} Preempt ,\, MsgInv ,\, TypeOK ,\,
 Messages ,\, MsgInv1b ,\, MsgInv2a ,\, MsgInv2b}%
 \@x{\@s{8.0}\@pfstepnum{4}{}\  {\QED} {\BY}\@pfstepnum{4}{1}
 ,\,\@pfstepnum{4}{2} ,\,\@pfstepnum{4}{3} ,\,\@pfstepnum{4}{4}
 ,\,\@pfstepnum{4}{5} ,\,\@pfstepnum{2}{1}\  {\DEF} Next}%
 \@x{\@s{4.0}\@pfstepnum{3}{}\  {\QED} {\BY}\@pfstepnum{3}{1}
 ,\,\@pfstepnum{3}{2} ,\,\@pfstepnum{3}{3}\  {\DEF} Inv ,\, vars ,\, Next}%
 \@x{\@s{4.0}\@pfstepnum{2}{2.} {\CASE} {\UNCHANGED} vars\@s{2.0}
 {\BY}\@pfstepnum{2}{2}\  {\DEF} vars ,\, Inv ,\, TypeOK ,\, AccInv ,\,
 MsgInv ,\, VotedForIn ,\,}%
 \@x{\@s{22.44} SafeAt ,\, WontVoteIn ,\, MaxBalInSlot ,\, MsgInv1b ,\,
 MsgInv2a ,\, MsgInv2b}%
 \@x{\@s{4.0}\@pfstepnum{2}{}\  {\QED} {\BY}\@pfstepnum{2}{1}
 ,\,\@pfstepnum{2}{2}\ }%
 \@x{\@pfstepnum{1}{}\  {\QED} {\BY}\@pfstepnum{1}{1} ,\,\@pfstepnum{1}{2} ,\,
 \texttt{PTL} {\DEF} Spec}%
\@pvspace{8.0pt}%
\begin{lcom}{0}%
\begin{cpar}{0}{F}{F}{0}{0}{}%
\ensuremath{Safety} asserts that \ensuremath{Spec} implies that \ensuremath{Safe} always holds.%It uses (1) theorem \ensuremath{Invariant} which states that \ensuremath{Spec} implies all the invariants in \ensuremath{Inv} always hold and (2) \ensuremath{Inv \Rightarrow Safe}.
\end{cpar}%
\end{lcom}%
\@x{ {\THEOREM} Safety \.{\defeq} Spec \.{\implies} {\Box} Safe}%
\@x{\@pfstepnum{1}{}\  {\USE} {\DEF} {\mathcal{B}}}%
\@x{\@pfstepnum{1}{1.}\  Inv \.{\implies} Safe}%
 \@x{\@s{4.0}\@pfstepnum{2}{}\  {\SUFFICES} {\ASSUME} Inv ,\, {\NEW} v1
 \.{\in} {\mathcal{V}} ,\,\@s{2.0} {\NEW} v2 \.{\in} {\mathcal{V}} ,\, {\NEW} s \.{\in}
 {\mathcal{S}} ,\, {\NEW} b1 \.{\in} {\mathcal{B}} ,\, {\NEW} b2 \.{\in} {\mathcal{B}} ,\,}%
 \@x{\@s{82.91} ChosenIn ( b1 ,\, s ,\, v1 ) ,\, ChosenIn ( b2 ,\, s ,\, v2 )
 ,\, b1 \.{\leq} b2}%
\@x{\@s{52.31} {\PROVE} v1 \.{=} v2 \@s{4.0} {\BY} {\DEF} Safe ,\, Chosen}%
\@x{\@s{4.0}\@pfstepnum{2}{1.} {\CASE} b1 \.{=} b2}%
 \@x{\@s{8.0}\@pfstepnum{3}{1.}\  \E\, a \.{\in} {\mathcal{A}} \.{:} VotedForIn (
 a ,\, b1 ,\, s ,\, v1 ) \.{\land} VotedForIn ( a ,\, b1 ,\, s ,\, v2 )}%
\@x{\@s{27.1} {\BY}\@pfstepnum{2}{1} ,\, QuorumAssumption {\DEF} ChosenIn}%
 \@x{\@s{8.0}\@pfstepnum{3}{}\  {\QED} {\BY}\@pfstepnum{3}{1} ,\, VotedOnce
 {\DEF} Inv}%
\@x{\@s{4.0}\@pfstepnum{2}{2.} {\CASE} b1 \.{<} b2}%
 \@x{\@s{8.0}\@pfstepnum{3}{1.}\  SafeAt ( b2 ,\, s ,\, v2 )\@s{2.0} {\BY}
 VotedInv ,\, QuorumAssumption {\DEF} ChosenIn ,\, Inv}%
 \@x{\@s{8.0}\@pfstepnum{3}{2.}\  {\PICK} Q1 \.{\in} {\mathcal{Q}} \.{:} \A\, a
 \.{\in} Q1 \.{:} VotedForIn ( a ,\, b1 ,\, s ,\, v1 )\@s{2.0} {\BY} {\DEF}
 ChosenIn}%
 \@x{\@s{8.0}\@pfstepnum{3}{3.}\  {\PICK} Q2 \.{\in} {\mathcal{Q}} \.{:} \A\, a
 \.{\in} Q2 \.{:} VotedForIn ( a ,\, b1 ,\, s ,\, v2 ) \.{\lor} WontVoteIn (
 a ,\, b1 ,\, s )\@s{2.0} {\BY}\@pfstepnum{3}{1} ,\,\@pfstepnum{2}{2}\ 
 {\DEF} SafeAt}%
 \@x{\@s{8.0}\@pfstepnum{3}{}\  {\QED} {\BY}\@pfstepnum{3}{2}
 ,\,\@pfstepnum{3}{3} ,\, QuorumAssumption ,\, VotedOnce {\DEF} WontVoteIn
 ,\, Inv}%
 \@x{\@s{4.0}\@pfstepnum{2}{}\  {\QED} {\BY}\@pfstepnum{2}{1}
 ,\,\@pfstepnum{2}{2}\ }%
\@x{\@pfstepnum{1}{}\  {\QED} {\BY} Invariant ,\,\@pfstepnum{1}{1} ,\, \texttt{PTL}}%
\@x{}\bottombar\@xx{}%

}

%\includepdf[pages=-]{PaxosTLAPS-tla.pdf}
%\includepdf{./MultiPaxos.tla.pdf}
%\includepdf[pages={1-7}]{./MultiPaxos.tla.pdf}
%\includepdf[pages=-]{myfile.pdf}
%\includepdf[pages={1,3,5}]{myfile.pdf}

\notforThesis{
\end{document}
}